\documentclass[superscriptaddress, onecolumn,showpacs,
amssymb,amsmath,nobibnotes,aps,prd,
nofootinbib]{revtex4-2}
\pdfoutput=1
\usepackage{graphicx,subfigure,bm,color,psfrag,hyperref}
\usepackage{amsfonts}
\usepackage{lipsum}
\usepackage{mathtools}
\usepackage{verbatim}
\usepackage[normalem]{ulem}
\usepackage[dvipsnames]{xcolor}
\hypersetup{colorlinks,linkcolor={blue},citecolor={red},urlcolor={magenta}}   
\newcommand{\SP}[1]{\textcolor{blue}{[{\bf SP}: #1]}}

\begin{document}

\title{Phase space analysis of sign-shifting interacting dark energy models}

\author{Sudip Halder}
\email{sudiphalder197@gmail.com}
\affiliation{Department of Mathematics, Presidency University, 86/1 College Street, Kolkata 700073, India}

\author{Jaume  de Haro}
\email{jaime.haro@upc.edu}
\affiliation{Departament de Matem\`atiques, Universitat Polit\`ecnica de Catalunya, Diagonal 647, 08028 Barcelona, Spain}

\author{Tapan Saha}
\email{tapan.maths@presiuniv.ac.in}
\affiliation{Department of Mathematics, Presidency University, 86/1 College Street, Kolkata 700073, India}

\author{Supriya Pan}
\email{supriya.maths@presiuniv.ac.in}
\affiliation{Department of Mathematics, Presidency University, 86/1 College Street, Kolkata 700073, India}
\affiliation{Institute of Systems Science, Durban University of Technology, PO Box 1334, Durban 4000, Republic of South Africa}

\begin{abstract}
\noindent The theory of non-gravitational interaction between a pressure-less dark matter (DM) and dark energy (DE) is a phenomenologically rich cosmological domain which has received magnificent attention in the community. In the present article we have considered some interacting scenarios with some novel features: the interaction functions do not depend on the external parameters of the universe, rather, they depend on the intrinsic nature of the dark components;  the assumption of unidirectional flow of energy between DM and DE has been extended by allowing the possibility of bidirectional energy flow characterized by some sign shifting interaction functions; and  
the DE equation of state has been considered to be either constant or dynamical in nature. These altogether add new ingredients in this context, and, we performed the phase space analysis of each interacting scenario in order to understand their global behaviour. According to the existing records in the literature, this combined picture has not been reported elsewhere.  From the analyses, we observed that the DE equation of state as well as the coupling parameter(s) of the interaction models can significantly affect the nature of the critical points. It has been found that within these proposed sign shifting interacting scenarios, it is possible to obtain stable late time attractors which may act as global attractors corresponding to an accelerating expansion of the universe. The overall outcomes of this study clearly highlight that the sign shifting interaction functions are quite appealing in the context of cosmological dynamics and they deserve further attention. 

\end{abstract}

\maketitle

\section{Introduction}
\label{sec-introduction}

\noindent Over the last two decades, dynamics of the universe has been surprisingly thrilling due to the availability of a large amount of observational data.  At the end of nineties, observations from Type Ia supernovae first reported that our universe is passing through a phase of accelerated expansion \cite{SupernovaSearchTeam:1998fmf,SupernovaCosmologyProject:1998vns}  and this accelerated expansion is supposed to be driven by the presence of some exotic matter sector in our universe sector having large negative pressure. This exotic matter can be described in various ways. Two well known approaches are the modification of the matter sector of the universe in the context of Einstein's General Relativity, dubbed as dark energy  (DE) \cite{Copeland:2006wr,Frieman:2008sn,Bamba:2012cp}, or, the modification of the gravitational sector of the universe in various ways \cite{Nojiri:2006ri,Nojiri:2010wj,DeFelice:2010aj,Capozziello:2011et,Clifton:2011jh,Koyama:2015vza,Cai:2015emx,Nojiri:2017ncd,CANTATA:2021ktz,Bahamonde:2021gfp}, known as geometrical DE.  Apart from DE or geometrical DE, our universe sector also contains a non-luminous dark matter (DM)  fluid responsible for the observed structure formation of the universe.  And according to the high precision  
data from several astronomical missions, nearly, 68\% of the total energy density of the universe is occupied by either DE or geometrical DE and more or less 28\% of the total energy density of the universe is occupied by DM, that means,  nearly 96\% of the total energy budget of the universe is comprised by DE and DM. Thus, the dynamics of the universe is heavily dependent on the dark sector (DE$+$DM) of the universe. However, despite many astronomical missions, the  nature, origin and the evolution of the dark sector have remained mysterious so far and probing the physics of the dark sector has been one of the challenges for modern cosmology at the present moment. 
In order to describe the present universe, several cosmological models have been proposed and  investigated by several investigators.  Among these models, the $\Lambda$-Cold Dark Matter ($\Lambda$CDM) cosmological model constructed in the framework of GR in which the cosmological constant $\Lambda$ plugged into the Einstein's gravitational equations acts as the source of DE and DM is cold (pressure-less), has been found to be extremely successful in the light of a number of observational datasets. Nevertheless, $\Lambda$CDM  faces several theoretical and observational challenges, and therefore, a revision of the standard $\Lambda$CDM cosmology has been suggested in recent times.

A popular revision of the $\Lambda$CDM cosmology is the theory of non-gravitational interaction between DE and DM where an energy exchange phenomenon between these dark sectors, widely known as the interacting DE-DM, also known as interacting DE (IDE) or Coupled DE (CDE) cosmology. This particular theory received massive attention in the scientific community for many interesting consequences \cite{Amendola:1999er,Amendola:2003eq,Guo:2004xx,Gumjudpai:2005ry,Cai:2004dk,Wang:2005ph,Barrow:2006hia,Berger:2006fk,Setare:2007at,Rosenfeld:2007ri,Feng:2007wn,Boehmer:2008av,Olivares:2008bx,Valiviita:2008iv,MohseniSadjadi:2008na,Cruz:2008er,Wang:2008te,Caldera-Cabral:2008yyo,Gavela:2009cy,Caldera-Cabral:2009hoy,Valiviita:2009nu,LopezHonorez:2010esq,Yu:2010kh,Wei:2010uh,Duran:2010hi,Chen:2010ws,Cao:2010fb,Clemson:2011an,Marulli:2011jk,Baldi:2011wy,He:2011qn,Avelino:2012tc,Harko:2012za,Sun:2013pda,Chimento:2013rya,Li:2013bya,Yang:2014gza,Yang:2014okp,Yang:2014hea,Li:2014eha,Tamanini:2015iia,Goncalves:2015eaa,Pan:2012ki,Yang:2016evp,Nunes:2016dlj,Yang:2017yme,Dutta:2017kch,DiValentino:2017iww,Santos:2017bqm,Mifsud:2017fsy,Yang:2017ccc,Linton:2017ged,Yang:2017zjs,Yang:2017iew,Yang:2018euj,vonMarttens:2018iav,Yang:2018uae,Li:2019loh,Mishra:2019uqm,Pan:2019gop,DiValentino:2019ffd,vonMarttens:2019ixw,Cheng:2019bkh,DiValentino:2019ffd,Pan:2020zza,Pan:2020mst,DiValentino:2020kpf,BeltranJimenez:2020qdu,Yang:2021hxg,Gao:2021xnk,Lucca:2021eqy,Linton:2021cgd,Guo:2021rrz,Chatzidakis:2022mpf,Landim:2022jgr,Zhao:2022ycr,Gao:2022ahg,Hou:2022rvk,Zhai:2023yny,Li:2023fdk,Teixeira:2023zjt,Rodriguez-Benites:2023otm} (also see \cite{Bolotin:2013jpa,Wang:2016lxa,Wang:2024vmw}), such as, alleviating the cosmic coincidence problem \cite{Amendola:1999er,Cai:2004dk,Pavon:2005yx,Huey:2004qv,delCampo:2008sr,delCampo:2008jx}; crossing the phantom divide line without invoking any scalar field with negative sign in the kinetic part \cite{Wang:2005jx,Sadjadi:2006qb,Pan:2014afa}; weakening/solving  the Hubble constant tension \cite{Kumar:2016zpg, Kumar:2017dnp, DiValentino:2017iww,Yang:2018euj,Kumar:2019wfs, Pan:2019gop} between  Planck (within $\Lambda$CDM paradigm) \cite{Aghanim:2018eyx} and SH0ES (Supernovae and $H_0$ for the Equation of State of dark energy) \cite{Riess:2021jrx,Riess:2022mme}; and the clustering tension \cite{Pourtsidou:2016ico,An:2017crg,Kumar:2019wfs} between Planck (within $\Lambda$CDM model) and other astronomical probes at low redshifts, e.g. weak gravitational lensing and galaxy clustering~\cite{Asgari:2019fkq,KiDS:2020suj,Joudaki:2019pmv,DES:2021wwk,DES:2021bvc,DES:2021vln,KiDS:2021opn,Heymans:2020gsg,DES:2020ahh,Philcox:2021kcw}.  In IDE, the dynamics of the dark sector is mainly governed by the choice of a coupling/interacting function, $Q$, that controls the transfer of energy between DE and DM and this coupling function is taken from the phenomenological ground\footnote{Some attempts have been made to derive the coupling functions from an action integral \cite{Gleyzes:2015pma,Boehmer:2015kta,Boehmer:2015sha,vandeBruck:2015ida,DAmico:2016jbm,Pan:2020zza}, however, the final destination is yet to be discovered. Thus, at this moment, there is no reason to exclude any possible approach to study the theory of DM-DE interaction, even the approach adopts a phenomenological route.}.  Now, in the choice of the coupling functions, as they represent the transfer of energy and/or momentum  between the DM and DE sectors, $Q$ is usually assumed to be the functions of the energy densities of DE and DM.   In general, two varieties of the interaction functions are considered in the literature: (i) the interaction functions where  the Hubble rate, $H$, of the Friedmann-Lema\^{i}tre-Robertson-Walker (FLRW) universe explicitly appears, see for instance, \cite{Cai:2004dk,Wang:2005ph,Barrow:2006hia}; 
(ii) the interaction functions where $H$ does not appear explicitly, e.g. \cite{Valiviita:2008iv,Clemson:2011an,Yang:2021oxc}.  Concerning the above two approaches,  even though the interaction between these dark sectors is viewed as a  local phenomenon \cite{Valiviita:2008iv}, and the presence of the global expansion factor may be avoided, however, this debate is still unending and it is very hard to prefer the first approach over the other (see \cite{Pan:2020mst}).  On the other hand, one can put another question mark on the direction of energy flow between the dark sectors which is characterized by the choice of the interaction function. In a large class of interaction models, the flow of energy between the dark sectors is assumed to be unidirectional, that means, throughout the period of energy exchange mechanism between the dark sectors, the energy transfer can happen either from ``DE to DM'' or from ``DM to DE''. According to the theoretical and observational grounds, there are evidences of the energy  transfer from DE to DM \cite{He:2011qn,Yang:2018euj,Rodriguez-Benites:2023otm}, while the direction of energy flow can be reversed in future \cite{Joseph:2022khn}, but this conclusion depends on the underlying interaction model, properties of DE, and the observational data \cite{Wang:2021kxc,Hou:2022rvk}\footnote{We note that in Refs. \cite{Wang:2021kxc,Hou:2022rvk}, the authors have considered various interaction models and constrained them using different datasets and they reported that both the possibilities, that means the transfer of energy from DE to DM and from DM to DE are 
allowed according to the observational datasets. We further mention that the properties of DE (i.e. whether it is quintessence or phantom) and the direction of energy transfer between the dark components are connected with the stability of the interaction model at the level of perturbations, see for instance, Ref. \cite{Yang:2021hxg}. }, hence, this is one of the interesting questions in the context of interacting DE scenarios. Although these unidirectional interacting scenarios are simple by construction and they have been widely used in the community, however, it is very natural to examine  
whether the direction of energy transfer may alter during the course of energy exchange between the dark sectors. These kind of interaction models are known as sign changeable or sign shifting interaction functions and such models are appealing since they allow us to investigate whether the cosmologies with sign changeable interaction models are physically viable.  However, because of some unknown reasons, sign changeable interaction models did not get much 
 attention in the community \cite{Wei:2010fz,Wei:2010cs,Sun:2010vz,Li:2011ga,Guo:2017deu,Arevalo:2019axj,Pan:2019jqh,Arevalo:2022sne}, but such models are worth investigating in the light of current cosmological tensions \cite{Pan:2019jqh}. Interestingly,  model independent inference on the interaction between the dark sectors as performed in \cite{Escamilla:2023shf} hints for a sign shifting nature of the interaction function. This gives enough motivation to allow a sign changeable nature in the interaction functions and investigate the consequences.

In this article we therefore consider some sign shifting interaction models where the interaction functions depend only on the intrinsic nature of the dark fluids and perform their phase space analysis. As the choice of the interaction functions are not unique, thus, we have considered a variety of interaction functions that have been constructed using the known interaction functions in the literature. On the other hand, as the nature of DE is another unknown character to be discovered (hopefully) with the help of the  upcoming astronomical surveys, therefore, in this work,  in order to be inclusive we have considered that the equation of state of DE could be either constant or dynamical. Now, focusing on the dynamical equation of state of DE, one may have a cluster of possibilities since there is no unique route to determine its expression.  
Keeping this issue in mind, we have adopted a very well known dynamical equation of state of DE which depends only on its energy density and having only one free parameter which characterizes the nature  of the DE (i.e. where DE is quintessential or phantom) through its sign.  This equation of state has been extensively investigated in the cosmological dynamics and it recovers the usual barotropic equation of state of DE as a special case.   So far we are aware of the literature, the phase space analysis of the proposed sign shifting interacting functions considering both the constant and  dynamical equation of state parameters of DE has never been performed in the literature. This is the first time we are reporting the results in the literature.

The article is structured as follows. In section \ref{sec-basic-eqns} we introduce the basic equations of an interacting DM-DE model and then propose the models of interaction that we wish to study in this work. In section \ref{sec-dyn-systems} we construct the autonomous system corresponding to each interaction function and discuss the nature of the critical points obtained from the interaction functions and also their qualitative behaviour in terms of the cosmological parameters. 
Finally, in section \ref{sec-conclusion} we  close the article summarizing the key findings.

\section{Interacting Dark Energy}
\label{sec-basic-eqns}

\noindent We consider the homogeneous and isotropic universe where its gravitational sector is described by the Einstein's General Relativity (GR) and its matter distribution is minimally coupled to gravity. The matter sector consists of two heavy fluids of the universe, namely, a pressure-less (or cold) DM and a DE fluid which are interacting with each other in a non-gravitational way. In order to proceed with the mathematical structure of such scenario, we consider the spatially flat Friedmann-Lema\^{i}tre-Robertson-Walker (FLRW) metric

\begin{eqnarray}
	ds^2=-{dt}^2+{a^2(t)}{d\bf{x}}^2,
\end{eqnarray}
where $t$ is the co-moving time; $a(t)$ is the expansion scale factor of the universe; ${d\bf x}^2$ represents the 3-dimensional flat space line element. The Friedmann equations for the above line element  can be written as
\begin{eqnarray}
	3H^2={\kappa}^2(\rho_c+\rho_d),\label{friedmann-1}\\
	2\dot{H}+3H^2=-{\kappa}^2(p_c+p_d),\label{friedmann-2}
\end{eqnarray}
where an overhead dot denotes the derivative with respect to the cosmic time; $\kappa^2=8\pi G$, is the Einstein's gravitational constant; $H \equiv \dot{a} (t)/a (t)$, is the Hubble rate of the FLRW universe; ${\rho_d,~p_d}$ are respectively the energy density and pressure of the DE fluid obeying the barotropic equation-of-state $w_d= p_d/\rho_d<-1/3$; $\rho_c$, $p_c$ are respectively the energy density and pressure of DM in the form of dust, i.e., $p_c = 0$, henceforth, we call this DM as cold DM, abbreviated as CDM.  As CDM and DE are interacting with each other, therefore, the conservation equations  of these dark fluids can be represented as

\begin{eqnarray}
\dot{\rho}_c + 3 H \rho_c = -Q (\rho_c, \rho_d),\label{cons-CDM}\\
\dot{\rho}_d + 3 H (1+w_d) \rho_d = +Q (\rho_c, \rho_d),\label{cons-DE}
\end{eqnarray}
where $Q(\rho_c, \rho_d)$ denotes the real valued interaction function (also known as the interaction rate) that corresponds to the transfer of energy and (or) momentum between these dark fluids. For  $Q (\rho_c, \rho_d) > 0$, energy flow occurs from DM to DE, and for $Q (\rho_c, \rho_d) <0$, energy flow occurs in the reverse direction, that means from DE to DM. The interaction function $Q (\rho_c, \rho_d)$ is the key ingredient of this scenario because it controls the dynamics of the universe at the background and perturbation levels. We notice that the conservation equations (\ref{cons-CDM}) and (\ref{cons-DE}) can be put in a different format leading to

\begin{eqnarray}
\dot{\rho}_c + 3 H \left(1+ w_{\rm c, eff} \right) \rho_c = 0,\label{cons-CDM-non-int}\\
\dot{\rho}_d + 3 H \left(1+ w_{\rm d, eff} \right) \rho_d  = 0,\label{cons-DE-non-int}
\end{eqnarray} 
which represent a non-interacting scenario of DM and DE with the effective equation-of-state parameters 
\begin{eqnarray}
    w_{\rm c, eff} = \frac{Q (\rho_c, \rho_d)}{3H \rho_c},\quad  w_{\rm d, eff} = w_d - \frac{Q (\rho_c, \rho_d)}{3H \rho_d},
\end{eqnarray}
from which one can notice that the effective nature of the DM equation of state could be non-cold in the sense that the effective equation of state of DM could be non-zero (i.e. $w_{\rm c, eff} \neq 0$) for $Q (\rho_c, \rho_d) \neq 0$, see for instance \cite{Pan:2022qrr}, and additionally, the effective nature of the DE equation of state could be either quintessential ($w_{\rm d, eff} > -1$) or phantom ($w_{\rm d, eff} < -1$) depending on the sign of $Q (\rho_c, \rho_d)$.

Now since the interaction function affects the evolution of both CDM and DE, henceforth,  the expansion rate of the universe $H$ will be equally affected and as a result
the cosmological parameters will be influenced as well. We introduce  the equation of state of the total fluid $w_{\rm tot} = \frac{{\rm total~pressure}}{{\rm total~energy~density}} = \frac{p_c + p_d}{\rho_c +\rho_d} = \frac{p_d}{\rho_c + \rho_d}$ (since $p_c = 0$)  and the deceleration parameter of the universe, $q = - (1+ \dot{H}/H^2)$, which take the following forms 

\begin{eqnarray}
    w_{\rm tot} = w_d \Omega_d, \quad \quad q = \frac{1}{2} (1+ 3w_d \Omega_d),
\end{eqnarray}
where $\Omega_d = \kappa^2 \rho_d/ 3H^2$ is the density parameter for DE and from the Friedmann equation (\ref{friedmann-1}), one can derive the density parameter for CDM, $\Omega_c ~(= \kappa^2 \rho_c/ 3H^2)$ as $\Omega_c = 1- \Omega_d$.

\subsection{Models}\label{subsec-models}

\noindent In this work we propose several interaction functions having the sign changing property during the evolution of the universe. One of the important features of all the interaction functions that we are going to propose in this section is that, all of them do not depend on the external parameters of the universe, rather they all depend on the intrinsic nature of the dark sector. 
The first interaction function in this series has the following form

\begin{eqnarray}\label{model1}
\mbox{Model I:} \quad Q_{\rm I} = \Gamma (\rho_c -\rho_d),
\end{eqnarray}
where $\Gamma$ is the coupling parameter of the interaction function measuring the strength of the interaction and it has the dimension of the Hubble rate $H$. As argued, $Q$ does not depend on the external parameters of the universe, e.g. the scale factor of the universe or its expansion rate, rather it depends on the intrinsic properties of DM and DE, namely, their energy densities, $\rho_c$ and $\rho_d$.  Hence, one may expect that this interaction model could offer some inherent nature of the dark components. We further note that $Q_{\rm I}$ may change its sign depending on the dominating role played by one of the fluids, that means, if the dominating role played by DM over DE ($\rho_c > \rho_d$) is suddenly altered, i.e., DE starts dominating over DM as in the late time accelerating phase ($\rho_d > \rho_c$), then $Q_{\rm I}$ shifts its sign.

We generalize the sign-shifting interaction function of eqn. (\ref{model1})
as follows 

\begin{eqnarray}\label{model2}
\mbox{Model II:} \quad Q_{\rm II} = \Gamma_c \rho_c -\Gamma_d \rho_d,
\end{eqnarray}
where $\Gamma_c$ and $\Gamma_d$ are the coupling parameters of the interaction function having the dimension equal to the dimension of the Hubble parameter. 
For the interaction function $Q_{\rm II}$ to be sign changeable, both the coupling parameters $\Gamma_c$, $\Gamma_d$ should have the same sign, that means either $\Gamma_c >0$, $\Gamma_d > 0$ or $\Gamma_c < 0$, $\Gamma_d <0$, but never $\Gamma_c \Gamma_d <0$. Similar to the earlier interaction function, this model also does not include any external parameters of the universe.

The next model in this series that we introduce has the following form 

\begin{eqnarray}\label{model3}
\mbox{Model III:} \quad Q_{\rm III} =\Gamma \left(\rho_c - \rho_d - \frac{\rho_c \rho_d}{\rho_c + \rho_d} \right),  
\end{eqnarray}
where as already noted, $\Gamma$ is the coupling parameter of the interaction function having the dimension equal to the dimension of the Hubble parameter. 
Notice that the interaction function (\ref{model3}) is obtained by including a new function $Q_{\rm new} = - \Gamma \rho_c \rho_d (\rho_c + \rho_d)^{-1}$ with the model of eqn. (\ref{model1}), that means $Q_{\rm III} = Q_{\rm I} + Q_{\rm new}$.  Similar to the earlier two interaction functions, here too, we notice that this interaction function depends only on the intrinsic nature of the dark fluids.

Lastly, we introduce two new interactions of the forms
\begin{eqnarray}\label{model4}
\mbox{Model IV:} \quad  Q_{\rm IV} = \Gamma_c \rho_c -\Gamma_{cd} \frac{\rho_c\rho_d}{\rho_c + \rho_d}, 
\end{eqnarray}
and 
\begin{eqnarray}\label{model5}
\mbox{Model V:} \quad  Q_{\rm V} = \Gamma_d \rho_d -\Gamma_{cd} \frac{\rho_c\rho_d}{\rho_c + \rho_d}, 
\end{eqnarray}
where $\Gamma_c$, $\Gamma_d$ and $\Gamma_{cd}$ are the coupling parameters and they all have the dimension of the Hubble parameter. 
One can easily notice that the last term of both Model IV (eqn. (\ref{model4})) and Model V (eqn. (\ref{model5})) are same but the models differ in their first terms containing $\rho_c$ and $\rho_d$, respectively. We further note that $\Gamma_c$, $\Gamma_d$ and $\Gamma_{cd}$ are all constants in such a way so that the models allow sign shifting property. That means, for Model IV (eqn. (\ref{model4})), $\Gamma_c$ and $\Gamma_{cd}$ will enjoy the same sign and for Model V (eqn. (\ref{model5})), $\Gamma_d$ and $\Gamma_{cd}$ will enjoy the same sign but $\Gamma_c \neq \Gamma_{cd}$ for Model IV and $\Gamma_d \neq \Gamma_{cd}$ for Model V, otherwise $Q_{\rm IV}$ and $Q_{\rm V}$ will represent the energy flow only in one direction.   
As the choice of the interaction function is not unique, therefore, one can construct a variety of such models, however, it should be kept in mind that the interaction functions may lead to negative energy densities of the dark sectors as argued in \cite{Yang:2021oxc}, hence, working with an arbitrary interaction function needs precaution.

\section{Dynamical Analysis of Interacting models}
\label{sec-dyn-systems}

\noindent The dynamical analysis of the interaction models is the heart of this work. The dynamical analysis plays a crucial role in understanding the local and global dynamics of the underlying cosmological scenarios.  In order to perform the dynamical analysis of the underlying interacting scenarios, one needs to define a new set of dimensionless variables in terms of which one can study the behaviour of the system of differential equations.  We refer \cite{Curbelo:2005dh,Amendola:2006qi,Quartin:2008px,Chen:2008pz,Boehmer:2008av,Chen:2008ft,Boehmer:2009tk,Mahata:2015lja,Shahalam:2015sja,Singh:2015rqa,Khurshudyan:2015mva,Bernardi:2016xmb,Carneiro:2017evm,Shahalam:2017fqt,Odintsov:2018awm,Aljaf:2019ilr,Paliathanasis:2019hbi,Hernandez-Almada:2020ulm,Kaeonikhom:2020fqs,
Biswas:2020een,Chakraborty:2020yfe,Sa:2021eft,Khyllep:2021wjd,Sa:2023coi} (also see the review article \cite{Bahamonde:2017ize} and the references therein) where several interacting scenarios have been studied through the dynamical system analysis.
In this section we shall describe the dynamical systems for the proposed sign-shifting interaction models. Additionally, we shall also show that the choice of the dimensionless variables is extremely important because for a wrong choice of these variables, one may not be able explore the whole space of critical points.

\subsection{Model I}
\label{subsec-modelI}

\noindent We begin our analysis with the first interaction function $Q_{\rm I}$ of eqn.  (\ref{model1}) and 
we define  the following dimensionless variables
\begin{eqnarray}\label{eqn-dimless-variables}
	x=\frac{\kappa^2 \rho_c}{3H^2}, \quad \quad y=\frac{H_0}{H},
\end{eqnarray}
where $H_0$ ($>0$) is a constant and here it denotes the present value of the Hubble parameter.\footnote{Note that instead of taking $H_0$, one may consider any $\widetilde{H}$ which is also constant so that $y = \widetilde{H}/H$ becomes dimensionless. } 
The density parameters for CDM, DE, the equation state of the total fluid, $w_{\rm tot}$, and the deceleration parameter $q $   can be expressed in terms of the dimensionless parameters as follows 
\begin{eqnarray*}
&& \Omega_c=x, \quad \quad \quad \quad \quad \quad \Omega_d=1-x,\nonumber \\
&& w_{\rm tot}=w_d (1-x), \quad \quad  q=\frac{1}{2}[1+3w_d (1-x)]. 
\end{eqnarray*}
Now, inserting the dimensionless variables of eqn. (\ref{eqn-dimless-variables}) into the equations of motion  (\ref{friedmann-1}), (\ref{friedmann-2}), (\ref{cons-CDM}), and using the interaction function $Q_{\rm I}$ of eqn. (\ref{model1}), we have the following autonomous system: 
 \begin{eqnarray}
 && x'=-\gamma y(2x-1)+3w_d x(1-x), \label{eq1}\\
 && y'=\frac{3}{2}y \left(1+w_d (1-x) \right) \label{eq2},
 \end{eqnarray}
where a prime over a variable denotes its differentiation with respect to $N=\ln (a/a_0)$ (here $a_0$ is the present value of the scale factor) and $\gamma$ is defined as $\gamma=\Gamma/H_0$, hence, $\gamma$ becomes a dimensionless parameter. 
The critical points of the autonomous system (\ref{eq1}) $-$ (\ref{eq2}) are obtained by solving the equations $x' =0$ and $y' = 0$. In this case we have three critical points $A_1=(0,0)$, $A_2=(1,0)$ and $A_3= \left(\frac{1+w_d}{w_d},-\frac{3(1+w_d)}{\gamma(2+w_d)} \right)$.

In order to better understand the dynamical system when the Hubble rate is small, it will be useful to use the variables $(x,H)$. Now, using the cosmic time derivative, the dynamical system for the variables $(x,H)$ becomes
\begin{eqnarray}\label{system}
    \left\{\begin{array}{ccc}
      \dot{x}   & =& -\Gamma(2x-1)+3w_dHx(1-x),  \\
      \dot{H}  & = &-\frac{3}{2}H^2(1+w_d(1-x)).
    \end{array}\right.
\end{eqnarray}
On can see that the autonomous system (\ref{system}) has only two critical points, namely,  
$A_0=(\frac{1}{2}, 0)$ and $\left(\frac{1+w_d}{w_d}, -\frac{\Gamma (2+w_d)}{3 (1+w_d)}\right)$ which is actually $A_3$ if we consider this critical point in terms of $(x, y)$ coordinates.
Here, it is important to understand that the point $A_0$ does not appear in the coordinates $(x,y)$, because it corresponds to $y=\infty$, and for the same reason the points $A_1$ and $A_2$ do not appear in the coordinates $(x,H)$, because they correspond to $H=\infty$. This clearly indicates that the above two autonomous systems are not giving the complete information.

For this reason it is important to get a new coordinate system where all the critical points appear. This could be done by introducing a new variable 
\begin{eqnarray}\label{new-variable}
z=\frac{y}{1+y} = \frac{H_0}{H+H_0},
\end{eqnarray}
where $H=0$ corresponds to $z=1$ and $H=\infty$ corresponds  to $z=0$. So, in the new coordinates $(x,z)$
the dynamical system (\ref{eq1}) $-$ (\ref{eq2}) becomes
\begin{eqnarray}\label{new-dynamical-system-model-I}
    \left\{\begin{array}{ccc}
x'   &=& -\gamma \left(\frac{z}{1-z} \right)(2x-1)+3w_dx(1-x),  \\
z'    &=& \frac{3}{2}(1-z)z(1+w_d(1-x)).
    \end{array}\right.
\end{eqnarray}
But, note that, the dynamical system (\ref{new-dynamical-system-model-I}) is singular at $z=1$ (i.e., $H=0$). Thus, in order to regularize it, and taking into account that the phase portrait does not change topologically when one multiplies the vector field by a positive function, we regularize it by multiplying the factor $(1-z)$, which leads to the regular dynamical system
\begin{eqnarray}\label{dy-sys-xz}
    \left\{\begin{array}{ccc}
      x'   &=& -\gamma {z}(2x-1)+3(1-z)w_dx(1-x),  \\
     z'    &=& \frac{3}{2}(1-z)^2z(1+w_d(1-x)). 
    \end{array}\right.
\end{eqnarray}
Thus, finally one can investigate the dynamical system (\ref{dy-sys-xz}) through the stability analysis of the critical points. The physical domain, namely $R$, is the square $R=[0,1]^2$,
which in order to be positively invariant, i.e., to ensure that a solution of the dynamical system with initial conditions in $R$ never leaves it,  one has to impose that $\gamma>0$ (i.e., $\Gamma>0$). Effectively, it follows from the equation (\ref{new-dynamical-system-model-I}) that close to $x=0$, one has $x'\cong \gamma z>0$ and close to $x=1$, one has $x'\cong -\gamma z<0$.
For $z=0$ and $z=1$, one has $z'=0$, meaning that, for $\gamma>0$,  the dynamical system never crosses the lines $x=0$, $x=1$, $z=0$ and $z=1$, that is,  
the domain $R$ is positively invariant. As the dynamical system contains the DE equation of state which could be either constant or dynamical, therefore, we aim to investigate both the cases separately. In the following we present our analyses for constant $w_d$ and dynamical $w_d$.

\subsubsection{Constant $w_d$}

\noindent Considering that the DE equation of state, $w_d$, is constant, 
in Table~\ref{table-model-I} we present the critical points of the dynamical system (\ref{dy-sys-xz}), their existence, stability, and the values of the cosmological parameters evaluated at those critical points. Now depending on the nature of $w_d$, it is comprehensible that the nature of the critical points will certainly be affected. Thus, in order to be precise, 
we divide the entire parameter space  of $w_d$ into three disjoint regions, namely, quintessence or non-phantom ($w_d> -1$), cosmological constant ($w_d = -1$) and phantom ($w_d< -1$). 
In the following we present how $w_d$ affects the nature of the critical points.

\begin{figure*}[htp]
	\centering
	\includegraphics[width=0.6\textwidth]{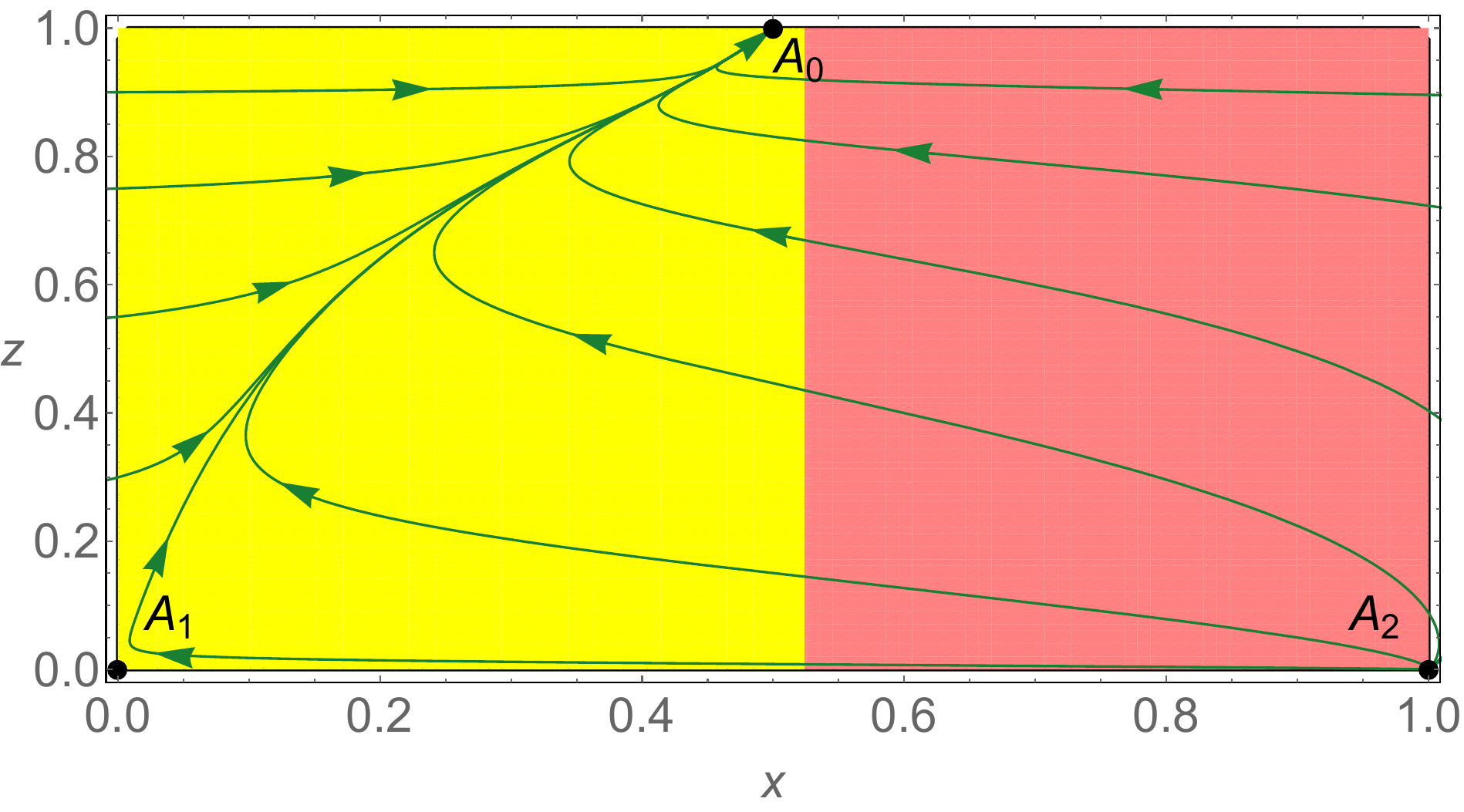}
	\caption{The phase portrait plot describing Model I (eqn. (\ref{model1})) with $w_d\geq-1$ and $\gamma>0$. In this case we have taken $w_d =-0.95$ and $\gamma = 0.4$. We note that one can take any value of $w_d \geq -1$ and any positive value of $\gamma$ in order to get similar graphics. Here, the yellow shaded region represents the accelerated region (i.e. $q<0$) and the pink shaded region corresponds to the decelerated region (i.e. $q>0$). } 
	\label{fig1:model-I}
\end{figure*}

\begin{figure*}
	\includegraphics[width=0.6\textwidth]{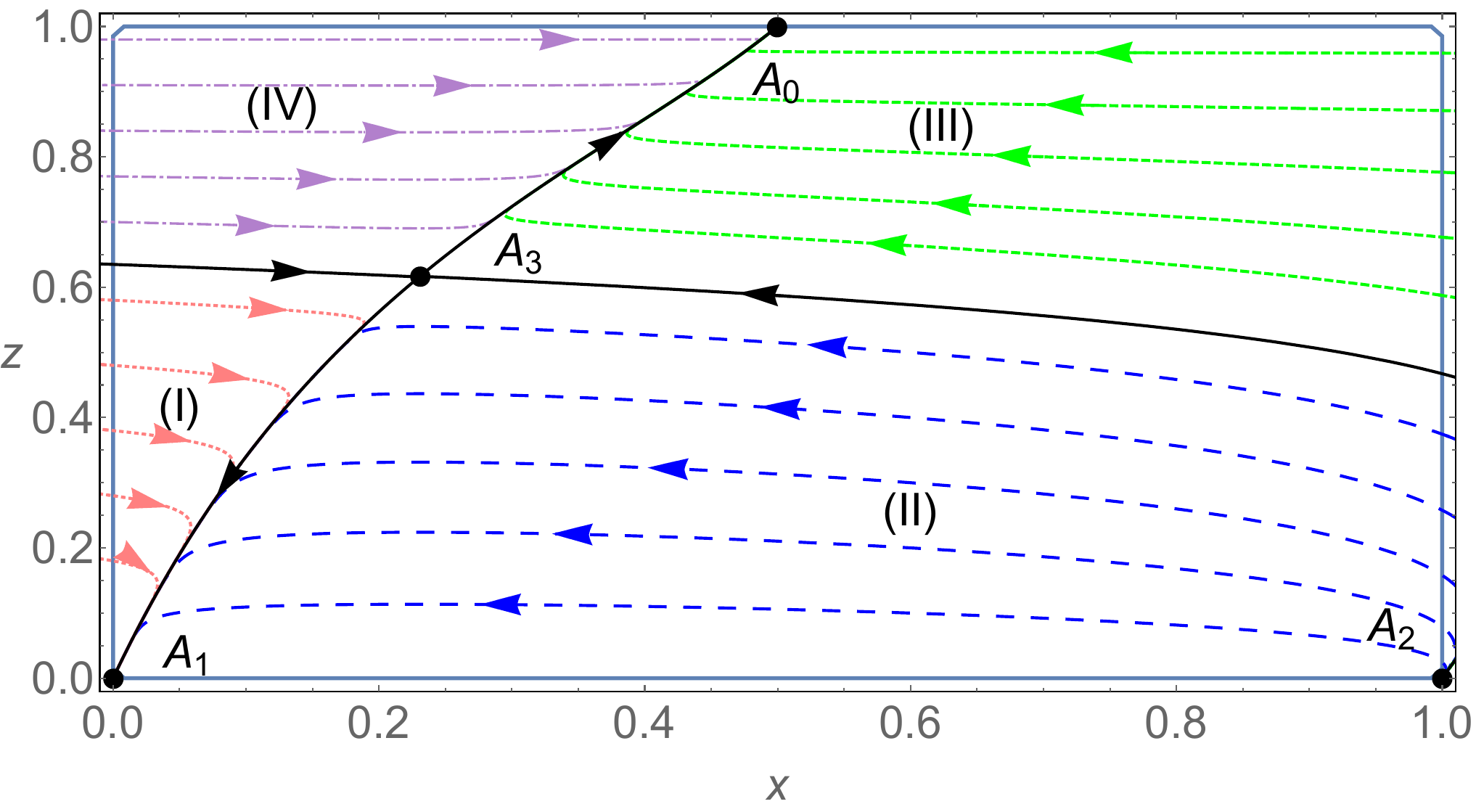}
	\caption{The phase portrait plot describing Model I (eqn. (\ref{model1})) for $-2< w_d < -1$ and $\gamma > 0$. In this case we have taken $w_d =-1.3$ and $\gamma = 0.8$. We note that one can take any specific value of $\gamma~(>0)$ to draw the plot, however, as long as $\gamma$ decreases, the regions I and IV become very small and they look indistinguishable from one another. } 
	\label{fig2:model-I}
\end{figure*}

\begin{figure*}
	\includegraphics[width=0.6\textwidth]{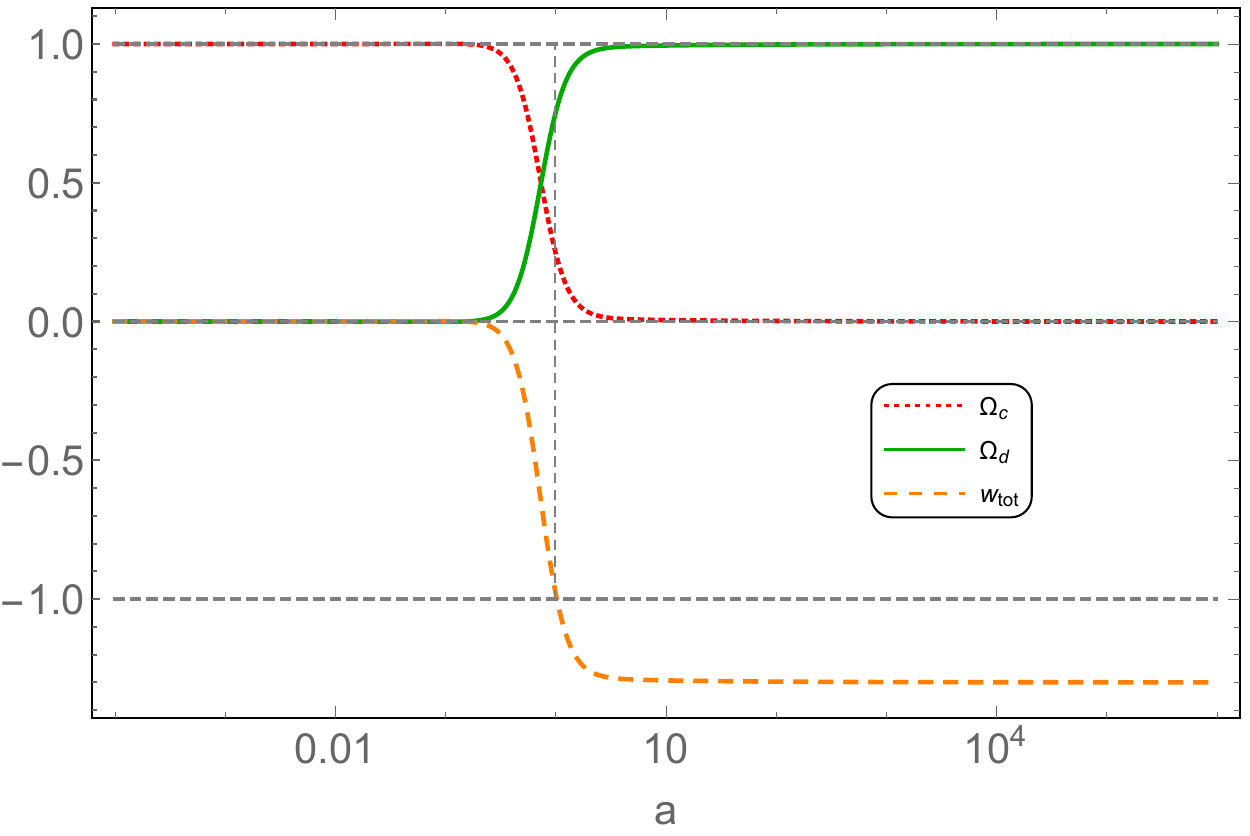}
	\caption{We show the evolution of the CDM density parameter ($\Omega_c$), dark energy density parameter ($\Omega_d$) and the total equation of state (EoS) parameter ($w_{\rm tot}$) for Model I (eqn. (\ref{model1})) for  $-2<w_d<-1$. We have taken $w_d=-1.3$, $\gamma=0.8$ with the initial conditions $x~(N =0) =0.25$, $z~(N =0) = 0.05$ taken from {\it region II} of Fig. \ref{fig2:model-I}. For the initial condition on $x(N)$ and $z(N)$ from the {\it region I} of Fig. \ref{fig2:model-I}, again we shall obtain $\Omega_c=0$ and $\Omega_d=1$ at late time. If we choose the initial conditions on $x(N)$ and $y(N)$ from {\it region III} and {\it IV} of Fig. \ref{fig2:model-I}, we shall reach $\Omega_c=\Omega_d=1/2$ in an asymptotic way. } 
	\label{fig3:model-I}
\end{figure*}

\begin{figure*}
	\includegraphics[width=0.6\textwidth]{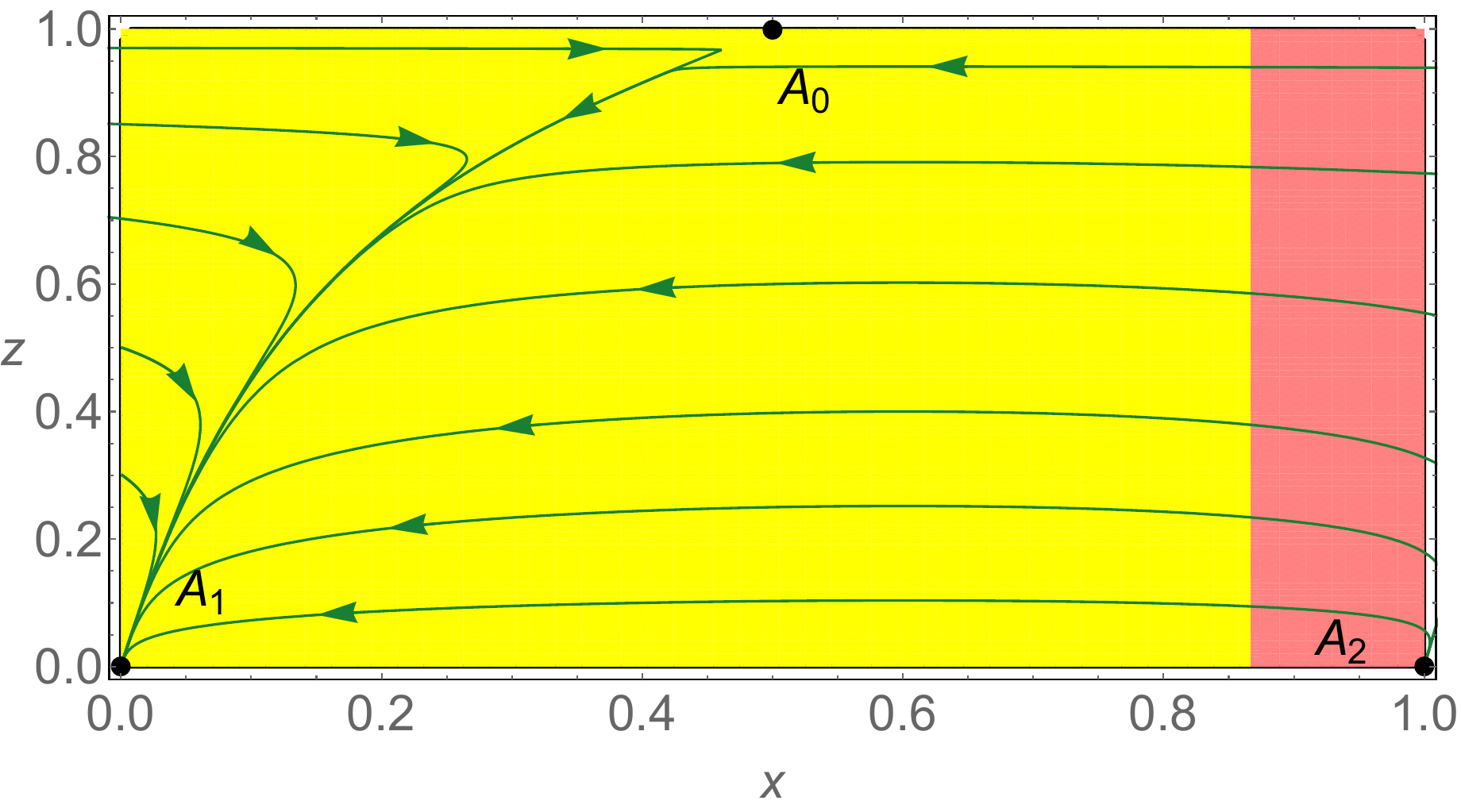}
	\caption{The phase portrait plot describing Model I of (\ref{model1}) with $w_d \leq -2$ and $\gamma>0$. In this case we have taken $w_d =-2.5$ and $\gamma = 0.8$. We note that one can take any value of $w_d \leq -2$ and any positive value of $\gamma$ in order to get similar graphics.  Here, the yellow shaded region represents the accelerated region (i.e. $q<0$) and the pink shaded region corresponds to the decelerated region (i.e. $q>0$) }
	\label{fig:model-I-wd-less-than-minus-2}
\end{figure*}

\begin{enumerate}

\item For non-phantom dark energy, i.e., when $w_d>-1$, the point $A_3$ does not belong to the physical domain $R$. In addition, 
since $w_{\rm tot}=w_d(1-x)$, this means that $w_{\rm tot}>-1$ and taking into account that 
$z'    = \frac{3}{2}(1-z)^2z(1+w_d(1-x))$, we have $z'>0$. So, $A_1$ and $A_2$ are unstable. On the line $z=1$, we have $z'=0$, $x'<0$ for $x>1/2$ and $x'>0$ for $x<1/2$. Thus, the critical point $A_0$ is stable, in fact, it is a global attractor. Therefore, at late times, the universe accelerates with $w_{\rm tot}=w_d/2>-1$, when $-1<w_d<-2/3$, and it decelerates when $-2/3<w_d<-1/3$. In addition, since $w_{\rm tot}>-1$ the Hubble rate decreases to zero, and $\Omega_d=\Omega_c=1/2$. The phase plot is shown in Fig. \ref{fig1:model-I}. This finishes the study for a non-phantom dark energy.

\item When  $w_d=-1$, one has $A_1=A_3$.
In this case, the linearization does not decide the nature of the critical point $A_1$. In fact, $A_1$ and $A_2$ are unstable because
$z'=\frac{3}{2}(1-z)^2zx>0$, for $0<z<1$ and $x>0$. Moreover, only the unphysical orbit $z=0$ ($H=\infty$) converges to $A_1$. Now, on $z=1$ line, we obtain $z'=0$, $x'<0$ for $x>1/2$ and $x'>0$ for $x<1/2$. Consequently, in this case $A_0$ is a global attractor. Again, the phase plot is given in Fig. \ref{fig1:model-I}.

\item For a phantom dark fluid, two cases arise:

\begin{enumerate}

\item When $-2<w_d< -1$: The critical point $A_3$ belongs to  the physical region  $R$. Clearly, we obtain $0<(1+w_d)/w_d<1/2$. Whenever $x<(1+w_d)/w_d$, we get $1+w_{\rm tot}<0$ and so, $z'$ is negative. Similarly, we have $1+w_{\rm tot}>0$ for $x>(1+w_d)/w_d$ and as a result $z'$ is positive. On $z=1$ line, we obtain $z'=0$, $x'<0$ for $x>1/2$ and $x'>0$ for $x<1/2$. Again, on $z=0$ line, $z'=0$ and $x'=3w_dx(1-x)$ which is negative. Therefore, our domain $R$ is divided into four regions. Thus, an orbit in the regions I and II of Fig. \ref{fig2:model-I}, at late times, converges to $A_1$. For an orbit in the regions III and IV, at late time, it converges to $A_0$. Note that $A_0$ means $H=0$ with $\Omega_c=\Omega_d=1/2$ and $w_{\rm tot}=w_d/2>-1$. On the contrary, $A_1$ means $H=\infty$ with $\Omega_d=1$ and $w_{\rm tot}=w_d<-1$. Fig. \ref{fig3:model-I} shows the evolution of the density parameters, namely, $\Omega_c$, $\Omega_d$, and the total equation of 
state parameter, $w_{\rm  tot}$.

\item When $w_d \leq -2$: For the particular case with $w_d=-2$, which represents a very high phantom regime, $A_3=A_0$. Here, $(1+w_d)/w_d=1/2$, so we have $z'<0$ for $x<1/2$ and $z'>0$ for $x>1/2$. On the line $z=0$, we obtain $z'=0$ and $x'<0$. Therefore, $A_0$, $A_2$ are unstable and $A_1$ becomes a global attractor. When $-2> w_d$, the critical point $A_3$ does not belong to the physical region $R$. Again, $(1+w_d)/w_d>1/2$ which implies that $z'<0$ if $x<(1+w_d)/w_d$ and $z'>0$ if $x>(1+w_d)/w_d$. Thus, $A_0$, $A_2$ are unstable and $A_1$ is a global attractor. Fig. \ref{fig:model-I-wd-less-than-minus-2} exhibits the behavior.

\end{enumerate}

\end{enumerate}

\begin{table*}
\begin{center}
	\begin{tabular}{|c|c|c|c|c|c|c|c|c|}
 \hline 
		Point &~  $x$ &~ $z$ &~ Existence  &~ Stability &~ Acceleration &~ $\Omega_{c}$ &~ $\Omega_{d}$ &~ $w_{\rm tot}$\\ 
  \hline 
  
		$A_0$ & $\frac{1}{2}$ & 1 &  for all $\gamma>0$ and $w_d<-1/3$ & stable for $-2<w_d<-1/3$ & $w_d< -2/3$ & $1/2$ & $1/2 $ & $w_d/2$\\ 
  \hline 
  
	$A_1$ & $0$ & $0$ & for all $\gamma>0$ and $w_d<-\frac{1}{3}$ & stable for $w_d<-1$ & $w_d<-\frac{1}{3}$ & $0$ & $1$ & $w_d$ \\ 	\hline 
 
       $A_2$ & $1$ & $0$ & for all $\gamma>0$ and $w_d<-\frac{1}{3}$ & unstable & no & $1$ & $0$ & $0$ \\ 

       \hline 
       
       $A_3$ & $\frac{1+w_d}{w_d}$ & $-\frac{3(1+w_d)}{2\gamma-3+w_d(\gamma-3)}$ & for all $\gamma>0$ and $-2\leq w_d \leq -1$ & unstable & yes & $\frac{1+w_d}{w_d}$ & $-\frac{1}{w_d}$ & $-1$ \\ 

       \hline 
	\end{tabular}
	\caption{The critical points, their existence, stability, and the values of the cosmological parameters evaluated at those points for the interacting scenario driven by the interaction function $Q_{\rm I} = \Gamma (\rho_c - \rho_d)$ of eqn. (\ref{model1})  are summarized. }
	\label{table-model-I}
\end{center}
\end{table*}


\subsubsection{Dynamical $w_d$}

\noindent The case with dynamical equation of state of DE, $w_d$, is interesting for two reasons. First of all,  the autonomous system (\ref{dy-sys-xz}) with constant $w_d$ is a special case of the dynamical $w_d$ case. On the other hand and most importantly, the dynamical $w_d$ scenario may offer a bigger space of critical points and hence one may expect new results in this context. In this article, we shall consider a parametric form for $w_d$ to investigate the autonomous system (\ref{dy-sys-xz}). The choice of the dynamical $w_d$ is not unique, and one can consider a variety of choices. 
The question then arises, what should be a possible choice for $w_d$ to proceed with the analysis? A possible choice for dynamical $w_d$ capturing a wide variety of models in this direction may take the
following form \cite{Nojiri:2005sx,Nojiri:2005sr}

\begin{eqnarray}\label{dynamical-eos}
    p_d = -\rho_d - f(\rho_d),
\end{eqnarray}
where $f$ is any analytic function of $\rho_d$. Notice from (\ref{dynamical-eos}) that, $w_d = p_d /\rho_d = -1 - f(\rho_d)/\rho_d$, describes a deviation from the cosmological constant $w_d =-1$ through the dynamical term $f(\rho_d)/\rho_d$.  
A general choice of this equation of state could be $p_d = -\rho_d - A \rho_d^n$, where $n$ and $A$ are constants in which $n$ is a dimensionless constant but $A$ has dimension. We restrict ourselves to $n =2$ in this article for which we have

\begin{eqnarray}\label{dynamical-eos-special}
    p_d = -\rho_d - A \rho_d^2 \quad \Longleftrightarrow \quad w_d = -1 - A \rho_d. 
\end{eqnarray}
 Now, for the equation of state (\ref{dynamical-eos-special}), the autonomous system (\ref{new-dynamical-system-model-I}) becomes,  
\begin{eqnarray}\label{autonomous-model-I-dyn-w1}
    \left\{\begin{array}{ccc}
      x'   &=& -\gamma \left(\frac{z}{1-z}\right)(2x-1)-3x(1-x)\left( 1+\nu \frac{(1-z)^2(1-x)}{z^2}\right),  \\
     z'    &=& \frac{3}{2}(1-z)z\left(x-\nu \frac{(1-z)^2(1-x)^2}{z^2}\right),
    \end{array}\right.
\end{eqnarray}
 where $\nu=\frac{3AH_0^2}{\kappa^2}$. We regularize the autonomous system (\ref{autonomous-model-I-dyn-w1}) by multiplying the factor $z^2(1-z)$ on the right hand sides of (\ref{autonomous-model-I-dyn-w1}) and finally obtain  
 \begin{eqnarray}\label{autonomous-model-I-dyn-w}
    \left\{\begin{array}{ccc}
      x'   &=& -\gamma {z^3}(2x-1)-3(1-z)x(1-x)\left( z^2+\nu (1-z)^2(1-x)\right),  \\
     z'    &=& \frac{3}{2}(1-z)^2z\left(x z^2-\nu (1-z)^2(1-x)^2\right).
    \end{array}\right.
\end{eqnarray}
Note that the qualitative behaviour for both the autonomous systems (\ref{autonomous-model-I-dyn-w1}) and (\ref{autonomous-model-I-dyn-w}) remain topologically equivalent. 
From the autonomous system (\ref{autonomous-model-I-dyn-w}), one can now find the critical points
by solving the equations $x' =0$ and $z' = 0$ and the critical points of the system (\ref{autonomous-model-I-dyn-w}) are

\begin{itemize}
    \item $\bar{A}_0  = (\frac{1}{2}, 1)$, \quad $\bar{A}_1  = (0, 0)$, \quad $\bar{A}_2  = (1, 0)$, \quad
    $S = \left\{\left(x_c, \frac{3 x_c}{3x_c-\gamma(2x_c-1)} \right)\right\}$,

    \end{itemize}
    where $S$ is the set of critical points in which 
    $x_c$ denotes a real root of $f(x) \equiv 9x^3-\nu {\gamma}^2 (2x-1)^2 (1-x)^2 = 0$.\footnote{We note that $f(x)$ can be obtained from the following two nullclines: 
\begin{align}
x z^2 - \nu (1-z)^2 (1-x)^2  = 0, \label{eqn-model-I-footnote-dyn}\\
- \gamma z^3 (2x-1) -3x (1-x) (1-z)\left[ z^2 + \nu (1-z)^2 (1-x)\right] = 0.
\end{align}} 
Since, $f(x)$ is a fourth degree equation in $x$, the set $S$ may contain maximum 4 critical points.

Now, as the physical domain in our case is $R = [0, 1]^2$, therefore, we are interested to investigate the number of roots of $f(x)$ in $[0,1]$.   
As $f(0) = -{\gamma}^2\nu<0$ (for $\nu >0$) and $f(1)=9>0$, hence, by the Bolzano's theorem\footnote{{Bolzano's theorem: Let $f$ be a real valued and continuous function in a compact interval $[k, l]$ in $\mathbb{R}$ and suppose that $f(k)$, $f(l)$ have opposite signs, that means $f(k) f(l) <0$. Then there is at least one point $m$ in $(k, l)$ such that $f(m) = 0$.} }\cite{Apostol:105425}, $f(x)$ will have at least one real root in $(0,1)$. Here we argue that $f$ will have only one real root in $(0, 1)$. It also follows that since $ 0 < x_c < 1$, therefore, the condition $z = \frac{3 x_c}{3x_c-\gamma(2x_c-1)} \leq 1$ leads to $x_c\leq 1/2$. 
Hence, our domain is slightly reduced and we need to check the number of roots of  $f(x)$ in $[0,1/2]$. Since, we have $f(1/2)=9/8>0$, it follows from the Bolzano's theorem \cite{Apostol:105425}, that there is at least one real root of $f(x)$ in $(0,1/2)$. Now, looking at the derivative of $f(x)$ with respect to $x$, given by $f' (x) = -16{\gamma}^2\nu (x-1)\left(x-\frac{3}{4}\right)\left(x-\frac{1}{2}\right)+27x^2$, one can check that $f'(x)>0$ for all $x\in(0,1/2)$. This shows that the function $f(x)$ is strictly increasing in $(0, 1/2)$ and consequently $f(x)$ has only one root in $(0,1/2)$. This concludes that the set $S$ has only one critical point in the physical domain $R$ and we label this critical point as $\bar{A}_3$. Now, in this case, we observe that the point $\bar{A}_3$ behaves qualitatively same as the point $A_3$ described earlier and correspondingly, the phase portrait is same as Fig. \ref{fig2:model-I}.

On the other hand, for $\nu<0$, the algebraic curve represented by the eqn. (\ref{eqn-model-I-footnote-dyn}) has no branches in the positive quadrant. Hence, for $\nu <0$, $S$ is an empty set, and as a result, the autonomous system for $\nu < 0$ admits only three critical points, namely, $\bar{A}_0$, $\bar{A}_1$, $\bar{A}_2$. The corresponding phase plot will be similar to Fig. \ref{fig1:model-I}.

\subsection{Model II}
\label{subsec-modelII}

\noindent We now consider the second interaction model in this series, i.e. $Q_{\rm II}$ of (\ref{model2}). Notice that $Q_{\rm II}$ has two coupling parameters $\Gamma_c$ and $\Gamma_d$. As demonstrated in section \ref{subsec-modelI}, here we shall consider the following dimensionless variables 

\begin{eqnarray}
	x=\frac{\kappa^2 \rho_c}{3H^2}, \quad \quad z= \frac{H_0}{H+H_0},
\end{eqnarray}
in order to understand the dynamics of the interacting scenario. With these choice of the dynamical variables, the 
autonomous system for this interaction function can be expressed as  
\begin{eqnarray}\label{autonomous-system-model-II-before-regularization}
\left\{
\begin{array}{ccc}
      x' &=&-\frac{z}{1-z}[\alpha x-\beta(1-x)]+3 w_d x(1-x),\\ 
	 
 z'    &=& \frac{3}{2}(1-z) z(1+w_d(1-x)),
 \end{array}\right.
\end{eqnarray}
where  $\alpha, \beta$ are the dimensionless parameters defined as $\alpha=\Gamma_c/H_0$, $\beta=\Gamma_d/H_0$. Now,  regularizing (\ref{autonomous-system-model-II-before-regularization}) we get 
\begin{eqnarray}\label{autonomous-system-model-II}
\left\{
\begin{array}{ccc}
      x' &=&-z[\alpha x-\beta(1-x)]+3(1-z)w_d x(1-x),\\ 
	 
 z'    &=& \frac{3}{2}(1-z)^2z(1+w_d(1-x)) .
 \end{array}\right.
\end{eqnarray}
The physical domain, namely $R$, is the square $R=[0,1]^2$, and it follows from the autonomous system (\ref{autonomous-system-model-II}) that along the lines $x=0$ and $x=1$, one has $x'=\beta z$ and $x'=-\alpha z$, and also the lines $z=0$ and $z=1$ remain invariant. Thus, to ensure that the physical domain $R$ is positively invariant we have to restrict our attention on $\alpha>0$ and $\beta>0$ (i.e., $\Gamma_c>0$ and $\Gamma_d>0$), and therefore, $Q_{\rm II}$ may allow a sign change during the evolution of the universe without exhibiting any unphysical properties in the energy densities of the dark sector.

Now, in a similar fashion we focus on two cases, namely, the constant $w_d$ and dynamical $w_d$. In the following we consider both the possibilities.

\subsubsection{Constant $w_d$}

\noindent Considering $w_d$ as a constant, in  Table~\ref{table-model-II} we summarize the critical points of the autonomous system (\ref{autonomous-system-model-II}), their existence, stability and  as well as the cosmological parameters evaluated at those critical points. Now we consider three different regions of $w_d$ as follows: 
if $w_d$ has quintessential nature (i.e. $w_d > -1$); if $w_d$ mimics a cosmological constant (i.e. $w_d =-1$); if $w_d$ has a phantom character ($w_d < -1$). In what follows we investigate each case. 

\begin{figure*}[htp]
	\centering
	\includegraphics[width=0.6\textwidth]{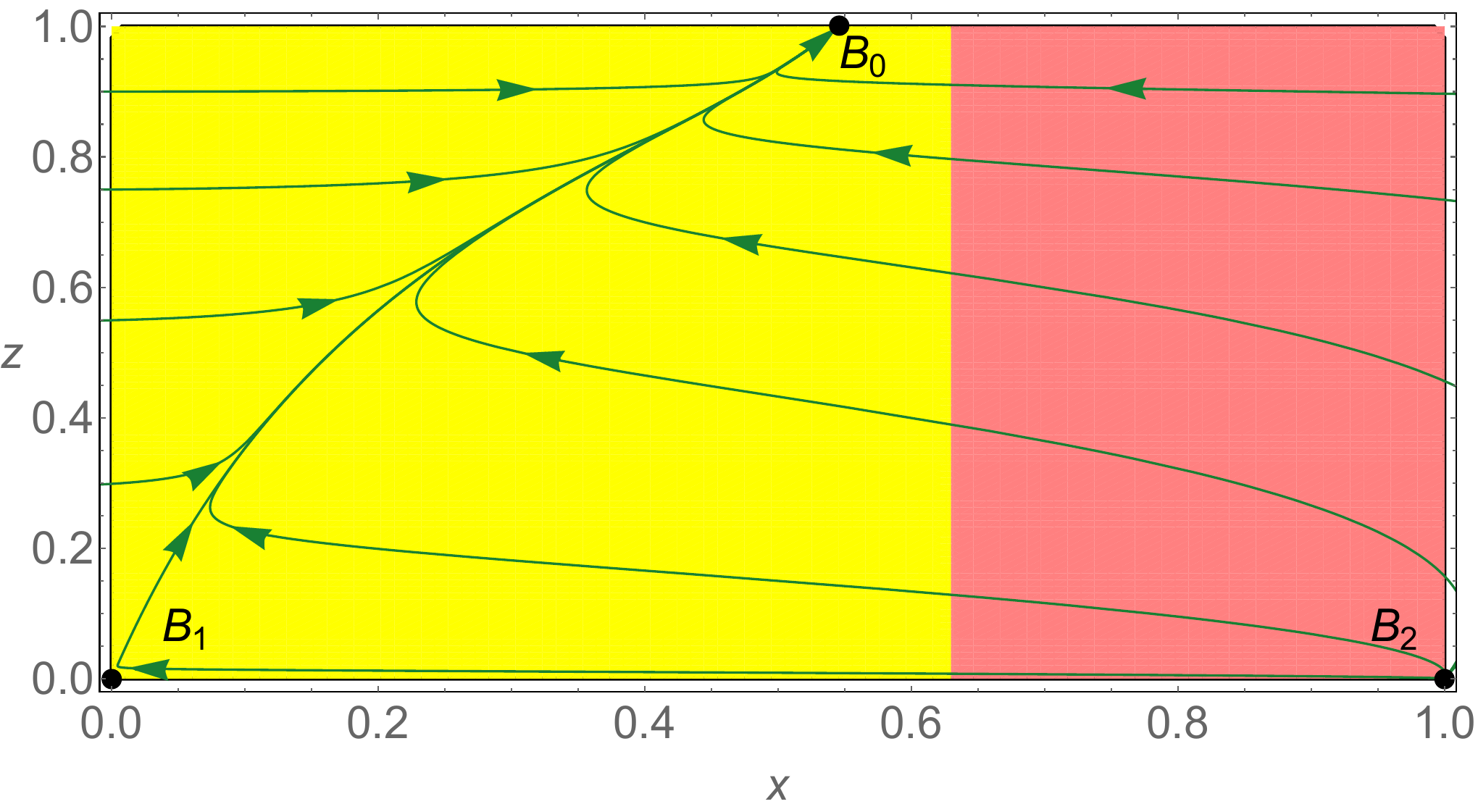}
	\caption{Phase portrait plot depicting Model II  (eqn. (\ref{model2}))  with $\alpha>0, \beta>0$ and $w_d\geq-1$. In this case we have taken $w_d =-0.9$, $\alpha = 0.5$ and $\beta=0.6$. We note that one can take any value of $w_d \geq -1$ and any positive value of $\alpha$ and $\beta$ in order to get similar graphics. Here, the yellow shaded region represents the accelerated region (i.e. $q<0$) and the pink shaded region corresponds to the decelerated region (i.e. $q>0$). }
	\label{fig5}
\end{figure*}

\begin{figure*}[htp]
	\centering
	\includegraphics[width=0.6\textwidth]{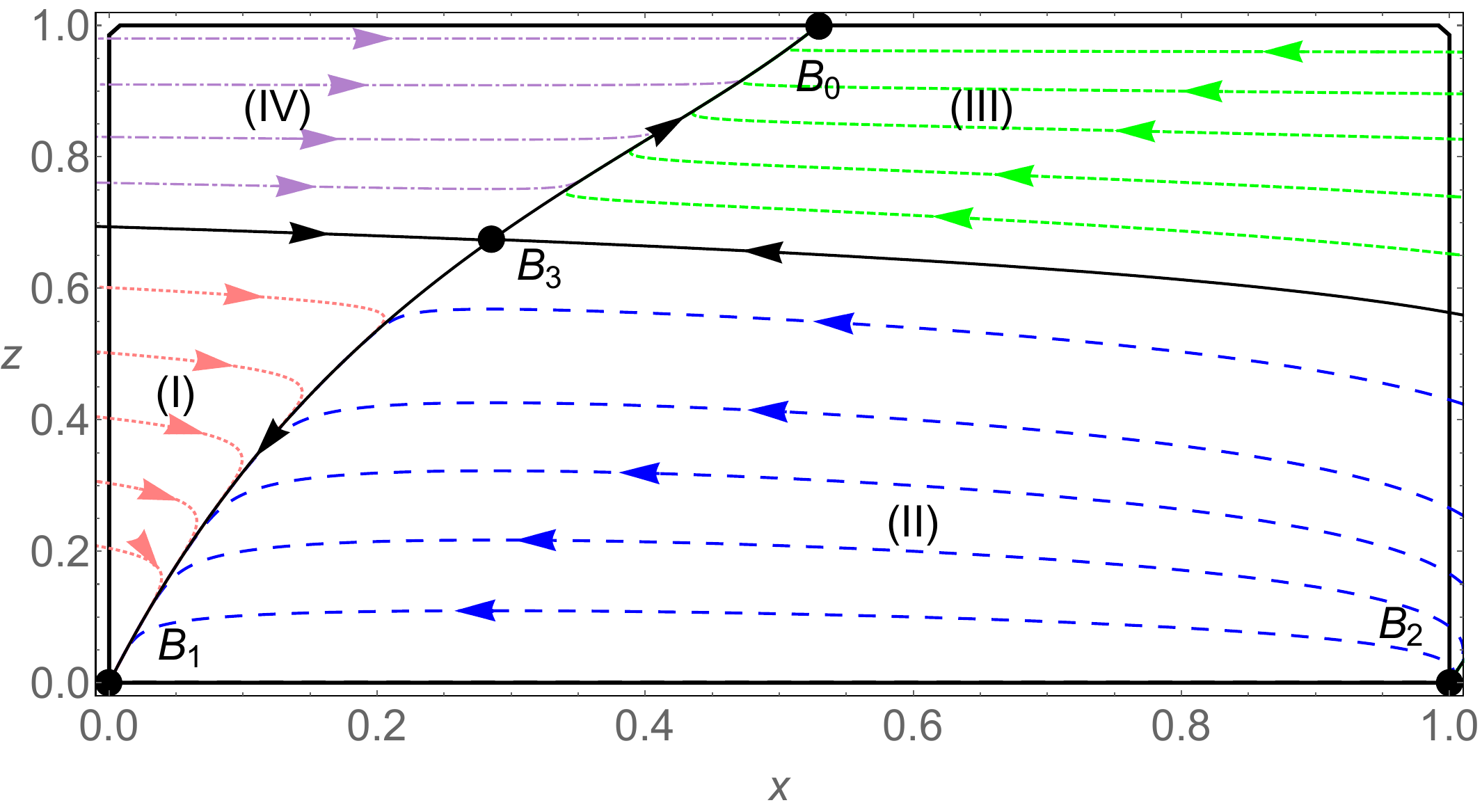}
	\caption{Phase plot for Model II (eqn. (\ref{model2})) with $\alpha>0, \beta>0$ and $-1-\frac{\beta}{\alpha}<w_d<-1$. Here, we have taken $w_d =-1.4$, $\alpha = 0.8$ and $\beta=0.9$. Note that one can take any specific value of $\alpha~( >0)$ and $\beta~(>0)$ to draw the plot, however, as long as $\alpha$ and $\beta$ decrease, the regions I and IV become very small and they look indistinguishable from one another. }
	\label{fig6}
\end{figure*}

\begin{figure*}[htp]
	\centering
	\includegraphics[width=0.6\textwidth]{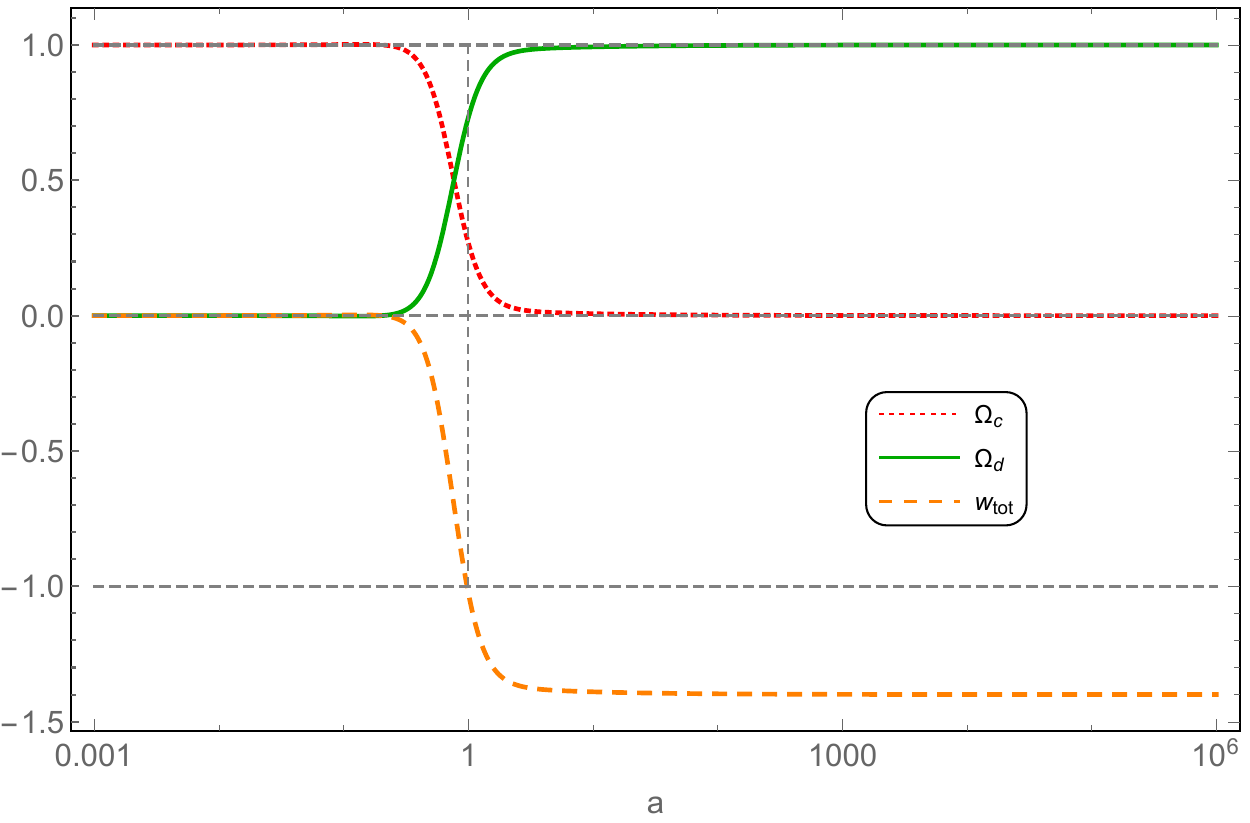}
	\caption{We show the evolution of the CDM density parameter ($\Omega_c$), dark energy density parameter ($\Omega_d$) and the total equation of state (EoS) parameter ($w_{\rm tot}$) for the Model II (eqn. (\ref{model2})). We have taken the following values of the parameters: $w_d=-1.4$, $\alpha=0.8$, $\beta=0.9$ and the following initial conditions $x~(N =0) =0.27$, $z~(N =0) = 0.08$ from the {\it region II} of Fig. \ref{fig6}. For the values of initial conditions on $x(N)$ and $y(N)$ from the {\it region I}, in a similar fashion we shall obtain $\Omega_c=0$ and $\Omega_d=1$ at late time. Again if we take initial conditions on $x(N)$ and $y(N)$ from the {\it regions III} and {\it IV} of Fig. \ref{fig6}, we shall reach $\Omega_c=\frac{\beta}{\alpha+\beta}$ and $\Omega_d=\frac{\alpha}{\alpha+\beta}$ in an asymptotic fashion. }
	\label{evo-plot-model2}
\end{figure*}

\begin{figure*}[htp]
	\centering
	\includegraphics[width=0.6\textwidth]{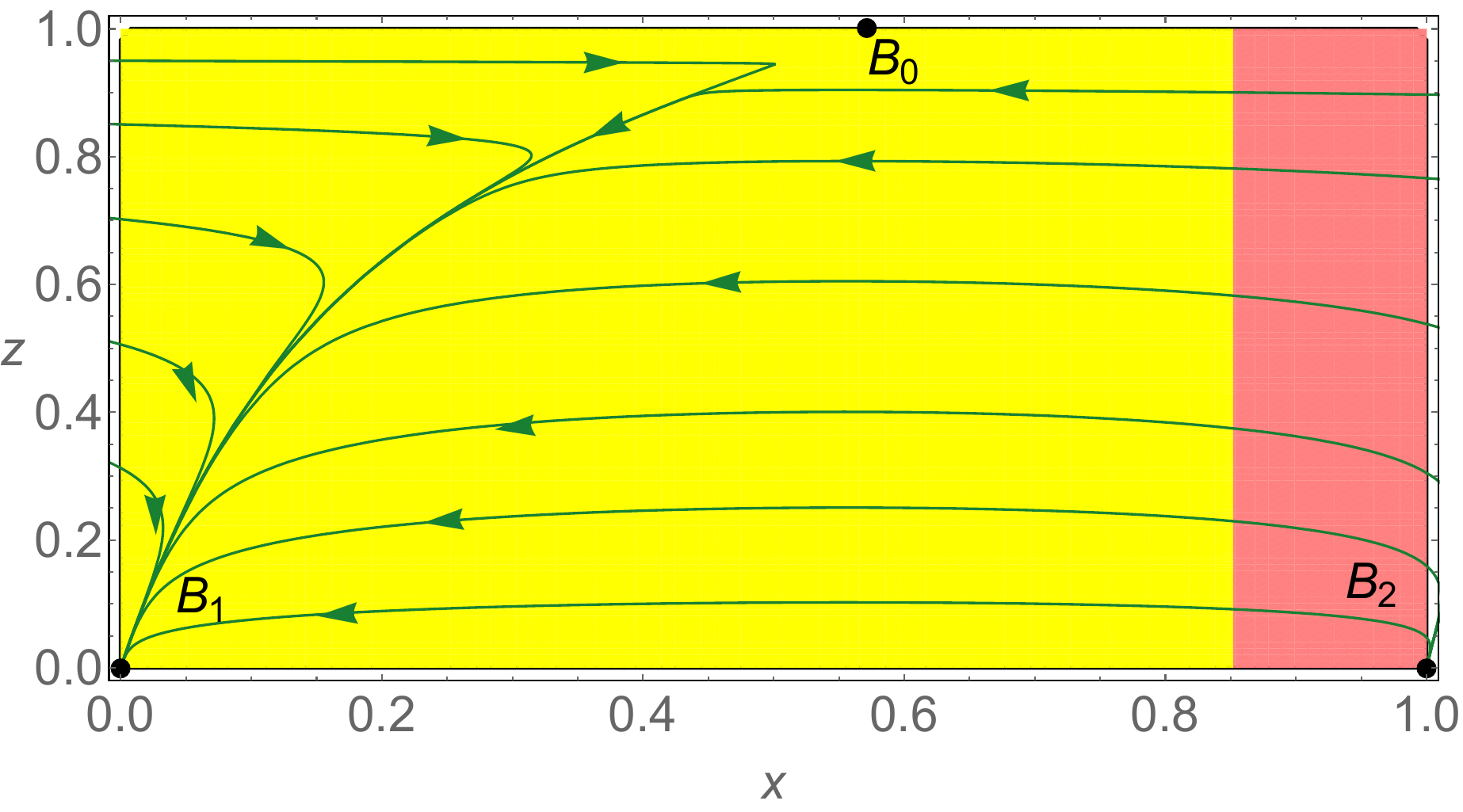}
	\caption{Phase portrait plot for Model II (eqn. (\ref{model2})) with $\alpha>0, \beta>0$ and $w_d\leq-1-\frac{\beta}{\alpha}$. In this case we have taken $w_d =-2.4$, $\alpha = 0.6$ and $\beta=0.8$. We note that one can take any value of $w_d \leq -1-\frac{\beta}{\alpha}$ and any positive value of $\alpha$ and $\beta$ in order to get similar graphics. Here, the yellow shaded region represents the accelerated region (i.e. $q<0$) and the pink shaded region corresponds to the decelerated region (i.e. $q>0$). }
	\label{fig7}
\end{figure*}

\begin{enumerate}

\item When $w_d>-1$, the point $B_3$ does not belong to the physical domain $R$. Now, we can see that on $z=1$ line, we have $z'=0$, $x'>0$ for $x<\frac{\beta}{\alpha+\beta}$ and $x'<0$ for $x>\frac{\beta}{\alpha+\beta}$. Also, we have $w_{\rm tot}=w_d(1-x)>-1$ which implies $z'=\frac{3}{2}z(1-z)^2(1+w_{\rm tot})>0$. So, $B_1$, $B_2$ are unstable critical points and $B_0$ is a global attractor. Note that, $B_0$ corresponds to $H=0$, $\Omega_c=\frac{\beta}{\alpha+\beta}$ and $\Omega_d=\frac{\alpha}{\alpha+\beta}$. The qualitative behaviour is displayed in Fig. \ref{fig5}.

\item When $w_d=-1$, we have $B_3=B_1$. Again, in this case, on the line $z=1$, one has $z'=0$, $x'>0$ for $x<\frac{\beta}{\alpha+\beta}$ and $x'<0$ for $x>\frac{\beta}{\alpha+\beta}$. Noting that $z'=\frac{3}{2} z (1-z)^2 x$ is positive. Thus, $B_1$, $B_2$ are unstable, and $B_0$ is a global attractor. Again, Fig. \ref{fig5} shows the phase plot.

\item Now dealing with a phantom dark energy fluid, we have the following results:
\begin{enumerate}
    \item When $-1-\frac{\beta}{\alpha}<w_d<-1$, the point $B_3$ belongs to the physical domain and we get $0<\frac{1+w_d}{w_d}<\frac{\beta}{\alpha+\beta}$. Also, we have $z'<0$ for $x<\frac{1+w_d}{w_d}$ and $z'>0$ for $x>\frac{1+w_d}{w_d}$. On $z=0$ line, $x'$ is negative. Again, on the line $z=1$, $x'$ is positive for $x<\frac{\beta}{\alpha+\beta}$ and $x'$ is negative for $x>\frac{\beta}{\alpha+\beta}$. Thus, we have two ``invariant stable orbits'' (see Fig. \ref{fig6}) which divide the physical domain in two parts.
    The orbits below these ``invariant manifolds'' converge to $B_1$ and the orbits above them converge to $B_0$. Fig. \ref{evo-plot-model2} displays the evolution of the density parameters, namely, $\Omega_c$, $\Omega_d$, and the total equation of 
    state parameter, $w_{\rm tot}$. 

    \item When $w_d=-1-\frac{\beta}{\alpha}$, we have $B_0=B_3$ and $\frac{1+w_d}{w_d}=\frac{\beta}{\alpha+\beta}$. For $x<\frac{1+w_d}{w_d}$, one has $z'<0$ and for $x>\frac{1+w_d}{w_d}$, $z'$ is positive. Since, $x'<0$ on $z=0$, $B_1$ is a global attractor. The qualitative behavior is given in Fig. \ref{fig7}.

    \item When $w_d<-1-\frac{\beta}{\alpha}$,
    the point $B_3$ does not belong to the physical domain $R$. Here, we obtain $\frac{1+w_d}{w_d}>\frac{\beta}{\alpha+\beta}$, $z'<0$ if $x<\frac{1+w_d}{w_d}$ and $z'>0$ if $x>\frac{1+w_d}{w_d}$. On $z=0$ line, we have $x'=3 w_d x (1-x)$ which is negative. Therefore, $B_0$, $B_1$ are unstable critical points and once again $B_1$ is a global attractor. Fig. \ref{fig7} exhibits the nature of phase portrait.    
    
    \end{enumerate}

\end{enumerate}

\begin{table*}
\begin{center}
	\begin{tabular}{|c|c|c|c|c|c|c|c|c|}\hline
 
		Point &  $x$ & $z$ & Existence  & Stability & Acceleration & $\Omega_{c}$ & $\Omega_{d}$ & $w_{\rm tot}$\\ \hline
  
		$B_0$ & $\frac{\beta}{\alpha+\beta}$ & 1 & $\alpha~(>0), \beta~(>0)$ and $w_d<-\frac{1}{3}$ &  $-1-\frac{\beta}{\alpha}<w_d<-\frac{1}{3}$ & $w_d<-\frac{1}{3}(1+\frac{\beta}{\alpha})$ & $\frac{\beta}{\alpha+\beta}$ & $\frac{\alpha}{\alpha+\beta} $ & $\frac{\alpha w_d}{\alpha+\beta}$\\ \hline 
  
	$B_1$ & $0$ & $0$ & $\alpha~(>0),\beta~(>0)$ and $w_d<- 
        \frac{1}{3}$ & $w_d<-1$ & $w_d<-\frac{1}{3}$ & $0$ & $1$ & $w_d$ \\ \hline
        
       $B_2$ & $1$ & $0$ & $\alpha~(>0),\beta~(>0)$ and $w_d<-\frac{1}{3}$ & unstable & no & $1$ & $0$ & $0$ \\ \hline
       
       $B_3$ & $\frac{1+w_d}{w_d}$ & $-\frac{3(1+w_d)}{(\alpha-3)w_d+(\alpha+\beta-3)}$ & $\alpha~(>0), \beta~(>0)$ and $-1-\frac{\beta}{\alpha}\leq w_d \leq -1$ & unstable & yes & $\frac{1+w_d}{w_d}$ & $-\frac{1}{w_d}$ & $-1$ \\ \hline
       
	\end{tabular}
	\caption{The critical points, their existence, stability, and the values of the cosmological parameters evaluated at those points for the interacting scenario driven by the interaction function $Q_{\rm II} = \Gamma_c \rho_c - \Gamma_d \rho_d$ of eqn. (\ref{model2}) 
 are summarized. }
	\label{table-model-II}
\end{center}
\end{table*}

\subsubsection{Dynamical $w_d$}

\noindent Now, consider the dynamical $w_d$ as in eqn. (\ref{dynamical-eos-special}), for which the autonomous system (\ref{autonomous-system-model-II-before-regularization}) takes the form:  
\begin{eqnarray}\label{autonomous-model-II-dyn-w1}
    \left\{\begin{array}{ccc}
      x'   &=& -\left(\frac{z}{1-z}\right)\left(\alpha x-\beta(1-x)\right)-3x(1-x)\left( 1+\nu \frac{(1-z)^2(1-x)}{z^2}\right),  \\
     z'    &=& \frac{3}{2}(1-z)z\left(x-\nu \frac{(1-z)^2(1-x)^2}{z^2}\right),
    \end{array}\right.
\end{eqnarray}
 where $\nu=\frac{3AH_0^2}{\kappa^2}$. We regularize the autonomous system (\ref{autonomous-model-II-dyn-w1}) by multiplying the factor $z^2(1-z)$ on the right hand sides of (\ref{autonomous-model-II-dyn-w1}) and finally obtain the following autonomous system which is topologically equivalent to (\ref{autonomous-model-II-dyn-w1}).
 \begin{eqnarray}\label{reg-autonomous-model-II-dyn-w1}
    \left\{\begin{array}{ccc}
      x'   &=& -{z^3}\left(\alpha x-\beta(1-x)\right)-3(1-z)x(1-x)\left( z^2+\nu (1-z)^2(1-x)\right),  \\
     z'    &=& \frac{3}{2}(1-z)^2z\left(x z^2-\nu (1-z)^2(1-x)^2\right).
    \end{array}\right.
\end{eqnarray}
Now, the critical points of the system (\ref{reg-autonomous-model-II-dyn-w1}) are 
\begin{itemize}
    \item $\bar{B}_0  = \left(\frac{\beta}{\alpha+\beta},1\right)$, \quad $\bar{B}_1  = (0, 0)$, \quad $\bar{B}_2  = (1, 0)$, \quad $S = \left\{\left(x_c,\frac{3 x_c}{3x_c-(\alpha+\beta)x+\beta}\right) \right\},$

\end{itemize}
where $S$ represents the set of critical points in which 
$x_c$ denotes a real root of $ h(x) \equiv 9x^3-\nu ((\alpha+\beta)x-\beta)^2 (1-x)^2 = 0$.\footnote{Note that $h(x)$ can be obtained from the following two nullclines:
\begin{align}
 x z^2 - \nu (1-z)^2 (1-x)^2  = 0,\label{eqn-model-II-footnote-dyn}\\ 
 -z^3 [\alpha x-\beta (1-x)] -3x (1-x) (1-z)\left[ z^2 + \nu (1-z)^2 (1-x)\right] = 0. 
 \end{align}}
 As $h (x)$ is a fourth degree equation in $x$, therefore, the set $S$ may contain maximum four critical points. Now, since our physical domain is $R$, therefore, we are interested to investigate the number of roots of $h(x)$ within the interval $[0,1]$. As $h(0)=-{\beta}^2\nu<0$ (for $\nu > 0$) and $h(1)=9>0$, hence, by the Bolzano's theorem \cite{Apostol:105425}, $h(x)$ will have at least one real root in $(0,1)$. Note that the $z$ component of the critical point should satisfy $0 \leq z \leq 1$.  Now, for any $x_c$ in $(0,1)$, the condition on the $z$ component, $\frac{3 x_c}{3x_c-(\alpha+\beta)x+\beta} \leq 1$ leads to $x_c\leq \beta/(\alpha+\beta)$. Consequently, we need to check the number of roots 
  of $h(x)$  in $[0,\frac{\beta}{\alpha+\beta}]$. Again, we notice that  $h(0)<0$ and $h\left(\beta/(\alpha+\beta) \right)=9 \left((\beta/(\alpha+\beta)\right)^3>0$ (since $\alpha >0$, $\beta >0$). Thus, from the Bolzano's theorem \cite{Apostol:105425}, we claim that there is at least one root of $h(x)$ in $\left(0, \beta/(\alpha+\beta) \right)$. Now, looking at the derivative of $h (x)$ given by 
    \begin{align*}
        h'(x)
         &= -4(\alpha+\beta)^2 \nu (x-1) \left(x-\frac{\beta}{\alpha+\beta} \right) \left(x-\frac{\alpha+2\beta}{2\alpha+2\beta} \right)+27x^2,
    \end{align*}
we can see that $h'(x)$ is positive in $\left(0, \beta/(\alpha+\beta) \right)$ i.e., $h(x)$ is strictly increasing in $\left(0, \beta/(\alpha+\beta) \right)$. Hence, $h(x)$ has only one root in $\left(0, \beta/(\alpha+\beta) \right)$ and correspondingly, the set $S$ contains only one critical point and we label this critical point as $\bar{B}_3$. 
In this case, we also observe that the qualitative nature of the critical point $\bar{B}_3$ is same as $B_3$ which has been described earlier and therefore, the phase portrait will be same as Fig. \ref{fig6}.

On the other hand, for $\nu<0$,  the algebraic curve represented by the eqn. (\ref{eqn-model-II-footnote-dyn}) has no branches in the positive quadrant. Hence, for $\nu <0$, $S$ does not have any critical point, that means, the autonomous system in this case has only three critical points, namely, $\bar{B}_0$, $\bar{B}_1$, $\bar{B}_2$ and the phase plot will be similar to Fig. \ref{fig5}.

\subsection{Model III}
\label{subsec-modelIII}

\noindent In this section we describe the dynamical analysis for the interacting scenario driven by the interaction function 
$Q_{\rm III}$ of (\ref{model3}). Using the same dynamical variables ($x$, $z$) defined as 

\begin{eqnarray}
	x=\frac{\kappa^2 \rho_c}{3H^2}, \quad \quad z= \frac{H_0}{H+H_0},
\end{eqnarray}
the autonomous system for this interacting scenario takes the form

\begin{eqnarray}\label{autonomous-system-model-III}
\left\{
\begin{array}{ccc}
	 x' &=&-\gamma \left(\frac{z}{1-z}\right) \left(x^2 + x - 1\right)+3w_d x(1-x),\\ 
	 z' &=&\frac{3}{2}(1-z)z\left(1+w_d (1-x)\right),
 \end{array}\right.
\end{eqnarray}
where $\gamma=\Gamma/H_0$ is the dimensionless parameter. Now, regularizing the vector fields, as we have described in \ref{subsec-modelI}, the autonomous system (\ref{autonomous-system-model-III}) can be reduced to the form 
\begin{eqnarray}\label{reg-autonomous-system-model-III}
\left\{
\begin{array}{ccc}
	 x' &=&-\gamma z \left(x^2 + x - 1\right)+3w_d(1-z) x(1-x),\\ 
	 z' &=&\frac{3}{2}(1-z)^2z\left(1+w_d (1-x)\right).
 \end{array}\right.
\end{eqnarray}
Here, $R=[0,1]^2$ is the physical domain and proceeding as in earlier we see that $R$ will be positively invariant if we restrict the parameter $\gamma$ by $\gamma>0$ (i.e. $\Gamma> 0$). 
We, now, investigate the autonomous system (\ref{reg-autonomous-system-model-III}) in terms of the nature of the critical points and their implications for both constant and dynamical $w_d$.

\subsubsection{Constant $w_d$}

\noindent For constant $w_d$, the critical points of the autonomous system (\ref{reg-autonomous-system-model-III}), their existence, stability and  as well as the cosmological parameters evaluated at those critical points are summarized in Table~\ref{table-model-III}. In the following we investigate the nature of the critical points for three different regions of $w_d$, namely, quintessence (i.e. $w_d > -1$); cosmological constant (i.e. $w_d =-1$); phantom (i.e. $w_d < -1$).

\begin{figure*}[htp]
	\centering
	\includegraphics[width=0.6\textwidth]{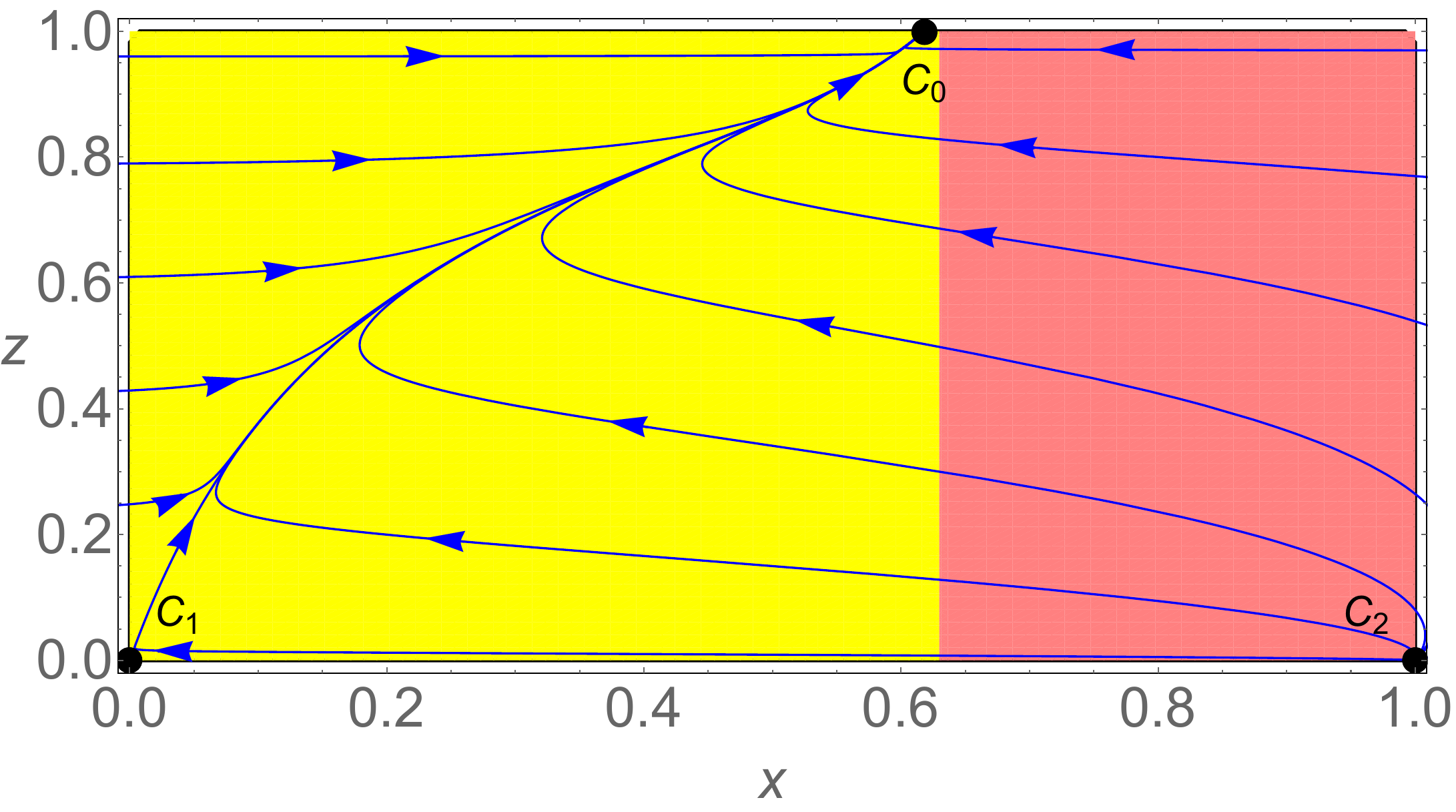}
	\caption{Phase plot for Model III (eqn. (\ref{model3})) with $\gamma>0$ and $w_d\geq-1$. In this case we have taken $w_d =-0.9$ and $\gamma = 0.5$. We note that one can take any value of $w_d \geq -1$ and any positive value of $\gamma$ in order to get similar graphics. Here, the yellow shaded region represents the accelerated region (i.e. $q<0$) and the pink shaded region corresponds to the decelerated region (i.e. $q>0$).  }
	\label{fig9}
\end{figure*}

\begin{figure*}[htp]
	\centering
	\includegraphics[width=0.6\textwidth]{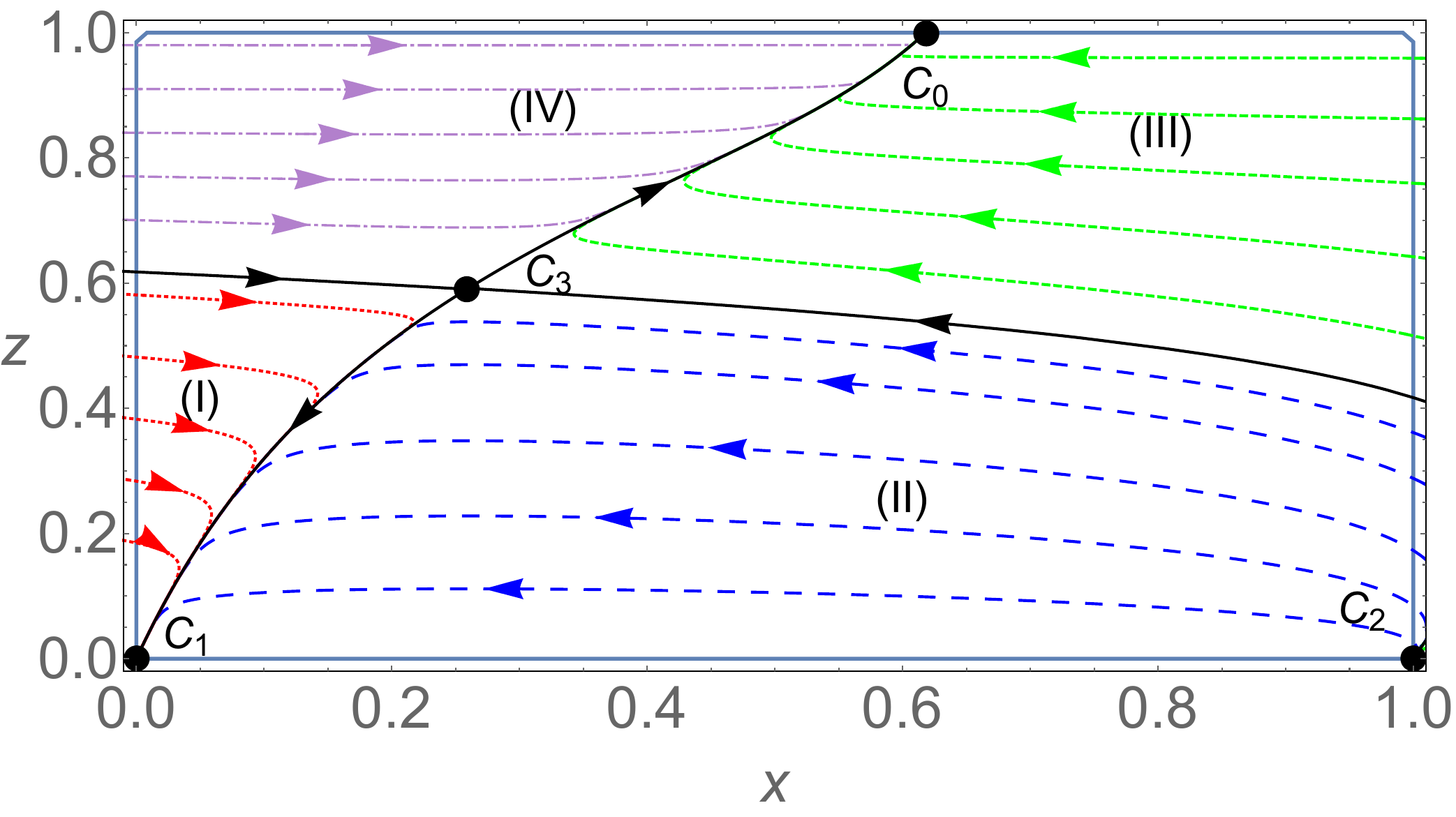}
	\caption{Phase plot for Model III (eqn. (\ref{model3})) with $\gamma>0$ and $-\frac{3+\sqrt{5}}{2}<w_d<-1$. In this case we have taken $w_d =-1.35$ and $\gamma = 0.8$. We note that one can take any specific value of $\gamma~( >0)$ to draw the plot, however, as long as $\gamma$ decreases, the regions I and IV become very small and they look indistinguishable from one another.  }
	\label{fig10}
\end{figure*}

\begin{figure*}
	\includegraphics[width=0.6\textwidth]{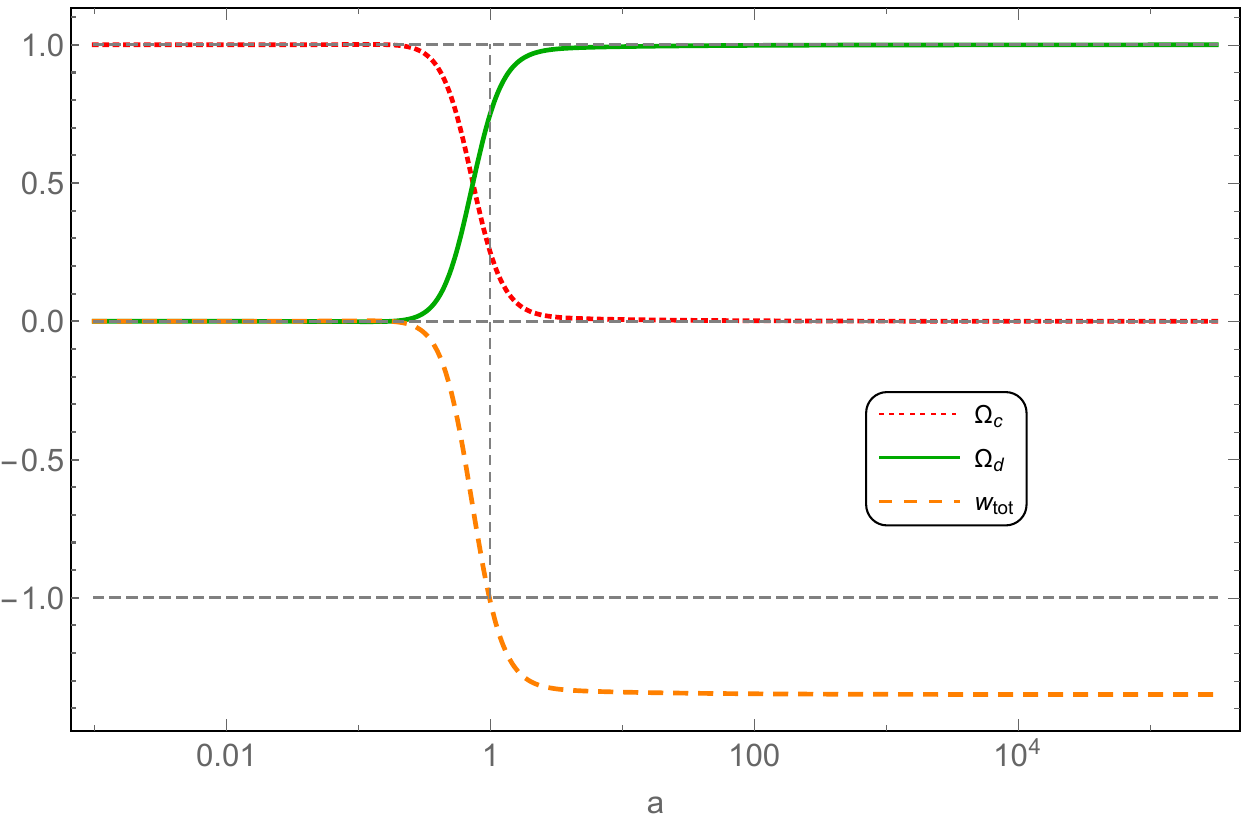}
	\caption{We display the evolution of the CDM density parameter $(\Omega_c)$, dark energy density parameter $(\Omega_d)$ and the total equation of state parameter $(w_{\rm tot})$ for the Model III (eqn. (\ref{model3})). We have chosen the following values of the parameters: $w_d=-1.35$, $\gamma=0.8$ and the following initial conditions: $x~(N= 0) =0.25$, $z~(N = 0) =0.07$ from the {\it region II} of Fig. \ref{fig10}. In addition, if we start any trajectory from {\it region I} of Fig. \ref{fig10}, it converges to the critical point $C_1$. So, for any initial conditions from {\it region I}, we shall get $\Omega_c=0$ and $\Omega_d=1$ at late time. Any trajectories starting from {\it regions III} and {\it IV} of Fig. \ref{fig10} will converge to the critical point $C_0$. Therefore, if we take initial conditions on $x(N)$ and $z(N)$ from {\it regions III} and {\it IV}, we shall reach $\Omega_c=\frac{\sqrt{5}-1}{2}$ and $\Omega_d=\frac{3-\sqrt{5}}{2}$ in an asymptotic fashion. }
	\label{evo-plot-model3}
\end{figure*}

\begin{figure*}[htp]
	\centering
	\includegraphics[width=0.6\textwidth]{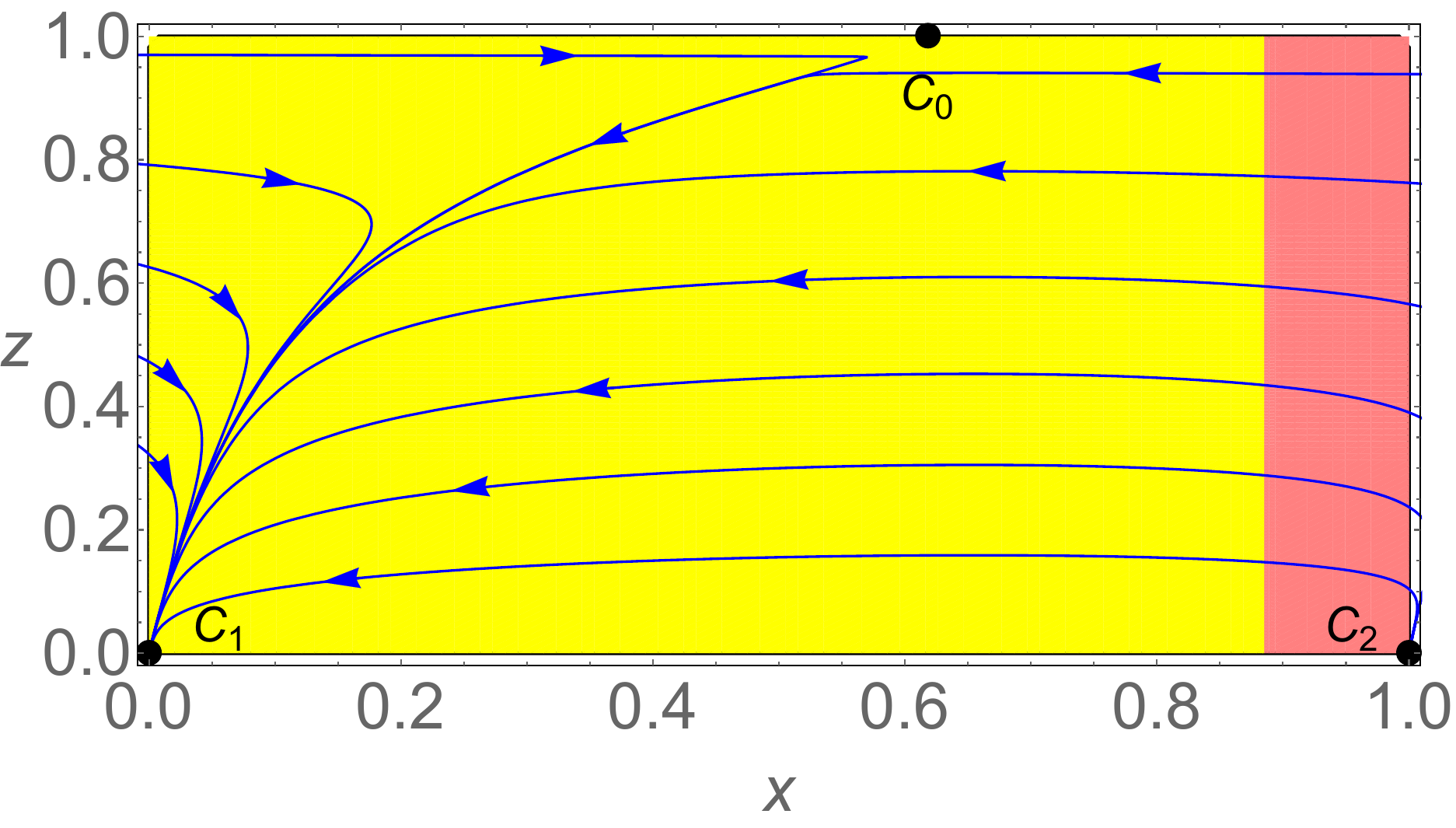}
	\caption{Phase plot for Model III (eqn. (\ref{model3})) with $\gamma>0$ and $w_d\leq-\frac{3+\sqrt{5}}{2}$. In this case we have taken $w_d =-2.9$ and $\gamma = 0.7$. We note that one can take any value of $w_d$ satisfying $w_d \leq -\frac{3+\sqrt{5}}{2}$ and any positive value of $\gamma$ in order to get similar graphics. Here, the yellow shaded region represents the accelerated region (i.e. $q<0$) and the pink shaded region corresponds to the 
 decelerated region (i.e. $q>0$). }
	\label{fig11}
\end{figure*}

\begin{enumerate}

\item  When $w_d>-1$, the point $C_3$ does not belong to the physical domain and $w_{\rm tot}=w_d(1-x)>-1$, which means $z'=\frac{3}{2} z (1-z)^2(1+w_{\rm tot})>0$. Now, on $z=1$, $x'$ is positive for $x<\frac{\sqrt{5}-1}{2}$ and $x'$ is negative for $x>\frac{\sqrt{5}-1}{2}$. So, $C_1$, $C_2$ are unstable and $C_0$ is a global attractor. Note that $C_0$ corresponds to $H=0$, $\Omega_c=\frac{\sqrt{5}-1}{2}$ and 
$\Omega_d=\frac{3-\sqrt{5}}{2}$. The phase plot is displayed in Fig. \ref{fig9}.

\item   When $w_d=-1$,  we have $C_1=C_3$, and $z'=\frac{3}{2} z (1-z)^2 x$ which is positive. On $z=1$ line, one has $x'>0$ for $x<\frac{\sqrt{5}-1}{2}$ and $x'<0$ for $x>\frac{\sqrt{5}-1}{2}$. Once again, $C_1$, $C_2$ are unstable critical points and the point $C_0$ continues being a global attractor. Again, the phase plot is shown in Fig. \ref{fig9}.

\item For $w_d < -1$, the parameter space can be categorized in two ways: 

\begin{itemize}
    \item When $-\frac{\sqrt{5}+3}{2}<w_d<-1$, the point $C_3$ enters in the physical domain and we obtain $0<\frac{1+w_d}{w_d}<\frac{\sqrt{5}-1}{2}$. Now, if $x<\frac{1+w_d}{w_d}$, implies $1+w_{\rm tot}<0$ which gives $z'=\frac{3}{2} z (1-z)^2(1+w_{\rm tot})<0$. Similarly, $z'>0$ for $x>\frac{1+w_d}{w_d}$. Again, on $z=1$, we obtain $x'>0$ for $x<\frac{\sqrt{5}-1}{2}$ and $x'<0$ for $x>\frac{\sqrt{5}-1}{2}$. Also, $x'$ is negative on $z=0$. Thus, the physical region $R$ is divided into four regions. Trajectories from regions I and II converge to $C_1$ and trajectories from regions III and IV converge to $C_0$. The Fig. \ref{fig10} shows the qualitative nature and Fig. \ref{evo-plot-model3} displays the evolution of $\Omega_c$, $\Omega_d$ and $w_{\rm tot}$.

    \item When $w_d\leq -\frac{\sqrt{5}+3}{2}$: For the special case with $w_d= -\frac{\sqrt{5}+3}{2}$, one has $C_0=C_3$ and $\frac{1+w_d}{w_d}=\frac{\sqrt{5}-1}{2}$. Now, $z'$ is positive for $x>\frac{1+w_d}{w_d}$ and $z'$ is negative for $x<\frac{1+w_d}{w_d}$. On $z=0$, $x'$ is negative which shows $C_1$ is a global attractor. Again for $w_d< -\frac{\sqrt{5}+3}{2}$, the point $C_3$ leaves the physical domain and one obtains $\frac{\sqrt{5}-1}{2}<\frac{1+w_d}{w_d}<1$. Also, one has $z'>0$ for $x>\frac{1+w_d}{w_d}$ and $z'<0$ for $x<\frac{1+w_d}{w_d}$. Thus, $C_1$ is a global attractor. In this case, the qualitative behavior is given in the Fig. \ref{fig11}.

\end{itemize}

\end{enumerate}

\begin{table*}
\begin{center}
	\begin{tabular}{|c|c|c|c|c|c|c|c|c|}\hline
 
		Point &  $x$ & $z$ & Existence  & Stability & Acceleration & $\Omega_{c}$ & $\Omega_{d}$ & $w_{\rm tot}$\\ \hline
  
		$C_0$ & $\frac{\sqrt{5}-1}{2}$ & 1 & $\gamma~(>0)$ and $w_d<-\frac{1}{3}$ & $-\frac{3+\sqrt{5}}{2}<w_d<-\frac{1}{3}$ & $w_d< -\frac{3+\sqrt{5}}{6}$ & $\frac{\sqrt{5}-1}{2}$ & $\frac{3-\sqrt{5}}{2} $ & $\frac{w_d(3-\sqrt{5})}{2}$\\ \hline 
  
	$C_1$ & $0$ & $0$ & $\gamma~(>0)$ and $w_d<-\frac{1}{3}$ & 
        $w_d<-1$ & $w_d<-\frac{1}{3}$ & $0$ & $1$ & $w_d$ \\ \hline	
 
       $C_2$ & $1$ & $0$ & $\gamma~(>0)$ and $w_d<-\frac{1}{3}$ & unstable & no & $1$ & $0$ & $0$ \\ \hline

       $C_3$ & $\frac{1+w_d}{w_d}$ & $\frac{3w_d(1+w_d)}{(3-\gamma){w_d}^2+3w_d(1-\gamma)-\gamma}$ & $\gamma~(>0)$ and $-\frac{3+\sqrt{5}}{2}\leq w_d \leq -1$ & unstable & yes & $\frac{1+w_d}{w_d}$ & $-\frac{1}{w_d}$ & $-1$ \\ \hline
       
	\end{tabular}
	\caption{The critical points, their existence, stability, and the values of the cosmological parameters evaluated at those points for the interacting scenario driven by the interaction function $Q_{\rm III} = \Gamma \left(\rho_c - \rho_d - \frac{\rho_c \rho_d}{\rho_c + \rho_d} \right)$ of eqn. (\ref{model3})  are summarized. }
	\label{table-model-III}
\end{center}
\end{table*}

\subsubsection{Dynamical $w_d$}

\noindent We consider the dynamical $w_d$ of the form (\ref{dynamical-eos-special}), for which the autonomuous system (\ref{autonomous-system-model-III}) becomes:  
\begin{eqnarray}\label{autonomous-model-III-dyn-w1}
    \left\{\begin{array}{ccc}
      x'   &=& -\gamma \left(\frac{z}{1-z}\right)\left(x^2+x-1\right)-3x(1-x)\left( 1+\nu \frac{(1-z)^2(1-x)}{z^2}\right),  \\
     z'    &=& \frac{3}{2}(1-z)z\left(x-\nu \frac{(1-z)^2(1-x)^2}{z^2}\right),
    \end{array}\right.
\end{eqnarray}
 where $\nu=\frac{3AH_0^2}{\kappa^2}$. The topologically equivalent autonomous system after regularization is given by the following:
 \begin{eqnarray}\label{reg-autonomous-model-III-dyn-w1}
    \left\{\begin{array}{ccc}
      x'   &=& -\gamma {z^3}\left(x^2+x-1\right)-3(1-z)x(1-x)\left( z^2+\nu (1-z)^2(1-x)\right),  \\
     z'    &=& \frac{3}{2}(1-z)^2z\left(x z^2-\nu (1-z)^2(1-x)^2\right).
    \end{array}\right.
\end{eqnarray}
The critical points of the system (\ref{reg-autonomous-model-III-dyn-w1}) are 

\begin{itemize}
    \item $\bar{C}_0  = \left(\frac{\sqrt{5}-1}{2}, 1\right)$, \quad $\bar{C}_1  = (0, 0)$, \quad $\bar{C}_2  = (1, 0)$, \quad $S = \left\{\left(x_c, \frac{3x_c}{3x_c-\gamma\left({x_c}^2+x_c-1\right)} \right) \right\},$

\end{itemize}
where $S$ is the set of critical points in which 
$x_c$ is a real root of $ g(x) \equiv 9 x^3 -{\gamma}^2 {\nu} \left(x^2+x-1\right)^2 (1-x)^2 = 0$.\footnote{Here $g(x)$ can be obtained from the following two nullclines:
\begin{align}
 x z^2 - \nu (1-z)^2 (1-x)^2  = 0,\label{eqn-model-III-footnote-dyn}\\ 
  - \gamma z^3 \left(x^2+x-1\right) -3x (1-x) (1-z)\left[ z^2 + \nu (1-z)^2 (1-x)\right] = 0. \end{align}} 
Note that as $g (x)$ is a sixth degree equation, therefore, the set $S$ may have maximum six critical points. Thus, it is now important to calculate the number of critical points in order to understand the phase space of the interacting scenario for this dynamical $w_d$. As $g(0) = -{\gamma}^2\nu<0$ (for $\nu >0$) and $g(1)=9>0$, it follows by the Bolzano's theorem \cite{Apostol:105425} that $g(x)$ has at least one real root in $(0,1)$. 
Now, for any $x_c$ lying in $(0,1)$, the condition $z_c=\frac{3x_c}{3x_c-\gamma\left({x_c}^2+x_c-1\right)} \leq 1$ requires $x_c\leq \frac{\sqrt{5}-1}{2}$. We also observe that $g \left(\frac{\sqrt{5}-1}{2} \right)= 9(-2+\sqrt{5}) >0$, and hence, from the Bolzano's theorem \cite{Apostol:105425}, there is at least one root of $g(x)$ in $(0,\frac{\sqrt{5}-1}{2})$. We now claim that $g(x)$ has only one root in the interval $[0, \frac{\sqrt{5}-1}{2}]$, and in that case, it is straightforward to conclude that there is only one critical point in the set $S$. Now, from the derivative of $g(x)$ given below
\begin{align*}
        g'(x) &= -6{\gamma}^2\nu \left(x-1 \right) \left(x-\sqrt{\frac{2}{3}}~\right) \left(x+\sqrt{\frac{2}{3}}~\right) \left(x- \frac{\sqrt{5}-1}{2} \right) \left(x+ \frac{\sqrt{5}+1}{2} \right)+27x^2,
    \end{align*}
we notice that $g'(x)>0$ for $x\in(0,\frac{\sqrt{5}-1}{2})$ which means that  $g(x)$ is strictly increasing in $(0,\frac{\sqrt{5}-1}{2})$. Therefore, the set $S$ contains only one critical point and we label this critical point as $\bar{C}_3$. In this case, the topological nature of the critical point $\bar{C}_3$ is same as $C_3$ described earlier and hence, the phase plot is same as the Fig. \ref{fig10}.

On the other hand, looking at eqn. (\ref{eqn-model-III-footnote-dyn}), and following the similar arguments as in earlier for $\nu <0$, we claim that the set $S$ does not have any critical point for $\nu<0$. Thus, in this case, we have only three critical points, namely, $\bar{C}_0$, $\bar{C}_1$, and $\bar{C}_2$, and the phase portrait looks like same as the Fig. \ref{fig9}.

\subsection{Model IV}
\label{subsec-modelIV}

\noindent In this section we describe the dynamical analysis for the interaction function $Q _{\rm IV}$ of eqn. (\ref{model4}). Considering the $(x,z)$ variables defined as 

\begin{eqnarray}
	x=\frac{\kappa^2 \rho_c}{3H^2}, \quad \quad z= \frac{H_0}{H+H_0},
\end{eqnarray}
the autonomous system for this interaction model takes the form 

\begin{eqnarray}\label{autonomous-system-model-IV}
\left\{\begin{array}{ccc}
    x'&=&-\left(\frac{z}{1-z}\right)\left(\alpha x -\beta x (1-x)\right) + 3 w_d x (1-x),\\ 
	 z'&=&\frac{3}{2} z (1-z) \left(1+w_d (1-x)\right),
 \end{array}\right.
 \label{4eq2} 
\end{eqnarray}
where $\alpha$, $\beta$ are defined as $\alpha = \Gamma_c/H_0$ and $\beta = \Gamma_{cd}/H_0$ respectively and they are dimensionless. Note again that $\alpha \neq \beta$ for the interaction model to be sign shifting. Here, we shall regularize the autonomous system (\ref{autonomous-system-model-IV}) similar to the procedure we have applied in (\ref{subsec-modelI}). The regularized autonomous system will be of the form
\begin{eqnarray}\label{reg-autonomous-system-model-IV}
\left\{\begin{array}{ccc}
    x'&=&-z \left(\alpha x -\beta x (1-x)\right) + 3 w_d (1-z) x (1-x),\\ 
	 z'&=&\frac{3}{2} z (1-z)^2 \left(1+w_d (1-x)\right).
 \end{array}\right.
 \label{4eq2} 
\end{eqnarray}
Also we note that $R=[0,1]^2$ is the physical domain and the lines $x=0$, $z=0$, $z=1$ are invariant under the flow generated by the autonomous system (\ref{reg-autonomous-system-model-IV}). We also see that along the line $x=1$, one obtains $x'=-\alpha z$. Hence, the physical domain $R$ will be positively invariant if we restrict the parameter $\alpha$ by $\alpha>0$ (i.e., $\Gamma_c>0$). It has been mentioned in (\ref{subsec-models}) that the interaction function $Q_{\rm IV}$ will exhibit sign shifting property provided $\Gamma_c$ and $\Gamma_{cd}$ are of the same sign. Consequently, to be consistent with the positively invariance of the physical domain $R$ and the sign shifting nature of the interaction function $Q_{\rm IV}$, we will restrict our attention here on $\alpha >0,  \beta>0$.

In the  following we investigate the nature of the critical points and their implications for the constant and non-constant nature of $w_d$.

\subsubsection{Constant $w_d$}

\noindent In this section we explore the nature of the critical points for different regions of $w_d$. 
In Table~\ref{table-model-IV}, we summarize the critical points of the autonomous system (\ref{reg-autonomous-system-model-IV}), their existence, stability and  as well as the cosmological parameters evaluated at those critical points. In the following we investigate the nature of the critical points for three different regions of $w_d$, namely, quintessence (i.e. $w_d > -1$); cosmological constant (i.e. $w_d =-1$); phantom (i.e. $w_d < -1$).

\begin{figure*}[htp]
	\centering
	\includegraphics[width=0.4\textwidth]{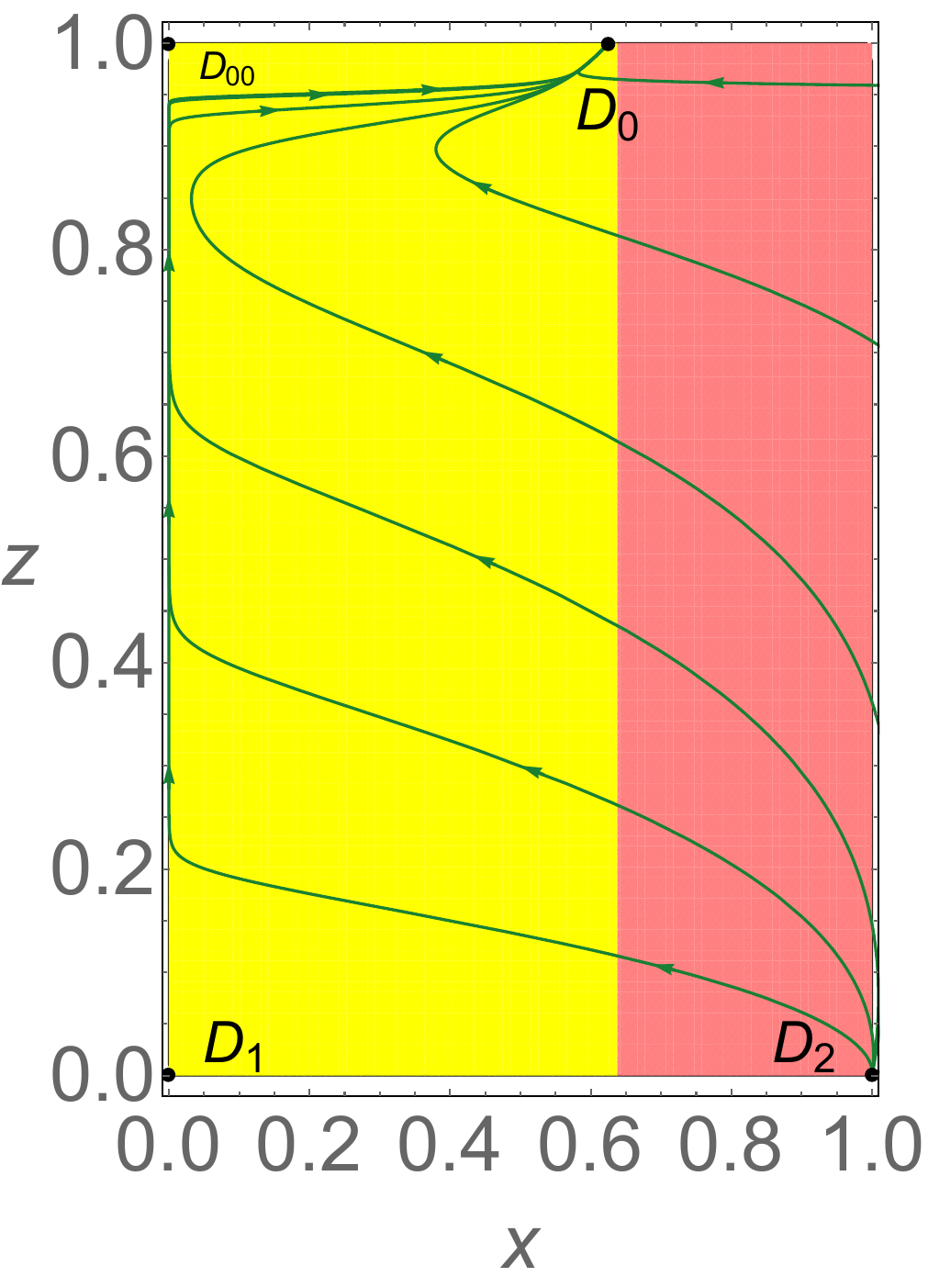}
 \includegraphics[width=0.4\textwidth]{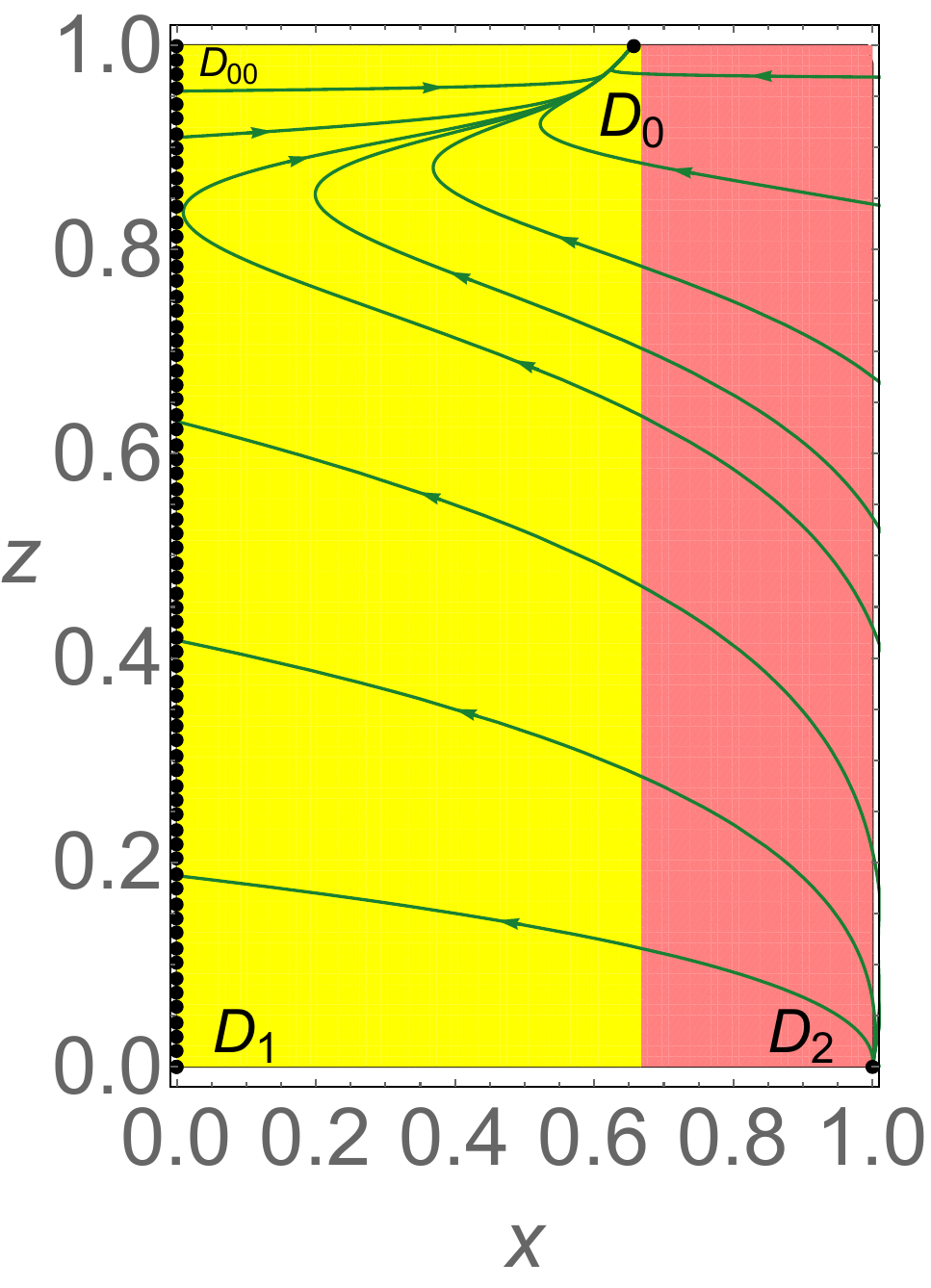}
	\caption{{\bf Left plot:} Phase plot for Model IV (eqn. (\ref{model4}))  with $w_d>-1$ and $\beta>\alpha>0$. In this case we have used $w_d=-0.92$, $\alpha=0.3$ and $\beta=0.8$. {\bf Right plot:} Phase plot for Model IV (eqn. (\ref{model4})) with $w_d=-1$ and $\beta>\alpha>0$. Particularly we have taken $w_d=-1$, $\alpha=0.31$ and $\beta=0.9$. We note that the yellow shaded region represents the accelerated region (i.e. $q<0$) and the pink shaded region corresponds to the decelerated region (i.e. $q>0$).  }
 \label{fig13}
\end{figure*}

\begin{figure*}[htp]
	\centering
	\includegraphics[width=0.4\textwidth]{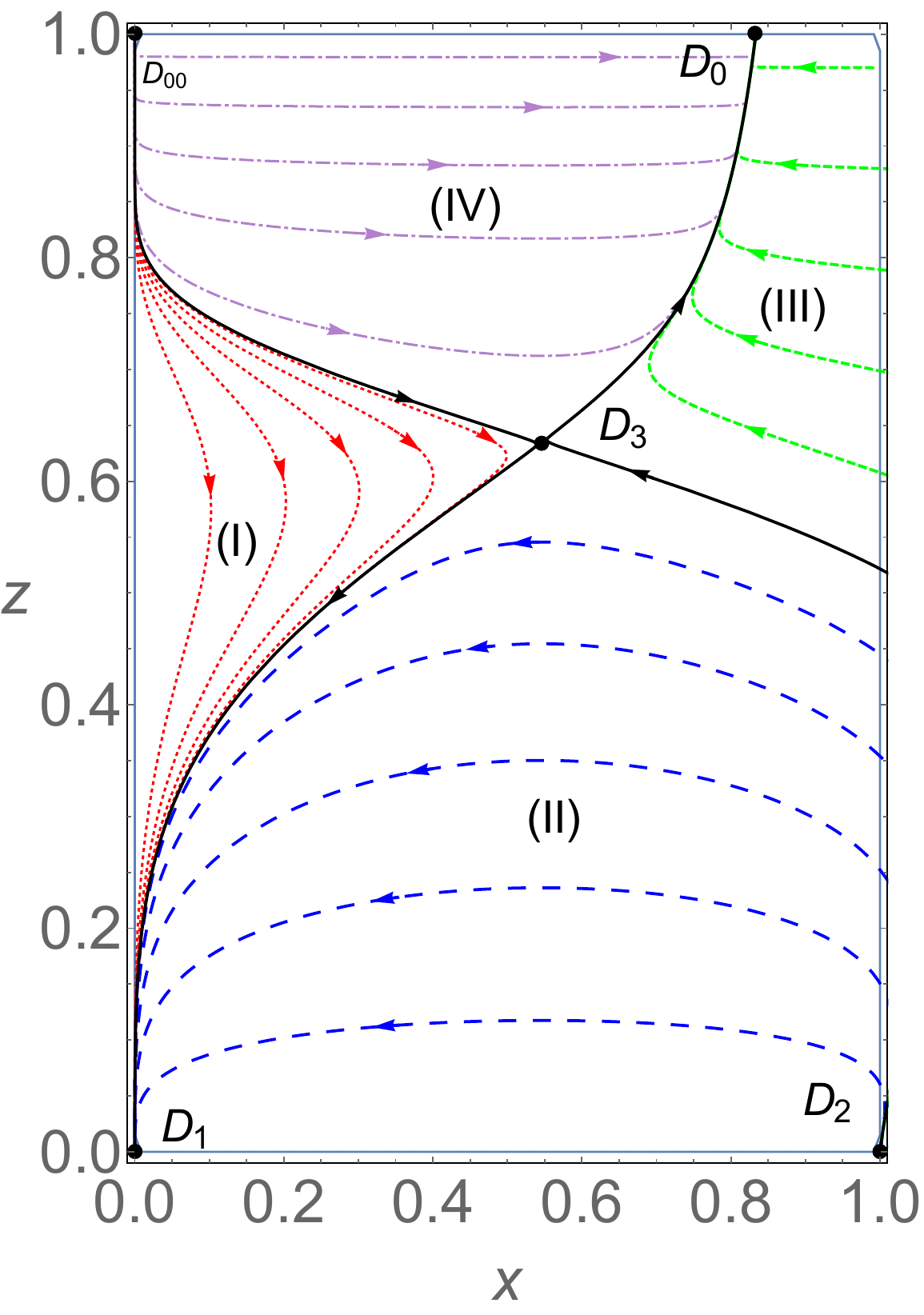}
    \includegraphics[width=0.4\textwidth]{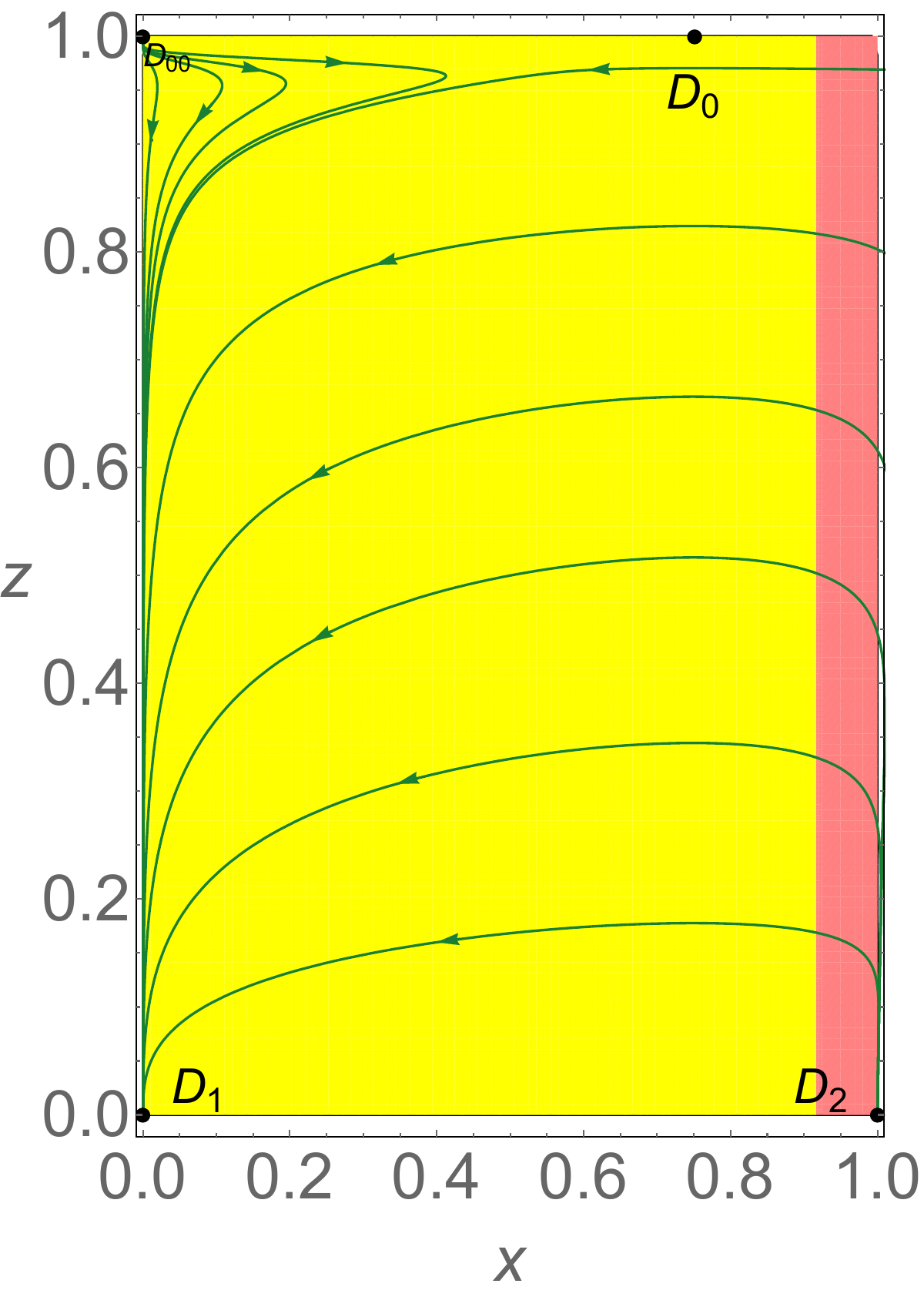}
	\caption{{\bf Left Plot:} Phase plot for Model IV (eqn. (\ref{model4})) with $-\frac{\beta}{\alpha}<w_d<-1$ and $\beta>\alpha>0$. For, numerical simulation we have used $w_d=-2.2$, $\alpha=1$ and $\beta=6$. {\bf Right Plot:} Phase plot for Model IV (eqn. (\ref{model4})) with $w_d\leq-\frac{\beta}{\alpha}$ and $\beta>\alpha>0$. For numerical simulation, we have used $w_d=-4.1$, $\alpha=0.2$ and $\beta=0.8$ from the parameter space. In this plot, the yellow shaded region corresponds to the accelerated region (i.e. $q<0$) and the pink shaded region corresponds to the decelerated region (i.e. $q>0$).  }
 \label{fig14}
\end{figure*}

\begin{figure*}
	\includegraphics[width=0.6\textwidth]{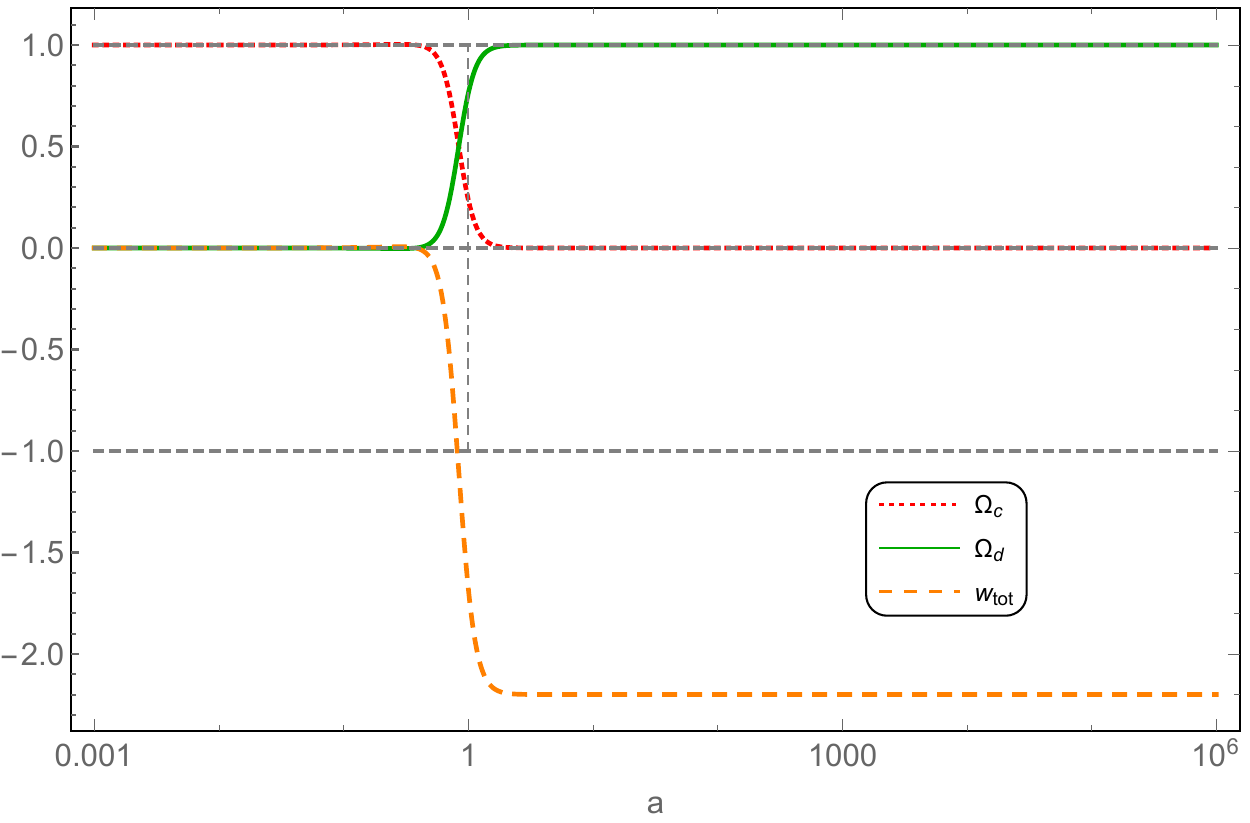}
	\caption{We display the evolution of the CDM density parameter $(\Omega_c)$, dark energy density parameter $(\Omega_d)$ and the total equation of state parameter $(w_{\rm tot})$ for the Model IV (eqn. (\ref{model4})). We have chosen the following values of the parameters: $w_d=-2.2$, $\alpha=1$, $\beta=6$ and the following initial conditions: $x~(N= 0) = 0.24$, $z~(N = 0) =0.01$ from the {\it region II} of the left plot of Fig. \ref{fig14}. In addition, if we start any trajectory from {\it region I} of the left plot of Fig. \ref{fig14}, it converges to the critical point $D_1$. So, for any initial conditions from {\it region I}, we shall get $\Omega_c=0$ and $\Omega_d=1$ at late time. Any trajectory starting from {\it regions III} and {\it IV} of the left plot of Fig. \ref{fig14} will converge to the critical point $D_0$. Therefore, if we take initial conditions on $x(N)$ and $z(N)$ from {\it regions III} and {\it IV}, we shall reach $\Omega_c=\frac{\beta-\alpha}{\beta}$ and $\Omega_d=\frac{\alpha}{\beta}$ in an asymptotic fashion.}
	\label{evo-plot-model4}
\end{figure*}

\begin{figure*}[htp]
	\centering
	\includegraphics[width=0.32\textwidth]{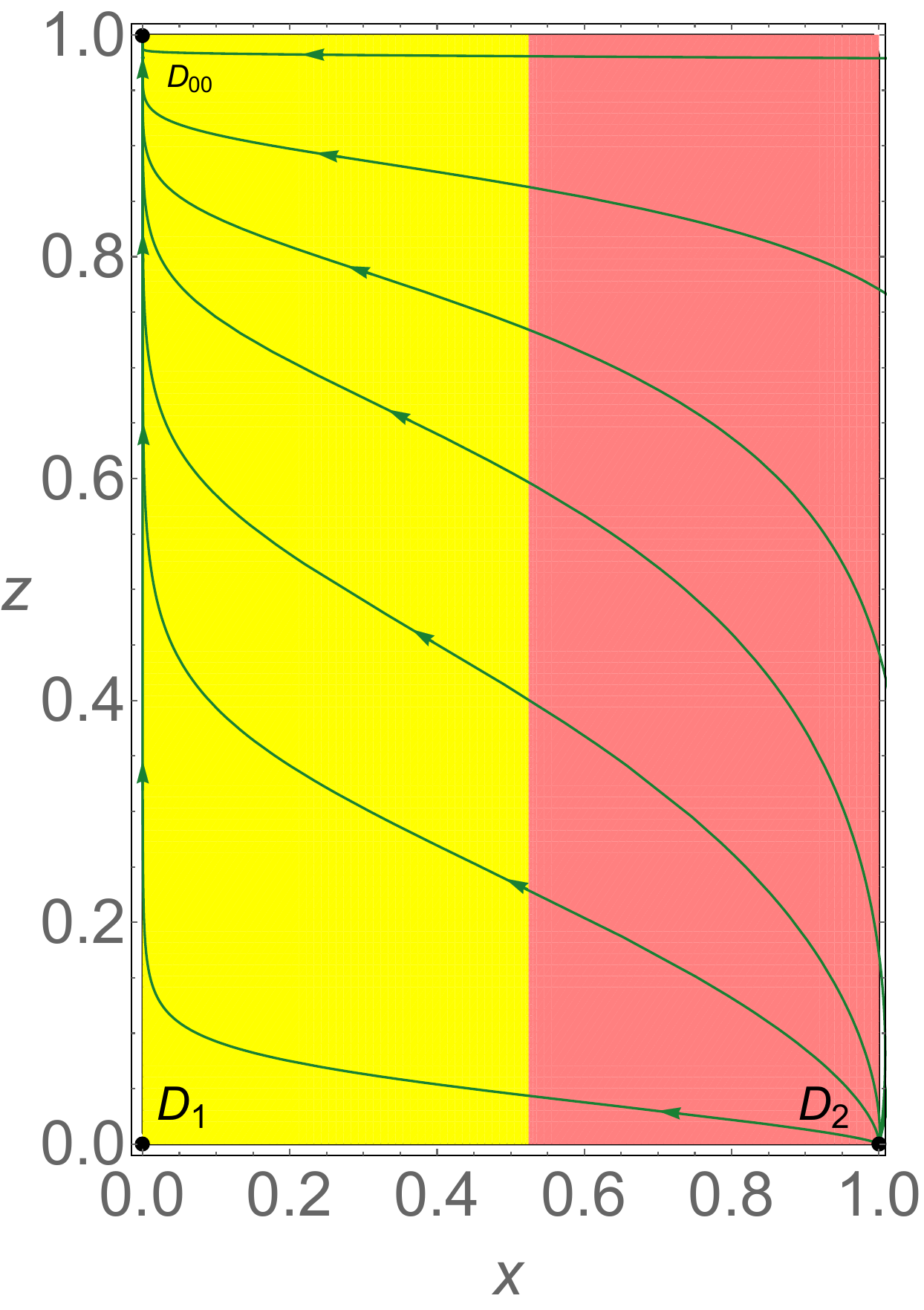}
 \includegraphics[width=0.32\textwidth]{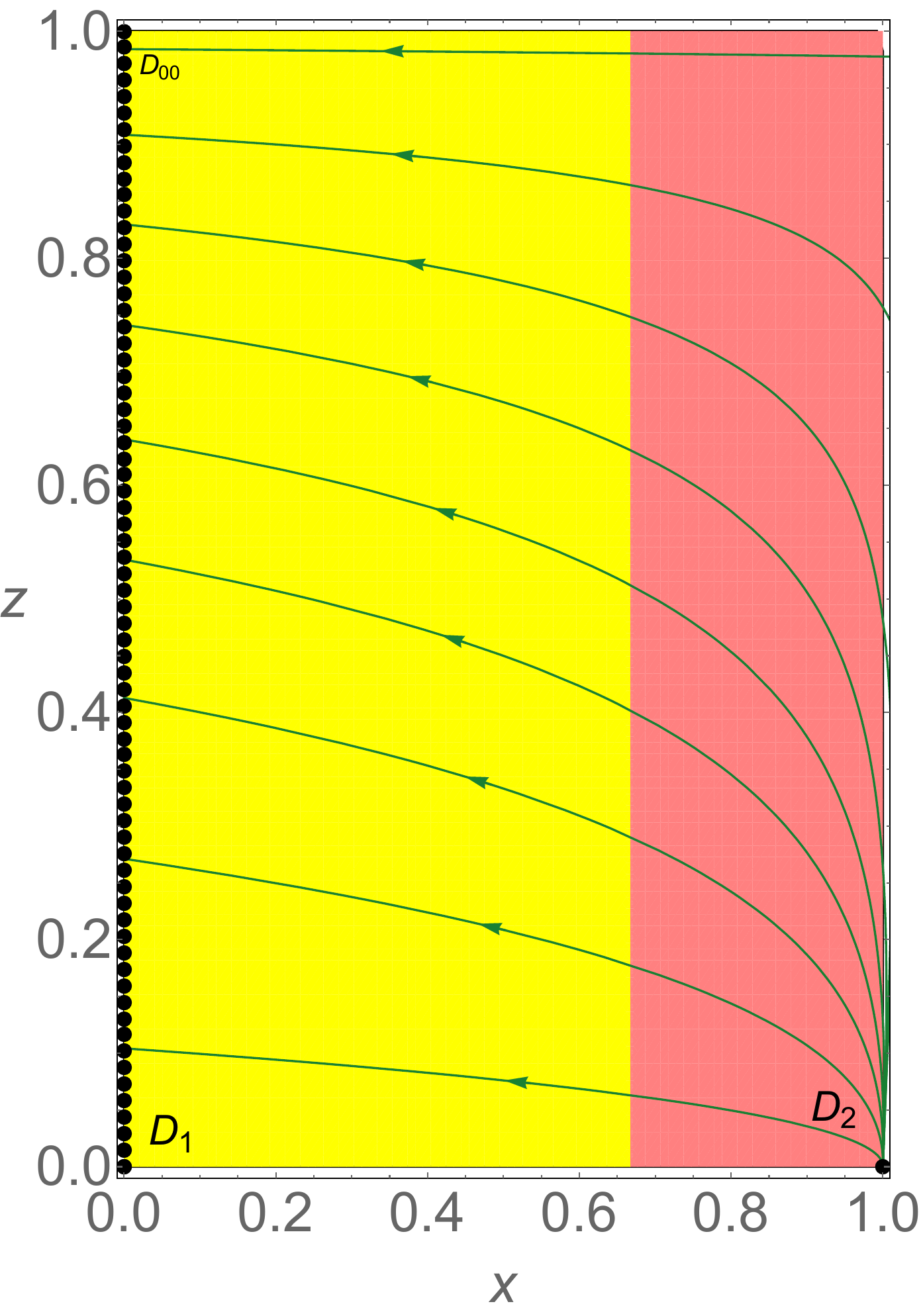}
 \includegraphics[width=0.32\textwidth]{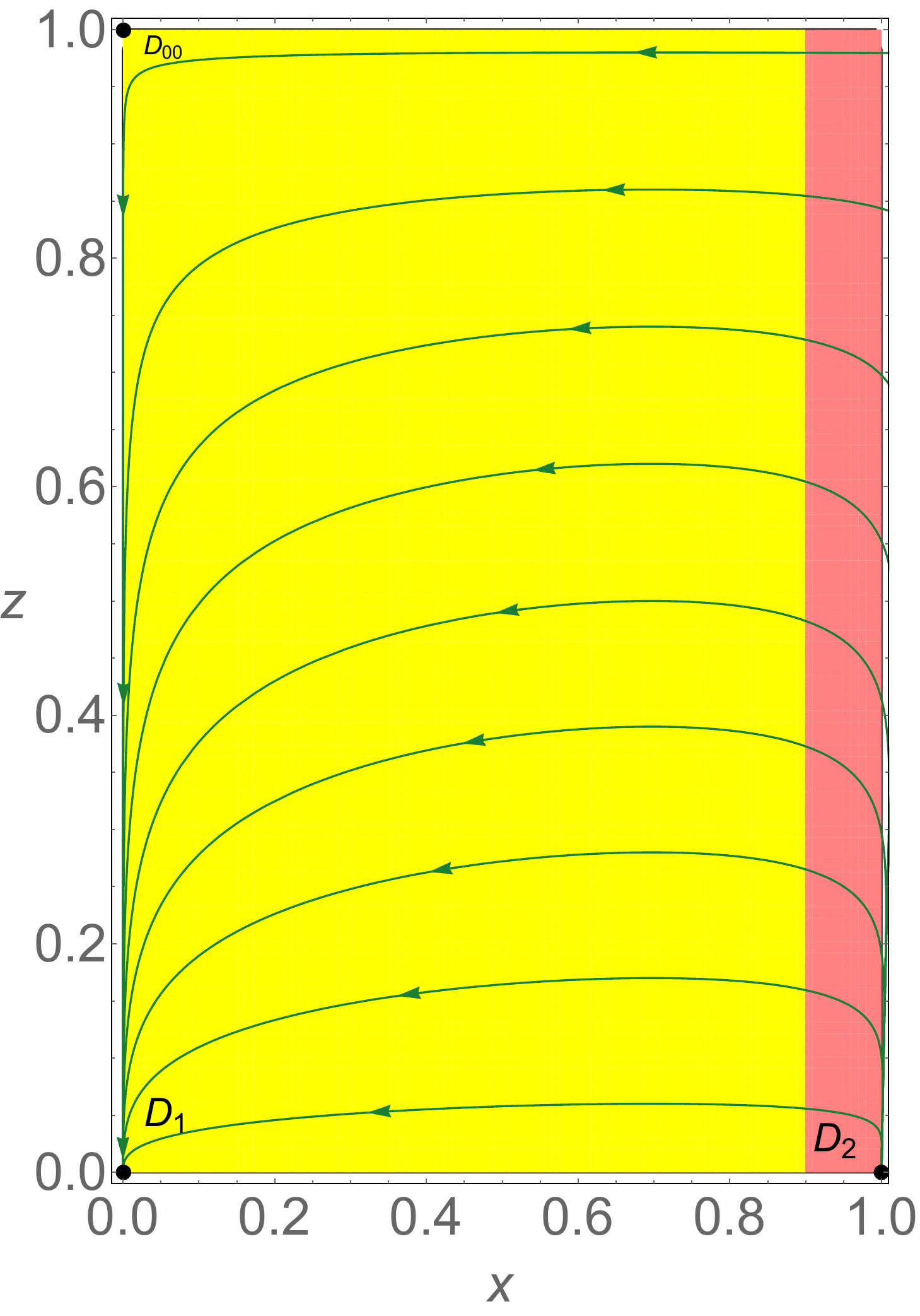}
	\caption{{\bf Left plot:} Phase plot for Model IV (eqn. (\ref{model4})) with $w_d>-1$ and $\alpha>\beta>0$. In this case we have used $w_d=-0.7$, $\alpha=0.2$ and $\beta=0.1$. {\bf Middle plot:} Phase plot for Model IV (eqn. (\ref{model4})) with $w_d=-1$ and $\alpha>\beta>0$. Particularly, we have chosen $w_d=-1$, $\alpha=0.08$ and $\beta=0.05$. {\bf Right plot:} Phase plot for Model IV (eqn. (\ref{model4})) with $w_d<-1$ and $\alpha>\beta>0$. For numerical simulation we have used $w_d=-3.3$, $\alpha=0.2$ and $\beta=0.1$. Here, the yellow shaded region represents the accelerated region (i.e. $q<0$) and the pink shaded region corresponds to the decelerated region (i.e. $q>0$). }
 \label{fig17}
\end{figure*}

\begin{itemize} 
\item When  $\beta>\alpha>0$, the following situations arise:

\begin{enumerate}
    \item $w_d>-1$: When $w_d>-1$, the point $D_3$ does not belong to the physical region $R$ and also $w_{\rm tot}=w_d(1-x)>-1$ which implies $z'=\frac{3}{2}z(1-z)^2(1+w_{\rm tot})>0$. In addition as  $\beta>\alpha>0$, the point $D_0$ enters in the region $R$. On $z=1$ line, the value of $x'$ is positive if $x<\frac{\beta-\alpha}{\beta}$ and $x'$ is negative if $x>\frac{\beta-\alpha}{\beta}$. So, $D_{00}$, $D_1$ and $D_2$ are unstable critical points and $D_0$ is a global attractor. At the point $D_0$, we have $H=0$, $\Omega_c=\frac{\beta-\alpha}{\beta}$ and $\Omega_d=\frac{\alpha}{\beta}$. Left plot of Fig. \ref{fig13} shows the behavior.

    \item $w_d=-1$: When $w_d=-1$, all points in the {\bf $OZ$} axis are critical points and also point $D_3$ lies on {\bf $OZ$} axis. Again, one gets $z'=\frac{3}{2}z(1-z)^2x$ which is positive. At $z=1$, one has $x'>0$ for $x<\frac{\beta-\alpha}{\beta}$ and $x'<0$ for $x>\frac{\beta-\alpha}{\beta}$. In this case, $D_0$ is an attractor but it is not a global attractor. The qualitative nature is displayed in right plot of Fig. \ref{fig13}.

    \item When $-\frac{\beta}{\alpha}<w_d<-1$: The critical point $D_3$ enters in the physical domain. We have $0<\frac{1+w_d}{w_d}<\frac{\beta-\alpha}{\beta}$. At $z=0$, $x'=3w_d x (1-x)$ which is negative and at $z=1$, we obtain $x'>0$ for $x<\frac{\beta-\alpha}{\beta}$ and $x'<0$ for $x>\frac{\beta-\alpha}{\beta}$. If $x<\frac{1+w_d}{w_d}$, implies $1+w_{\rm tot}<0$. So, $z'$ is negative. Similarly, $z'$ is positive for $x>\frac{1+w_d}{w_d}$. Thus, all orbits in the regions I and II of left plot of Fig. \ref{fig14}, at late time, converge to $D_1$. For an orbit in the regions III and IV of left plot Fig. \ref{fig14}, at late time, it converges to $D_0$. Note that $D_0$ means $H=0$ with $\Omega_c=\frac{\beta-\alpha}{\beta}, \Omega_d=\frac{\alpha}{\beta}$ and $w_{\rm tot}=\frac{\alpha w_d}{\beta}>-1$. On the contrary, $D_1$ means $H=\infty$ with $\Omega_d=1$ and $w_{\rm tot}=w_d<-1$. The Fig. \ref{evo-plot-model4} presents the evolution of $\Omega_c$, $\Omega_d$, and  $w_{\rm tot}$. 
    
   \item $w_d\leq -\frac{\beta}{\alpha}$: When $w_d=-\frac{\beta}{\alpha}$, one obtains $D_0=D_3$. Now, one has $z'<0$ for $x<\frac{1+w_d}{w_d}$ and $z'>0$ for $x>\frac{1+w_d}{w_d}$. Also, $x'$ is negative on $z=0$. Thus, $D_1$ is a global attractor. For $w_d<-\frac{\beta}{\alpha}$, the critical point $D_3$ leaves the physical region and one gets $\frac{\beta-\alpha}{\beta}<\frac{1+w_d}{w_d}<1$ which gives $z'<0$ for $x<\frac{1+w_d}{w_d}$ and $z'>0$ for $x>\frac{1+w_d}{w_d}$. Here also, we have $x'<0$ on $z=0$. Once again, $D_1$ is a global attractor. Right plot of Fig. \ref{fig14} exhibits the phase plot.  
\end{enumerate}

\item $\alpha>\beta > 0$: 
In this case the critical points $D_0$ and $D_3$ never belong to the physical domain.

\begin{itemize}

\item When $w_d>-1$, $z'$ is positive as $1+w_{\rm tot}>0$. At $z=1$ line, $x'$ is negative. Thus, $D_{00}$ is a global attractor. The point $D_{00}$ corresponds to $H=0$, $\Omega_c=0$ and $\Omega_d=1$. 

\item When $w_d=-1$, all points in the {\bf $OZ$} axis are critical points. Now, one has $z'=\frac{3}{2}z(1-z)^2x$ which is positive. Also, $x'$ is negative on both $z=0$ and $z=1$ lines. Here, we have no attractor. 

\item For {$w_d<-1$}, we have $0<\frac{1+w_d}{w_d}<1$. On $z=0$ line, $x'=3w_dx(1-x)$ which is negative. Here, we have $z'<0$ for $x<\frac{1+w_d}{w_d}$ and $z'>0$ for $x>\frac{1+w_d}{w_d}$. Thus, $D_1$ is the only global attractor.   
In all the cases, qualitative behaviors are shown in Fig. \ref{fig17}.

\end{itemize}

\end{itemize}

\begin{table*}
\begin{center}
	\begin{tabular}{|c|c|c|c|c|c|c|c|c|}\hline
 
		Point &  $x$ & $z$ & Existence  & Stability & Acceleration & $\Omega_{c}$ & $\Omega_{d}$ & $w_{\rm tot}$\\ \hline
  
		$D_0$ & $\frac{\beta-\alpha}{\beta}$ & 1 & $ \beta > \alpha > 0$ and $w_d<-\frac{1}{3}$ &  $ -\frac{\beta}{\alpha}<w_d<-\frac{1}{3}$ & $w_d<-\frac{\beta}{3\alpha} $ & $\frac{\beta-\alpha}{\beta}$ & $\frac{\alpha}{\beta} $ & $\frac{\alpha w_d}{\beta}$\\ \hline

        $D_{00}$ & $0$ & $1$ & $\alpha >0$, $\beta > 0$ and $ w_d<-\frac{1}{3} $  & $\alpha~(>0)>\beta$ with $-1<w_d<-\frac{1}{3}$  & $w_d<-\frac{1}{3}$ & $0$  & $1$ & $w_d$   \\ \hline
  
	$D_1$ & $0$ & $0$ & $\alpha >0$, $ \beta > 0$ and $w_d<-\frac{1}{3}$ &  $w_d<-1$ & $w_d<-\frac{1}{3}$ & $0$ & $1$ & $w_d$ \\ \hline
        
       $D_2$ & $1$ & $0$ & $\alpha >0$, $\beta > 0 $ and $w_d<-\frac{1}{3}$ & unstable & no & $1$ & $0$ & $0$ \\ \hline
       
       $D_3$ & $\frac{1+w_d}{w_d}$ & $ -\frac{3 w_d}{(\alpha-3) w_d +\beta}$ & $ \beta>\alpha>0 $ with $-\frac{\beta}{\alpha}\leq w_d \leq -1$ & unstable for $w_d\ne-1$ & yes & $\frac{1+w_d}{w_d}$ & $-\frac{1}{w_d}$ & $-1$  \\ 
       \hline
       
	\end{tabular}
	\caption{The critical points, their existence, stability, and the values of the cosmological parameters evaluated at those points for the interacting scenario driven by the interaction function $Q_{\rm IV} = \Gamma_c \rho_c -\Gamma_{cd} \frac{\rho_c\rho_d}{\rho_c + \rho_d}$ of eqn. (\ref{model4})  are summarized.  }
	\label{table-model-IV}
\end{center}
\end{table*}

\subsubsection{Dynamical $w_d$}

\noindent For the equation of state (\ref{dynamical-eos-special}), the autonomuous system (\ref{autonomous-system-model-IV}) becomes:
\begin{eqnarray}\label{autonomous-model-IV-dyn-w1}
    \left\{\begin{array}{ccc}
      x'   &=& -\left(\frac{z}{1-z}\right)x\left(\alpha-\beta(1-x)\right)-3x(1-x)\left( 1+\nu \frac{(1-z)^2(1-x)}{z^2}\right),  \\
     z'    &=& \frac{3}{2}(1-z)z\left(x-\nu \frac{(1-z)^2(1-x)^2}{z^2}\right),
    \end{array}\right.
\end{eqnarray}
 where $\nu=\frac{3AH_0^2}{\kappa^2}$. Proceeding as in the earlier cosmological models for non-constant $w_d$, we regularize the autonomous system (\ref{autonomous-model-IV-dyn-w1}) and obtain the following: 
 \begin{eqnarray}\label{reg-autonomous-model-IV-dyn-w1}
    \left\{\begin{array}{ccc}
      x'   &=& -{z^3}x\left(\alpha-\beta(1-x)\right)-3(1-z)x(1-x)\left( z^2+\nu (1-z)^2(1-x)\right),  \\
     z'    &=& \frac{3}{2}(1-z)^2z\left(x z^2-\nu (1-z)^2(1-x)^2\right),
    \end{array}\right.
\end{eqnarray}
The critical points of the system (\ref{reg-autonomous-model-IV-dyn-w1}) are

\begin{itemize}
    \item $\bar{D}_0  = \left(\frac{\beta-\alpha}{\beta}, 1\right)$, \quad $\bar{D}_{00}  = (0, 1)$, \quad $\bar{D}_1  = (0, 0)$, \quad $\bar{D}_2 = (1, 0)$, \quad $S = \left\{\left(x_c,  \frac{3}{3-\alpha+\beta (1-x_c)}\right)\right\},$
\end{itemize}
where $x_c$ is a root of $\Theta(x) \equiv 9x-\nu (-\alpha+\beta(1-x))^2 (1-x)^2 = 0$.\footnote{Note again that $\Theta(x)$ is obtained from the following two nullclines: 
\begin{align}
x z^2 - \nu (1-z)^2 (1-x)^2  = 0, \label{eqn-model-IV-footnote-dyn}\\ 
-z^3 x [\alpha -\beta (1-x)] -3x (1-x) (1-z)\left[ z^2 + \nu (1-z)^2 (1-x)\right] = 0. 
\end{align}
} Since $\Theta(x) $ represents a fourth degree equation in $x$, therefore, $S$ may contain maximum four critical points. 
In a similar fashion, we proceed to investigate the number of roots of $\Theta(x)$ in $[0, 1]$. 
Now, since $\Theta (0) = \nu (\beta - \alpha)^2<0$ (for $\nu > 0$) and $\Theta(1) = 9 >0$, from the Bolzano's theorem \cite{Apostol:105425}, we can say that $\Theta(x)$ has at least one root in the interval $(0,1)$. However, as the physical domain in this case is $R$, therefore, 
for valid critical point in this domain, we have to check whether the $z$ component of the critical point lies in $[0,1]$. Now, for any $x_c$ in $[0,1]$, the condition $\frac{3}{3-\alpha+\beta (1-x_c)} \leq 1$  demands that $\beta x_c \leq \beta-\alpha$. Now, as $\alpha>0$, $\beta>0$, we consider the following cases: 

\begin{itemize}
    \item $\beta > \alpha>0$:  In this case, the condition $\beta x_c \leq \beta-\alpha$ reduces to $x_c \leq \frac{\beta-\alpha}{\beta}$. We see that $\Theta(0)<0$ and $\Theta\left(\frac{\beta-\alpha}{\beta}\right) = 9 \frac{\beta-\alpha}{\beta} >0$, hence, from the Bolzano's theorem \cite{Apostol:105425}, there is at least one root of $\Theta(x)$ in $\left(0,\frac{\beta-\alpha}{\beta} \right)$.   We also observe that 
    \begin{align*}
        \Theta'(x) &= -4 {\beta}^2\nu \left(x-1 \right) \left(x-\frac{\beta-\alpha}{\beta} \right) \left(x-\frac{2\beta-\alpha}{2\beta} \right)+9>0,
    \end{align*}
  for $x\in \left(0,\frac{\beta-\alpha}{\beta} \right)$. Hence, $\Theta(x)$ is strictly increasing in $x\in \left(0,\frac{\beta-\alpha}{\beta} \right)$ and as a result $\Theta(x)$ has only one root in $x\in \left(0,\frac{\beta-\alpha}{\beta} \right)$. Thus, the set $S$ contains only one critical point which we call as $\bar{D}_3$. The critical point $\bar{D}_3$ behaves qualitatively same as the point $D_3$ and consequently, the phase portrait is topologically same as the left plot of Fig. \ref{fig14}.  
  
 Again, following the same arguments as in the earlier models for $\nu<0$, we can show that $\Theta(x)$ has no root in $(0,1)$ for $\nu<0$. Hence, in this case the autonomous system has only four critical points, namely, $\bar{D}_0$,  $\bar{D}_{00}$, $\bar{D}_1$, $\bar{D}_2$. The phase space stability analysis has been shown before. The phase portrait is same as the left plot of Fig. \ref{fig13}.

\item $\alpha>\beta >0$:  In this case, we have $\beta - \alpha <0$, and hence, $\bar{D}_0$ does not belong to the physical domain $R$. Also, the condition $\beta x_c \leq \beta-\alpha$, leads to the claim that $x_c$ does not belong to the domain $[0,1]$.  Therefore, we have only three critical points, $\bar{D}_{00}$, $\bar{D}_1$, $\bar{D}_2$ and the phase portrait is same as that of the phase portrait of the right plot of Fig. \ref{fig17}. On the other hand, for $\nu < 0$, we do not have any critical point in $S$. Here also we have only three critical points, namely, $\bar{D}_{00}$, $\bar{D}_1$, $\bar{D}_2$ and we see that the corresponding phase portrait looks similar to left plot of Fig. \ref{fig17}.

\end{itemize}

\begin{figure}[htp]
\centering
\includegraphics[width=0.32\textwidth]{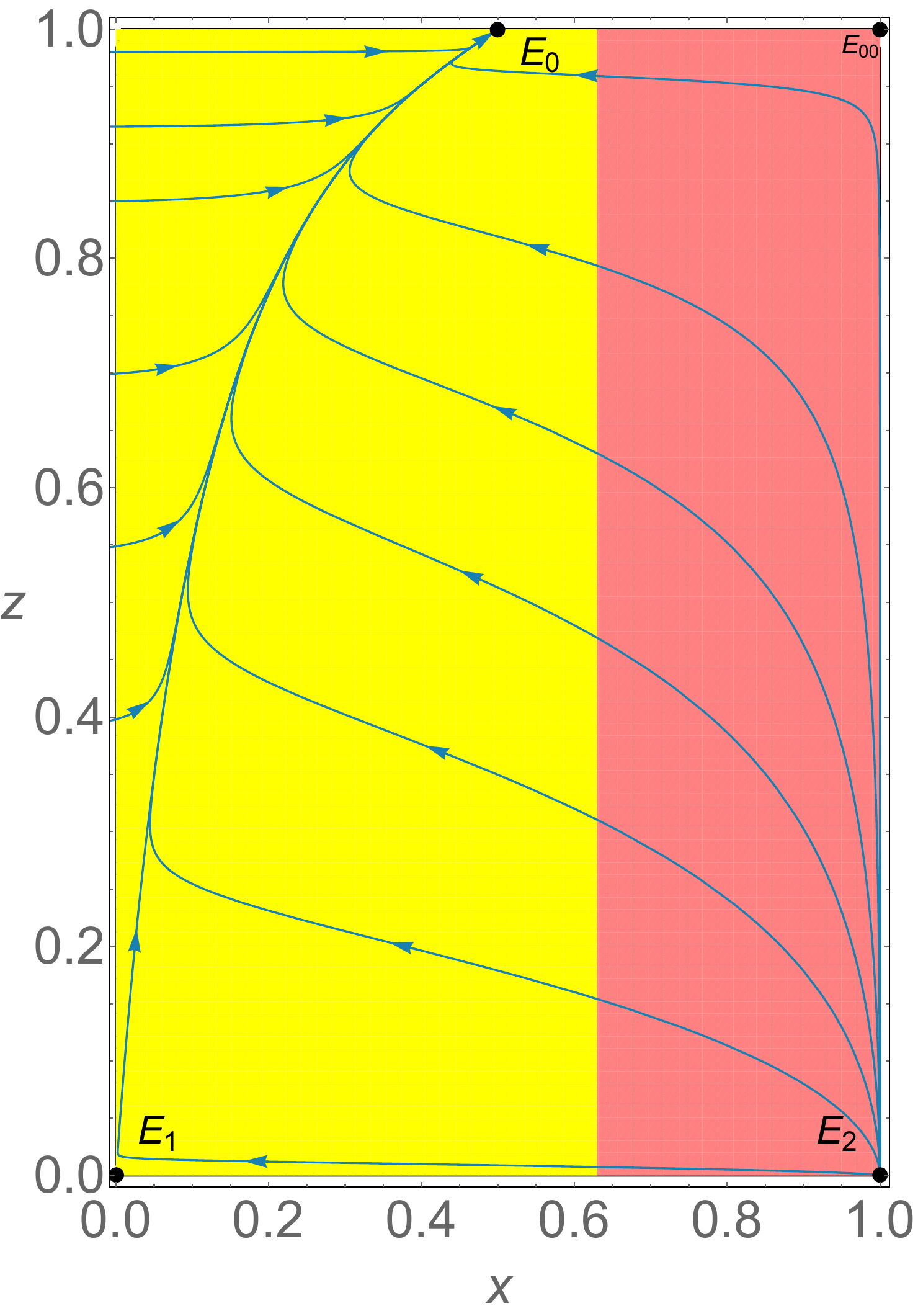}
 \includegraphics[width=0.32\textwidth]{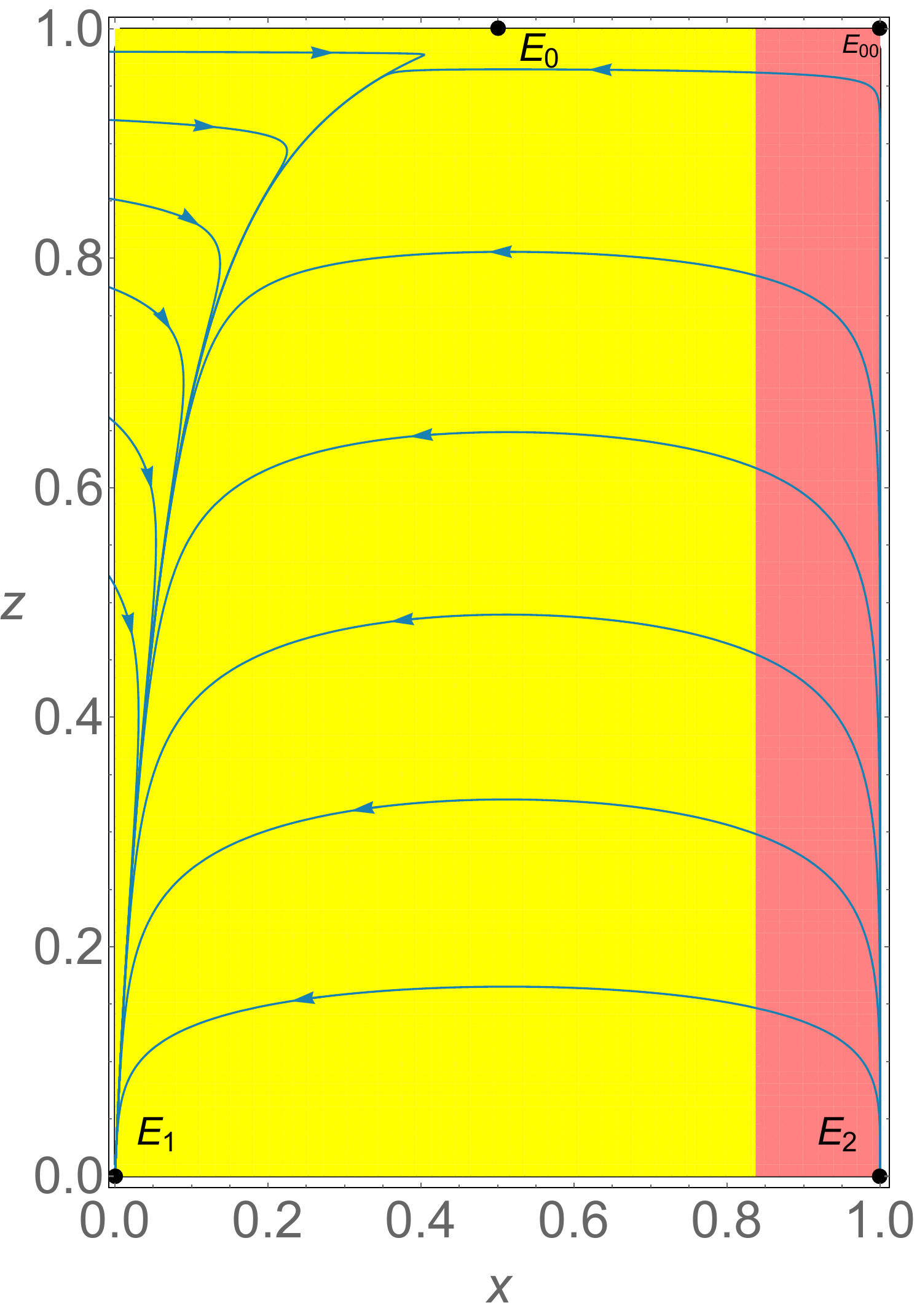}
  \includegraphics[width=0.32\textwidth]{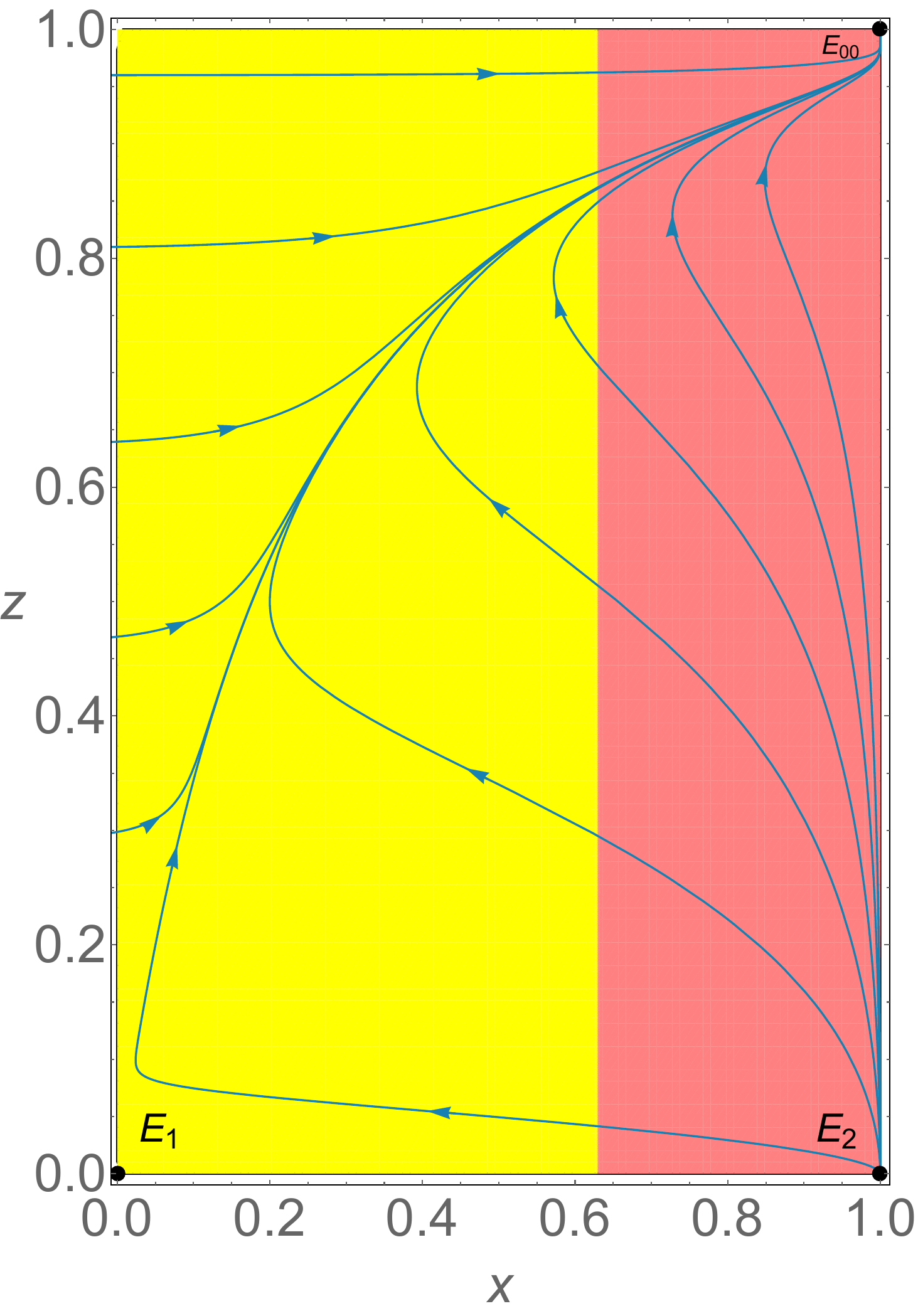}
 \caption{{\bf Left plot:} Phase plot for Model V (eqn. (\ref{model5})) with $w_d\geq-1$ and $\beta<\alpha<0$. In this case we have chosen $w_d =-0.9$, $\alpha = -0.3$ and $\beta=-0.6$. {\bf Middle plot:} Phase plot for Model V (eqn. (\ref{model5})) with $w_d\leq\frac{\beta}{\alpha-\beta}$ and $\beta<\alpha<0$. In this case we have chosen $w_d =-2.05$, $\alpha = -0.3$ and $\beta=-0.6$. {\bf Right plot:} Phase plot for Model V (eqn. (\ref{model5}))  with $w_d\geq -1$ and $\alpha<\beta<0$. In this case we have chosen $w_d =-0.9$, $\alpha = -0.6$ and $\beta=-0.3$. Here, the yellow shaded region represents the accelerated region (i.e. $q<0$) and the pink shaded region corresponds to the decelerated region (i.e. $q>0$).  }
 \label{fig18}
\end{figure}
\begin{figure}[htp]
\centering
 \includegraphics[width=0.4\textwidth]{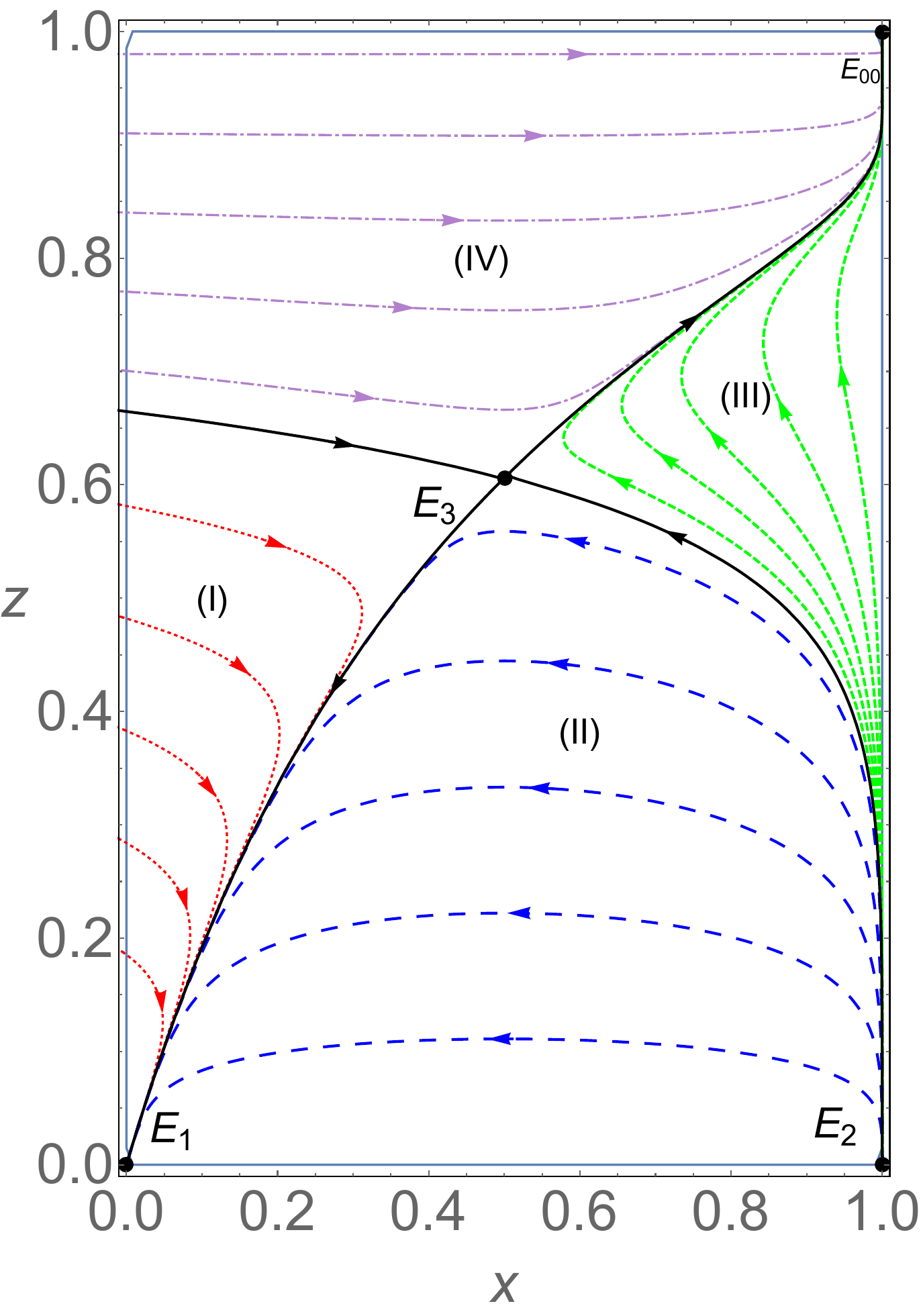}
\includegraphics[width=0.4\textwidth]{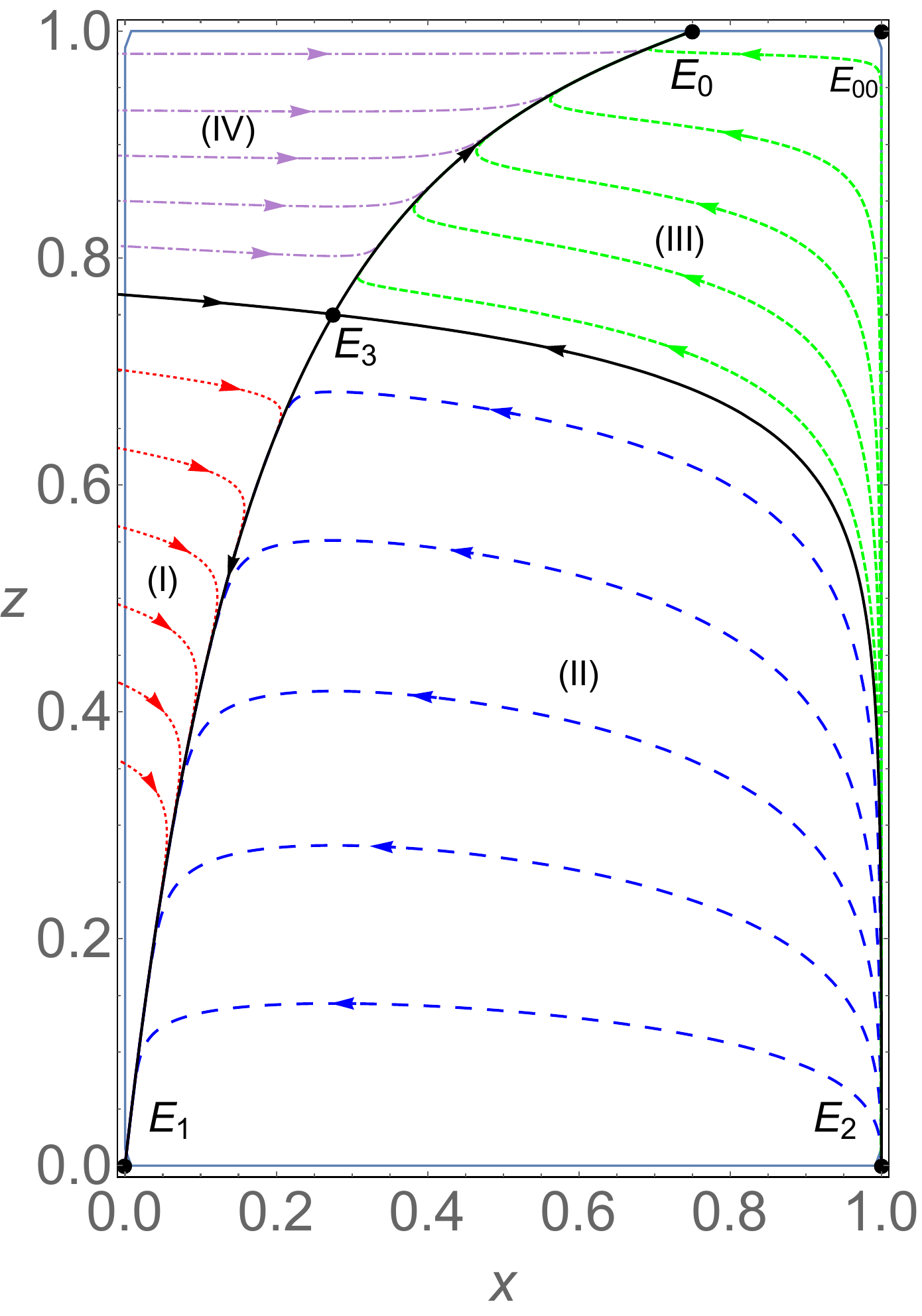}
\caption{{\bf Left plot:} Phase plot for Model V (eqn. (\ref{model5})) with $w_d<-1$ and $\alpha<\beta<0$. In this case we have chosen $w_d =-2$, $\alpha = -2$ and $\beta=-0.1$. {\bf Right plot:} Phase plot for Model V (eqn. (\ref{model5})) with $\frac{\beta}{\alpha-\beta}<w_d<-1$ and $\beta<\alpha<0$. In this case we have taken $w_d =-1.38$, $\alpha =- 0.6$ and $\beta=-0.8$. We note that one can take any specific value of $\alpha~( <0)$ and $\beta~(<0)$ with $\beta<\alpha<0$ to draw the plot, however, as long as $\alpha$ and $\beta$ increase, the regions I and IV become very small and they look indistinguishable from one another. }
\label{fig19}
\end{figure}
\begin{figure}
	\includegraphics[width=0.6\textwidth]{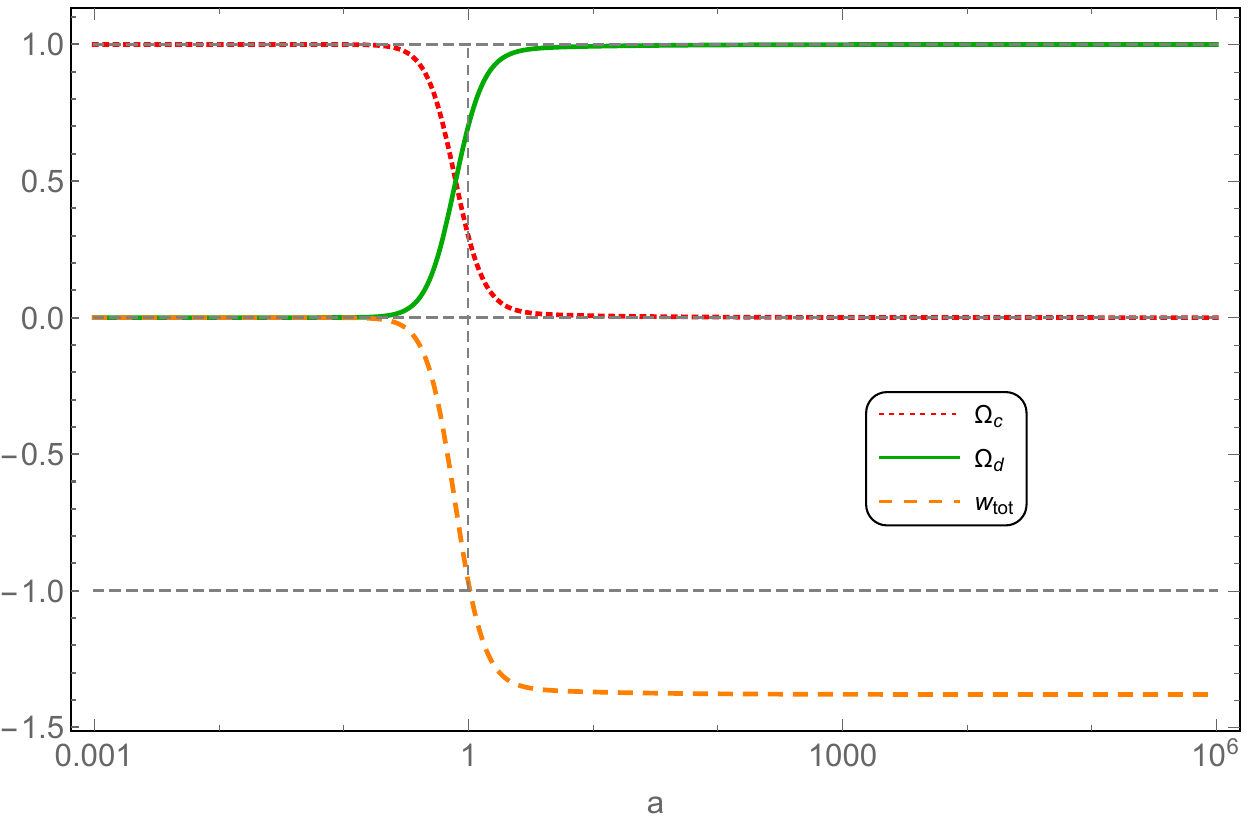}
	\caption{We display the evolution of the CDM density parameter $(\Omega_c)$, dark energy density parameter $(\Omega_d)$ and the total equation of state parameter $(w_{\rm tot})$ for Model V (eqn. (\ref{model5})). We have chosen the following values of the parameters: $w_d=-1.38$, $\alpha=-0.6$, $\beta=-0.8$ and the following initial conditions: $x~(N = 0) = 0.3$, $z~(N = 0) = 0.1$ from the {\it region II} of the right plot of Fig. \ref{fig19}. In addition, if we start any trajectory from {\it region I} of the right plot of Fig. \ref{fig19}, it converges to the critical point $E_1$. So, for any initial conditions from {\it region I}, we shall get $\Omega_c=0$ and $\Omega_d=1$ at late time. Any trajectory starting from {\it regions III} and {\it IV} of the right plot of Fig. \ref{fig19} will converge to the critical point $E_0$. Therefore, if we take initial conditions on $x(N)$ and $z(N)$ from {\it regions III} and {\it IV}, we shall reach $\Omega_c=\frac{\alpha}{\beta}$ and $\Omega_d=\frac{\beta-\alpha}{\beta}$ in an asymptotic fashion.}
	\label{evo-plot-model5}
\end{figure}

\subsection{Model V}
\label{subsec-modelV}

\noindent  In this section we discuss the dynamical analysis for the interacting scenario driven by the interaction function $Q_{\rm V}$ of eqn. (\ref{model5}). Using the dimensionless variables $(x,z)$ defined as 

\begin{eqnarray}
	x=\frac{\kappa^2 \rho_c}{3H^2}, \quad \quad z= \frac{H_0}{H+H_0},
\end{eqnarray}
we obtain the following autonomous system for the prescribed interacting scenario:

\begin{eqnarray}\label{autonomous-system-model-V}
\left\{\begin{array}{ccc}
   x'&=&-\left(\frac{z}{1-z}\right)(1-x)\left(\alpha -\beta x\right) + 3w_d x (1-x),\\ 
     z'&=&\frac{3}{2} z (1-z) \left(1+w_d (1-x)\right),
     \end{array}\right.
    \label{5eq2} 
\end{eqnarray}
where $\alpha$, $\beta$ are defined as $\alpha = \Gamma_d/H_0$ and $\beta = \Gamma_{cd}/H_0$ respectively and $\alpha \neq \beta$. After regularizing, in a similar way we have performed in (\ref{subsec-modelI}), the autonomous system (\ref{autonomous-system-model-V}) can be written of the form
\begin{eqnarray}\label{reg-autonomous-system-model-V}
\left\{\begin{array}{ccc}
     x'&=&-z (1-x)\left(\alpha -\beta x\right) + 3w_d (1-z) x (1-x),\\ 
     z'&=&\frac{3}{2} z (1-z)^2 \left(1+w_d (1-x)\right).
     \end{array}\right.
    \label{5eq2} 
\end{eqnarray}
The physical region is $R$ which is the square $R=[0,1]^2$ and following the similar arguments as in the case of earlier models, we observe that $R$ is positively invariant if $\alpha <0$ (i.e. $\Gamma_d<0$). The interaction function $Q_{\rm V}$ as mentioned in (\ref{model5}) is of sign shifting nature provided that the coupling parameters $\Gamma_d$ and $\Gamma_{cd}$ are of the same sign. Hence, for the dynamical analysis driven by the sign shifting interaction function $Q_{\rm V}$, and to keep the physical domain $R$ positively invariant, we assume the parametric condition $\alpha < 0, \beta<0$.

\subsubsection{Constant $w_d$}

\noindent For constant $w_d$, the critical points of the autonomous system (\ref{reg-autonomous-system-model-V}), their existence, stability and  as well as the cosmological parameters evaluated at those critical points are summarized in Table~\ref{table-model-V}. Similar to the earlier cases, here we consider various cases of $w_d$ depending on its parameter space. In what follows 
we consider various cases.

\begin{itemize}

    \item We consider the first case where the dimensionless coupling parameter satisfy the relation $\beta<\alpha<0$: 

\begin{enumerate}
    \item When $w_d>-1$, $E_3$ leaves the domain $R$ and one has $w_{\rm tot}=w_d(1-x)>-1$ which implies $z'>0$. On $z=1$ line, one gets $x'>0$ for $x<\frac{\alpha}{\beta}$ and $x'<0$ for $x>\frac{\alpha}{\beta}$. Thus, $E_0$ is a global attractor. Left plot of Fig. \ref{fig18} shows the behavior.
    \item For $w_d=-1$, one obtains $E_3=E_1$ and $z'=\frac{3}{2}z(1-z)^2x>0$. On $z=1$ line, one has $x'>0$ for $x<\frac{\alpha}{\beta}$ and $x'<0$ for $x>\frac{\alpha}{\beta}$. So, $E_0$ is a  global attractor. Again, the phase plot is given in left plot of Fig. \ref{fig18}.
    \item When $\frac{\beta}{\alpha-\beta}<w_d<-1$, then 
$E_3$ enters in the physical region and one can get $\frac{1+w_d}{w_d}<\frac{\alpha}{\beta}$. On $z=0$, one obtains $x'<0$. Also, on $z=1$, one has $x'>0$ for $x<\frac{\alpha}{\beta}$ and $x'<0$ for $x>\frac{\alpha}{\beta}$. Again, $z'$ is negative whenever $x<\frac{1+w_d}{w_d}$ and $z'$ is positive whenever $x>\frac{1+w_d}{w_d}$. Thus, all orbits in the regions I and II of the right plot of Fig. \ref{fig19}, at late time, converge to $E_1$. For an orbit in the regions III and IV of the right plot of Fig. \ref{fig19}, at late time, it converges to $E_{0}$. In Fig. \ref{evo-plot-model5} we show the evolution of $\Omega_c$, $\Omega_d$ and $w_{\rm tot}$.

    \item When $w_d= \frac{\beta}{\alpha-\beta}$, one can obtain $E_3=E_0$ and $x'$ is negative on $z=0$. Also $z'$ is negative for $x<\frac{1+w_d}{w_d}$ and $z'$ is positive for $x>\frac{1+w_d}{w_d}$. Thus, $E_1$ is a global attractor. The qualitative behavior is displayed in middle plot of Fig. \ref{fig18}.

     \item When $w_d< \frac{\beta}{\alpha-\beta}$, $E_3$ leaves the 
     physical domain and one has $\frac{1+w_d}{w_d}>\frac{\alpha}{\beta}$. As before, $x'$ is negative on $z=0$. Again, $z'$ is negative in left side of $x=\frac{1+w_d}{w_d}$ and $z'$ is positive in right side of $x=\frac{1+w_d}{w_d}$. Therefore, $E_1$ is a global attractor. The middle plot of Fig. \ref{fig18} exhibits the qualitative nature.

     \end{enumerate}

\item We consider the case when the dimensionless coupling parameters satisfy $\alpha<\beta<0$. In this case, $E_0$ leaves the domain as $\frac{\alpha}{\beta}>1$. In the following we discuss  the nature of the critical points for different values of $w_d$. 

  \begin{enumerate}
        \item For $w_d>-1$, $E_3$ leaves the domain $R$.  Now since $w_{\rm tot}=w_d(1-x)>-1$, hence, it implies that $z'$ is positive. On $z=1$ line, $x'$ is positive. As a result, $E_1$, $E_2$ are unstable and $E_{00}$ is a global attractor. Right plot of Fig. \ref{fig18} shows the qualitative behavior.

        \item When $w_d=-1$, we obtain $E_1=E_3$ and $z'=\frac{3}{2}z(1-z)^2x$ which is positive. At $z=1$ line, $x'$ is positive. Thus, $E_{00}$ is again a global attractor. Phase plot is displayed in right plot of Fig. \ref{fig18}.

        \item When $w_d<-1$, $E_3$ enters in the physical region $R$ and for this $w_d$, we also have $0<\frac{1+w_d}{w_d}<1$. On $z=0$ line, $x'=3w_d x (1-x)$ which is negative and on $z=1$ line, $x'$ is positive. Now left side of $x=\frac{1+w_d}{w_d}$, $z'$ is negative and right side of $x=\frac{1+w_d}{w_d}$, $z'$ is positive. Hence, the physical region $R$ is divided into four regions. Thus, all orbits in the regions I and II of the left plot of Fig. \ref{fig19}, at late time, converge to $E_1$. For an orbit in the regions III and IV of the left plot of Fig. \ref{fig19}, at late time, it converges to $E_{00}$.

    \end{enumerate}

\end{itemize}

\begin{table*}
\begin{center}
	\begin{tabular}{|c|c|c|c|c|c|c|c|c|}\hline
 
		Point &  $x$ & $z$ & Existence  & Stability & Acceleration & $\Omega_{c}$ & $\Omega_{d}$ & $w_{\rm tot}$\\ \hline
  
		$E_0$ & $\frac{\alpha}{\beta}$ & 1 & $ \beta < \alpha < 0$ and $w_d<-\frac{1}{3}$ &  $ \frac{\beta}{\alpha-\beta}<w_d<-\frac{1}{3}$ & $w_d<-\frac{1}{3} \frac{\beta}{\beta-\alpha} $ & $\frac{\alpha}{\beta}$ & $\frac{\beta-\alpha}{\beta} $ & $\frac{(\beta-\alpha) w_d}{\beta}$\\ \hline

       $E_{00}$ & $1$ & $1$ & $\alpha<0, \beta < 0$ and $ w_d<-\frac{1}{3} $  & $\alpha<\beta$ with $w_d<-\frac{1}{3}$  & no & $1$  & $0$ & $0$   \\ \hline
  
	$E_1$ & $0$ & $0$ & $\alpha<0,\beta< 0$ and $w_d<-\frac{1}{3}$ & $w_d<-1$ & $w_d<-\frac{1}{3}$ & $0$ & $1$ & $w_d$ \\ \hline
        
       $E_2$ & $1$ & $0$ & $\alpha<0,\beta <0 $ and $w_d<-\frac{1}{3}$ & unstable & no & $1$ & $0$ & $0$ \\ \hline
       
       $E_3$ & $\frac{1+w_d}{w_d}$ & $ \frac{3 w_d (1+w_d)}{3 {w_d}^2 + (\alpha-\beta+3) w_d -\beta}$ & $ \beta<\alpha<0 $ with $\frac{\beta}{\alpha-\beta}\leq w_d \leq -1$ &      & yes & $\frac{1+w_d}{w_d}$ & $-\frac{1}{w_d}$ & $-1$ \\
       
         &  &  & and & unstable &   &   &   &   \\ 
         
         &  &  & $\alpha<\beta<0$ with $w_d\leq -1$ &  &  &  &  & 
          \\ \hline
       
	\end{tabular}
	\caption{The critical points, their existence, stability, and the values of the cosmological parameters evaluated at those points for the interacting scenario driven by the interaction function $Q_{\rm V} = \Gamma_d \rho_d -\Gamma_{cd} \frac{\rho_c\rho_d}{\rho_c + \rho_d}$ of eqn. (\ref{model5})  are summarized.  }
	\label{table-model-V}
\end{center}
\end{table*}

\begin{table}[]
    \centering
    \resizebox{0.98\textwidth}{!}{%
    \begin{tabular}{|c|c|c|c|c|c|c|}
    
    \hline
    \multicolumn{7}{|c|}{Constant $w_d$}\\
    \hline
     &      & \multicolumn{3}{|c|}{Stability of Critical Points} & Deceleration  &     \\ \cline{3-5} 
     Interaction Models & Critical Points & Attractor & Global Attractor & Repeller & Parameter ($q$) & Figures    \\ \hline

     &       &       &      &     &  $q(A_0)=\frac{1}{2}\left(1+\frac{3w_d}{2}\right)<0 $    &     \\    
     
     &       &      & $A_0$ is a global attractor &  $A_0: w_d\leq-2$, $\gamma>0$    & if $w_d<-2/3$; &      \\

     $Q_{\rm I}=\Gamma(\rho_c-\rho_d)$ & $A_0$, $A_1$ & $A_0$, $A_1$ are attractors & for $w_d\geq-1$ and $\gamma>0$; &  $A_1: w_d\geq-1$, $\gamma>0$    & $q(A_1)=\frac{1}{2}(1+3w_d)<0$  & Figs. \ref{fig1:model-I}, \ref{fig2:model-I}, \ref{fig3:model-I}, \ref{fig:model-I-wd-less-than-minus-2}.     \\

     & $A_2$, $A_3$ & if $-2<w_d<-1$ and $\gamma>0$  & $A_1$ is a global attractor for & $A_2:$ always ($\gamma>0$)     &   if $w_d<-1/3$;  &         \\

     &      &       &  $w_d\leq-2$ and $\gamma>0$ &  $A_3: -2\leq w_d \leq -1$, $\gamma>0$    & $q(A_2)=1/2$; $q(A_3)=-1$ &         \\ 
     
     &        &       &      &       &         &        \\  \hline


       &         &       &         &  For $\alpha>0$ and $\beta>0$      &  $q(B_0)=\frac{1}{2}\left(1+\frac{3\alpha w_d}{\alpha+\beta}\right)<0$     &       \\

     $Q_{\rm II}=\Gamma_c\rho_c-\Gamma_d\rho_d$ & $B_0$, $B_1$  & $B_0$, $B_1$ are attractors  & $B_0$ is a global attractor  &  $B_0: w_d \leq -\left(1+\frac{\beta}{\alpha}\right)$    & if $w_d<-\frac{1}{3}\left(1+\frac{\beta}{\alpha}\right)$; &           \\

     & $B_2$, $B_3$  & for $-\left(1+\frac{\beta}{\alpha}\right)<w_d<-1$,  & for $w_d\geq-1$, $\alpha>0$ and $\beta>0$; &  $B_1: w_d\geq-1$      & $q(B_1)=\frac{1}{2}(1+3w_d)<0$ &  Figs. \ref{fig5}, \ref{fig6}, 
     \ref{evo-plot-model2}, \ref{fig7}.    \\

     &       &  $\alpha>0$ and $\beta>0$     & $B_1$ is a global attractor for   &  $B_2:$ always     & if $w_d<-1/3$; &   
      \\ 
     
     &        &       &   $w_d\leq-\left(1+\frac{\beta}{\alpha}\right)$, $\alpha>0$ and $\beta>0$    & $B_3: -\left(1+\frac{\beta}{\alpha}\right)\leq w_d \leq-1$   &  $q(B_2)=1/2$; $q(B_3)=-1$  &        \\ 
     
     &       &        &       &        &         &        \\      \hline


      &       &       &        & For $\gamma>0$      &   $q(C_0)=\frac{1}{2}\left(1+\frac{3w_d(3-\sqrt{5})}{2}\right)<0$   &         \\

     $Q_{\rm III}=\Gamma\left( \rho_c-\rho_d-\frac{\rho_c\rho_d}{\rho_c+\rho_d} \right)$ & $C_0$, $C_1$  & $C_0$, $C_1$ are attractors &  $C_0$ is a global attractor & $C_0: w_d\leq-(3+\sqrt{5})/2$     &  if $w_d<-(3+\sqrt{5})/6$; &         \\

     & $C_2$, $C_3$  & for $-(3+\sqrt{5})/2<w_d<-1$   & for $w_d\geq-1$ and $\gamma>0$;  &  $C_1: w_d\geq-1$     & $q(C_1)=\frac{1}{2}(1+3w_d)<0$ & Figs. \ref{fig9}, \ref{fig10}, \ref{evo-plot-model3}, \ref{fig11}.         \\

     &       &  and $\gamma>0$     & $C_1$ is a global attractor   &  $C_2:$ always    & if $w_d<-1/3$; &         \\ 
     
     &       &       &   for $w_d\leq-(3+\sqrt{5})/2$     &  $C_3: -(3+\sqrt{5})/2\leq w_d\leq-1$      &  $q(C_2)=1/2$; $q(C_3)=-1$  &         \\ 
     
     &        &       &        &        &          &         \\     \hline
     
     
     &      &        &         & For $\beta>\alpha>0$ &        &        \\

    $Q_{\rm IV}=\Gamma_c\rho_c-\Gamma_{cd}\frac{\rho_c\rho_d}{\rho_c+\rho_d}$ & $D_0$, $D_{00}$,  & $D_0$ is an attractor  & $D_0$ is a global attractor & $D_0: w_d\leq-\beta/\alpha$ & $q(D_0)=\frac{1}{2}\left(1+\frac{3\alpha w_d}{\beta}\right)<0$ &          \\

     &   $D_1$, $D_2$,    & if $-{\beta}/{\alpha}<w_d\leq-1$   & if $w_d>-1$ and $\beta>\alpha>0$; & $D_{00}: w_d\ne-1$       &  if $w_d<-\frac{\beta}{3\alpha}$; & Figs. \ref{fig13}, \ref{fig14}, \ref{evo-plot-model4}, \ref{fig17}.        \\

     &  $D_3$     &  and $\beta>\alpha>0$; & $D_1$ is a global attractor & $D_1: w_d>-1$  & $q(D_{00})=\frac{1}{2}(1+3w_d)<0$ &         \\

     &       &  $D_1$ is an attractor    & if $w_d\leq-\beta/\alpha$ with $\beta>\alpha>0$  &  $D_2:$ always  & if $w_d<-1/3$; &         \\

     &       &  if $-\beta/\alpha<w_d<-1$      & and if $w_d<-1$ with $\alpha>\beta>0$; & $D_3:-\beta/\alpha\leq w_d<-1$  &  $q(D_1)=\frac{1}{2}(1+3w_d)<0$     &        \\

     &       &  and $\beta>\alpha>0$     & $D_{00}$ is a global attractor   & For $\alpha>\beta>0$  &  if $w_d<-1/3$;     &       \\

     &      &       & if $w_d>-1$ and $\alpha>\beta>0$     &  $D_{00}: w_d<-1$     &  $q(D_2)=1/2$;       &          \\
     
     &       &       &      &  $D_1: w_d>-1$      &   $q(D_3)=-1$      &          \\
     
     &       &        &      &  $D_2:$ always      &         &      \\
     
     &        &         &       &       &         &       \\    \hline


     &       &        &        & For $\beta<\alpha<0$  &       &            \\

    $Q_{\rm V}=\Gamma_d\rho_d-\Gamma_{cd}\frac{\rho_c\rho_d}{\rho_c+\rho_d}$ &  $E_0$, $E_{00}$   & $E_0$, $E_1$ are attractors  & $E_0$ is a global attractor  & $E_0: w_d\leq \beta/(\alpha-\beta)$  & $q(E_0)=\frac{1}{2}\left(1+3w_d(1-\frac{\alpha}{\beta})\right)<0$ &          \\

     & $E_1$, $E_2$,  & if $\beta/(\alpha-\beta)<w_d<-1$  &if $w_d\geq-1$ with $\beta<\alpha<0$;  &  $E_{00}:$ always  & if $w_d<-\frac{1}{3}\frac{\beta}{\beta-\alpha}$; & Figs. \ref{fig18}, \ref{fig19}, \ref{evo-plot-model5}.          \\

     &   $E_3$    & with $\beta<\alpha<0$;  & $E_1$ is a global attractor  & $E_1: w_d\geq-1$       & $q(E_{00})=1/2$; &         \\

     &       & $E_{00}$, $E_1$ are attractors  & if $w_d\leq\beta/(\alpha-\beta)$ and $\beta<\alpha<0$;  &  $E_2:$ always      & $q(E_1)=\frac{1}{2}(1+3w_d)<0$ &   
           \\

     &       &   if $w_d<-1$ with $\alpha<\beta<0$     & $E_{00}$ is a global attractor  & $E_3: \beta/(\alpha-\beta)\leq w_d\leq-1 $ &   if $w_d<-1/3$;     &        \\

     &       &         &   if $w_d\geq-1$ with $\alpha<\beta<0$     &  For $\alpha<\beta<0$     &   $q(E_2)=1/2$;       &        \\ 

     &       &         &        &  $E_1: w_d\geq-1$  &   $q(E_3)=-1$      &      \\

     &       &         &        &   $E_2:$ always      &         &      \\

     &        &         &       &  $E_3: w_d\leq-1$       &         &      \\    
     
     &        &          &       &        &          &     \\   \hline

     \multicolumn{7}{|c|}{Dynamical $w_d$}\\
    \hline

      &      & \multicolumn{3}{|c|}{Stability of Critical Points} &  Deceleration     &     \\ \cline{3-5} 
     Interaction Models & Critical Points & Attractor & Global Attractor & Repeller & Parameter ($q$) & Figures    \\ \hline

     &       &       &      &     &      &     \\    
     
     &       &      &      &  $\bar{A}_1: \nu<0$ and $\gamma>0$    & $q(\bar{A}_0)=-1/4$; &      \\

     $Q_{\rm I}=\Gamma(\rho_c-\rho_d)$ & $\bar{A}_0$, $\bar{A}_1$ &  $\bar{A}_0$, $\bar{A}_1$ are attractors & $\bar{A}_0$ is a global attractor &  $\bar{A}_2:$ always ($\gamma>0$)    & $q(\bar{A}_1)\longrightarrow-\infty$ if $\nu>0$ & Figs. \ref{fig1:model-I}, \ref{fig2:model-I}.     \\

     &   $\bar{A}_2$, $\bar{A}_3$    &  if $\nu>0$ and $\gamma>0$      &   if $\nu<0$ and $\gamma>0$     &  $\bar{A}_3: \nu>0$ and $\gamma>0$ &  and $q(\bar{A}_1)\longrightarrow+\infty$ if $\nu<0$; &          \\

     &      &       &        &          &  $q(\bar{A}_2)=1/2$; $q(\bar{A}_3)=-1$  &         \\ 
     
     &        &       &      &       &         &        \\  \hline


       &         &       &       &         & $q(\bar{B}_0)=\frac{1}{2}\left(\frac{\beta-2\alpha}{\alpha+\beta}\right)<0$ &         \\

     $Q_{\rm II}=\Gamma_c\rho_c-\Gamma_d\rho_d$ & $\bar{B}_0$, $\bar{B}_1$  & $\bar{B}_0$, $\bar{B}_1$ are attractors  & $\bar{B}_0$ is a global attractor  &  $\bar{B}_1: \nu<0$, $\alpha>0$ and $\beta>0$    & if $\beta<2\alpha$; &  Figs. \ref{fig5}, \ref{fig6}.      \\

     & $\bar{B}_2$, $\bar{B}_3$  & if $\nu>0$, $\alpha>0$ and $\beta>0$  &  if $\nu<0$, $\alpha>0$ and $\beta>0$     &  $\bar{B}_2:$ always ($\alpha>0$, $\beta>0$)     & $q(\bar{B}_1)\longrightarrow-\infty$ if $\nu>0$ &       \\

     &       &       &     &  $\bar{B}_3:$ $\nu>0$, $\alpha>0$ and $\beta>0$     & and $q(\bar{B}_1)\longrightarrow+\infty$ if $\nu<0$; &   
      \\  
     
     &       &        &       &        &   $q(\bar{B}_2)=1/2$; $q(\bar{B}_3)=-1$      &        \\      

    &       &      &      &         &        &    \\ \hline  

      &       &       &       &        & $q(\bar{C}_0)= \frac{3\sqrt{5}-7}{4}<0$; &         \\

     $Q_{\rm III}=\Gamma\left( \rho_c-\rho_d-\frac{\rho_c\rho_d}{\rho_c+\rho_d} \right)$ & $\bar{C}_0$, $\bar{C}_1$  & $\bar{C}_0$, $\bar{C}_1$ are attractors & $\bar{C}_0$ is a global attractor & $\bar{C}_1:$ $\nu<0$ and $\gamma>0$     & $q(\bar{C}_1)\longrightarrow-\infty$ if $\nu>0$  &  Figs. \ref{fig9}, \ref{fig10}.      \\

     & $\bar{C}_2$, $\bar{C}_3$  & if $\nu>0$ and $\gamma>0$   &  if $\nu<0$ and $\gamma>0$     &  $\bar{C}_2:$ always ($\gamma>0$)     & and $q(\bar{C}_1)\longrightarrow+\infty$ if $\nu<0$;  &           \\

     &       &        &        &  $\bar{C}_3:$ $\nu>0$ and $\gamma>0$      & $q(\bar{C}_2)=1/2$; $q(\bar{C}_3)=-1$  &         \\ 
     
     &        &       &        &        &          &         \\     \hline
     
     
     &      &        &        & $\bar{D}_{00}: \nu\neq 0$ and $\beta>\alpha>0$ &  $q(\bar{D}_0)=\frac{\beta-3\alpha}{2\beta}<0$     &        \\

    $Q_{\rm IV}=\Gamma_c\rho_c-\Gamma_{cd}\frac{\rho_c\rho_d}{\rho_c+\rho_d}$ & $\bar{D}_0$, $\bar{D}_{00}$,  & $\bar{D}_0$, $\bar{D}_1$ are attractors  & $\bar{D}_0$ is a global attractor & $\bar{D}_1:$ $\nu<0$ and $\beta>\alpha>0$ & if $\beta<3\alpha$; &  Figs. \ref{fig13}, \ref{fig14}, \ref{fig17}.       \\

     &   $\bar{D}_1$, $\bar{D}_2$,    &  if $\nu>0$ and $\beta>\alpha>0$    & if $\nu<0$ and $\beta>\alpha>0$; & $\bar{D}_2:$ always ($\beta>\alpha>0$) & $q(\bar{D}_{00})=-1$; &          \\

     &  $\bar{D}_3$     &        & $\bar{D}_{00}$ is a global attractor & $\bar{D}_3:$ $\nu>0$ and $\beta>\alpha>0$  & $q(\bar{D}_1)\longrightarrow-\infty$ if $\nu>0$ &          \\

     &       &         & if $\nu<0$ and $\alpha>\beta>0$;  &  $\bar{D}_{00}:$ $\nu>0$ and $\alpha>\beta>0$  & and $q(\bar{D}_1)\longrightarrow+\infty$ if $\nu<0$; &         \\

     &       &         & $\bar{D}_1$ is a global attractor & $\bar{D}_1:$ $\nu<0$ and $\alpha>\beta>0$  &  $q(\bar{D}_2)=1/2$;     &        \\

     &       &       &  if $\nu>0$ and $\alpha>\beta>0$       & $\bar{D}_2:$ always ($\alpha>\beta>0$)  &   $q(\bar{D}_3)=-1$    &       \\
     
     &        &         &       &       &         &       \\    \hline


     &       &       &      & $\bar{E}_{00}:$ always ($\beta<\alpha<0$)  &   $q(\bar{E}_0)=\frac{3\alpha-2\beta}{2\beta}<0$    &            \\

    $Q_{\rm V}=\Gamma_d\rho_d-\Gamma_{cd}\frac{\rho_c\rho_d}{\rho_c+\rho_d}$ &  $\bar{E}_0$, $\bar{E}_{00}$   &  $\bar{E}_0$, $\bar{E}_1$ are attractors    &  $\bar{E}_0$ is a global attractor  & $\bar{E}_1:$ $\nu<0$ with $\beta<\alpha<0$  & if $3\alpha>2\beta$; &  Figs. \ref{fig18}, \ref{fig19}.       \\

     & $\bar{E}_1$, $\bar{E}_2$,  & if $\nu>0$ with $\beta<\alpha<0$; & if $\nu<0$ with $\beta<\alpha<0$;  &  $\bar{E}_2:$ always ($\beta<\alpha<0$)  & $q(\bar{E}_{00})=1/2$; &            \\

     &   $\bar{E}_3$    & $\bar{E}_{00}$, $\bar{E}_1$ are attractors  & $\bar{E}_{00}$ is a global attractor  & $\bar{E}_3:$ $\nu>0$ with $\beta<\alpha<0$   & $q(\bar{E}_1)\longrightarrow-\infty $ if $\nu>0$ &         \\

     &       &   if $\nu>0$ with $\alpha<\beta<0$     &  if $\nu<0$ with $\alpha<\beta<0$      &  $\bar{E}_1:$ $\nu<0$ with $\alpha<\beta<0$      & and $q(\bar{E}_1)\longrightarrow+\infty $ if $\nu<0$; &   
           \\

     &       &        &           & $\bar{E}_2:$ always ($\alpha<\beta<0$) &  $q(\bar{E}_2)=1/2$;  &        \\

     &       &         &        &  $\bar{E}_3:$ $\nu>0$ with $\alpha<\beta<0$     &  $q(\bar{E}_3)=-1$        &        \\   
     
     &        &          &       &        &          &     \\   \hline

    \end{tabular}%
    }
    \caption{Summary table describing all the interaction functions, the critical points and their nature for both constant and dynamical DE equation of state. }
    \label{tab:summary}
\end{table}

\subsubsection{Dynamical $w_d$}

\noindent  We work with the same dynamical $w_d$ of eqn. (\ref{dynamical-eos-special}) for which the autonomous system (\ref{autonomous-system-model-V}) takes the form:  
\begin{eqnarray}\label{autonomous-model-V-dyn-w1}
    \left\{\begin{array}{ccc}
      x'   &=& -\left(\frac{z}{1-z}\right)(1-x)(\alpha-\beta x)-3x(1-x)\left( 1+\nu \frac{(1-z)^2(1-x)}{z^2}\right),  \\
     z'    &=& \frac{3}{2}(1-z)z\left(x-\nu \frac{(1-z)^2(1-x)^2}{z^2}\right),
    \end{array}\right.
\end{eqnarray}
 where $\nu=\frac{3AH_0^2}{\kappa^2}$. Regularizing the autonomous system (\ref{autonomous-model-V-dyn-w1}), we get
 \begin{eqnarray}\label{reg-autonomous-model-V-dyn-w1}
    \left\{\begin{array}{ccc}
      x'   &=& -{z^3}(1-x)(\alpha-\beta x)-3(1-z)x(1-x)\left( z^2+\nu (1-z)^2(1-x)\right),  \\
     z'    &=& \frac{3}{2}(1-z)^2z\left(x z^2-\nu (1-z)^2(1-x)^2\right),
    \end{array}\right.
\end{eqnarray}
where the regularization amounts to the result that the autonomous systems (\ref{autonomous-model-V-dyn-w1}) and (\ref{reg-autonomous-model-V-dyn-w1}) are topologically equivalent. The critical points of the system (\ref{reg-autonomous-model-V-dyn-w1}) are 

\begin{itemize}
    \item $\bar{E}_0  = \left(\frac{\alpha}{\beta}, 1\right)$, \quad $\bar{E}_{00}  = (1, 1)$, \quad $\bar{E}_1  = (0, 0)$, \quad $\bar{E}_2 =(1, 0)$, \quad $S =\left\{\left(x_c, \frac{3x_c}{3x_c-(1-x_c)(\alpha-\beta x_c)} \right) \right \},$
\end{itemize}
where $S$ represents the set of critical points in which 
$x_c$ is a root of $\Psi(x) \equiv 9x^3-\nu (1-x)^4 (\alpha-\beta x)^2 = 0$.\footnote{Note that $\Psi(x)$ is obtained from the following two nullclines: 
\begin{align}
 x z^2 - \nu (1-z)^2 (1-x)^2  = 0,
 \\ -z^3 (1-x) [\alpha -\beta x] -3x (1-x) (1-z)\left[ z^2 + \nu (1-z)^2 (1-x)\right] = 0 .
    \end{align}}
As $\Psi(x)$ represents a six degree equation in $x$, therefore, $S$ may contain maximum six critical points.  Now we see that $\Psi(0) = -\nu \alpha^2 <0$ (for $\nu > 0$) and $\Psi(1)>0$, therefore, from the Bolzano's theorem \cite{Apostol:105425},  $\Psi(x)$ has at least one real root in the interval $(0,1)$. For real and physically meaningful critical points in our domain $R$, the $z$ component of the critical point i.e., $z_c=\frac{3x_c}{3x_c-(1-x_c)(\alpha-\beta x_c)}$ must belong to $[0,1]$. Thus, for any $0\leq x_c\leq 1$, the criterion $0\leq \frac{3x_c}{3x_c-(1-x_c)(\alpha-\beta x_c)}\leq 1$ leads to the condition that $(\alpha-\beta x_c) \leq 0$. Now, since $\alpha <0$, $\beta<0$, we consider the following cases: 

\begin{itemize}
    \item $\beta<\alpha<0$: In this case, we have $0\leq x_c \leq \alpha/\beta$. We also see that $\Psi(\alpha/\beta) =9\left(\alpha/\beta \right)^3>0$. Consequently, we conclude from the Bolzano's theorem \cite{Apostol:105425} that there is at least one root of $\Psi(x)$ in $\left(0, \alpha/\beta \right)$.  Now, from the derivative of $\Psi (x)$,
    \begin{align}\label{derivative-Psi-model-V}
        \Psi'(x)  &= -6{\beta}^2\nu (x-1)^3 \left(x-\frac{\alpha}{\beta} \right) \left(x-\frac{2\alpha+\beta}{3\beta} \right)+27x^2,
    \end{align}
  we notice that $\Psi'(x) > 0$ for $x\in \left(0,\alpha/\beta \right)$. Hence, the function $\Psi(x)$ is strictly increasing in $\left(0, \alpha/\beta \right)$ which finally concludes that there is only one root of $\Psi (x)$ in $\left(0, \alpha/\beta \right)$. Therefore, the set of critical points $S$ contains only one critical point and for convenience
  we label this critical point as $\bar{E}_3$. In this case, the point $\bar{E}_3$ qualitatively behaves like the point $E_3$ which is described earlier and correspondingly, the phase portrait looks same as the right plot of Fig. \ref{fig19}. 

  On the other hand, for $\nu<0$, following the earlier arguments, we can show that $\Psi(x)$ has no root in $(0,1)$. This shows that for $\nu <0$, we have only four critical points: $\bar{E}_0$, $\bar{E}_{00}$, $\bar{E}_1$, and $\bar{E}_2$ and the phase plot is same as the left plot of Fig. \ref{fig18}. 

  \item $\alpha<\beta<0$: We already know that $x_c$ satisfies the inequality $\alpha-\beta x_c \leq 0$, but in contrary to the earlier case, in this parameter space, we have  $\alpha/\beta>1$ and consequently, the critical point $\bar{E}_0$ does not belong to the physical domain $R$. We now investigate the number of real roots of $\Psi(x)$  in the interval $[0,1]$ for $\nu>0$ because for $\nu <0$, as already commented, there is no root of $\Psi (x)$ in $[0, 1]$. Here, we see that $\frac{2\alpha+\beta}{3\beta} > 1$.\footnote{One can check that $ \frac{2\alpha+\beta}{3\beta} = \frac{2}{3}(\frac{\alpha}{\beta})+ \frac{1}{3} > \frac{2}{3}+ \frac{1}{3}=1$ (since $\alpha/\beta > 1$). } Now, looking at the expression for $\Psi'(x)$ in eqn. (\ref{derivative-Psi-model-V}), we see that 
    \begin{itemize}
        \item $(x-1)<0$ as $x \in (0, 1)$,

         \item $\left(x-\frac{\alpha}{\beta} \right)<0$ in $(0, 1)$ since $\frac{\alpha}{\beta} > 1$,
        \item $\left(x-\frac{2\alpha+\beta}{3\beta} \right)<0$ in $(0, 1)$ as $\frac{2\alpha+\beta}{3\beta} > 1$, 
    \end{itemize}
    therefore, $\Psi'(x)$ is always positive in $(0,1)$. Thus, $\Psi(x)$ being a strictly increasing function in $(0,1)$ has only one root in $(0,1)$. Hence, the set $S$ contains only one critical point in the domain $R$. The phase space stability analysis of this critical point is same as that of the critical point $E_3$ and the phase plot is same as the left plot of Fig. \ref{fig19}. 

    As before, the polynomial $\Psi(x)$ has no real root in the interval $[0,1]$ for $\nu<0$. Therefore, in this case, we will have only three critical points: $\bar{E}_{00}$, $\bar{E}_1$, and $\bar{E}_2$, and the phase plot will be similar to the right plot of Fig. \ref{fig18}.

\end{itemize}

\section{Summary and concluding remarks}
\label{sec-conclusion}

\noindent Cosmology with non-gravitational interaction between DM and DE is the theme of this work. This particular theory, according to the existing records, has occupied a very decent place in the list of alternative cosmological models beyond $\Lambda$CDM. In this article we have raised some important questions regarding some not so usual constraints on the interaction functions and  performed a detailed phase space analysis of the interacting scenarios featuring some novel qualities that distinguish with the existing works in this direction.  
According to the existing records in the literature, in almost every interacting scenarios, some common (but not so natural) assumptions are considered, such as, the flow of energy should be either from DE to DM or in the reverse direction, that means either DE will be gainer or DM will be gainer. The question arises, why such unidirectional property of the interaction function should be obeyed given the fact that the nature of the interaction function is still an open question to the astrophysics and the cosmology community? On the other hand, there are ceaseless debates on the choice of the interaction functions $-$ whether the interaction functions should involve the (global) expansion rate explicitly or not. 
Moreover, in the context of the DE fluid, should we consider its equation of state to be dynamical or constant?

Considering these debates, in this work we have considered the following set-up: i) the assumption of  unidirectional interaction functions have been generalized by means of some sign shifting interaction functions which recover the unidirectional interaction functions as a special case, and thus, in this new picture of interacting dynamics, we allow the bidirectional energy flow; ii) the interaction functions do not depend on the external parameters of the universe, rather, they depend on the intrinsic nature of the dark components, and hence, they are expected to offer inherent nature of the dark sector at the fundamental level, and in addition, iii) along with the constant DE equation of state, we have considered a dynamical parametrization of the DE equation of state belonging to the class (\ref{dynamical-eos}) which adds a new ingredient in this context, and, so far we are aware of the existing literature, this is the first time we are reporting such analysis.

We begin the study by considering a very simple but elegant linear interaction function $Q_{\rm I} = \Gamma (\rho_c - \rho_d)$ of (\ref{model1}) and then considered its linear and nonlinear extensions in terms of other interaction functions given in eqns.  (\ref{model2}), (\ref{model3}), (\ref{model4}),  (\ref{model5}). With the choice of suitable dimensionless variables (this is very crucial in the analysis, since for some specific choices of the variables, one may not obtain all the critical points of the system), we have obtained all the critical points of the interacting scenarios. 
The detailed analyses of the interacting scenarios for both constant and dynamical $w_d$ are described in sections \ref{subsec-modelI}, \ref{subsec-modelII}, \ref{subsec-modelIII}, \ref{subsec-modelIV}, \ref{subsec-modelV}. Tables \ref{table-model-I}, \ref{table-model-II}, \ref{table-model-III}, \ref{table-model-IV}, \ref{table-model-V} summarize the critical points, their existence, stability and the values of the key cosmological parameters and 
in Figs. \ref{fig1:model-I} $-$  \ref{evo-plot-model5}, we have shown the nature of the critical points and the evolution of some key cosmological parameters.  For the sake of convenience, an overall summary of results extracted out of all the sign shifting interacting scenarios is given in Table~\ref{tab:summary} where mainly the nature of the critical points for the present sign shifting interacting scenarios has been shown for different nature of the DE equation of state.

Focusing on the constant DE equation of state, $w_d$, we found that, each sign shifting interacting scenario admits a variety of critical points which are qualitatively different, namely, the matter dominated  critical point which is unstable in nature; late time stable attractors corresponding to an accelerating expansion of the universe in which one stable attractor is completely DE dominated (i.e. $\Omega_d = 1$) and in one attractor both DE and DM exist and hence this attractor is physically more interesting according to the present observational results. 
Moreover, we found that all the sign shifting interacting scenarios also admit global attractors for different regions of the DE equation of state, namely, $w_d > -1$ (non-phantom), $w_d =-1$ (cosmological constant) and $w_d <-1$ (phantom), see Table~\ref{tab:summary}.  
At this point, it is important to mention that the sign shifting nature of the interaction functions can affect the space of critical points. In particularly, there is a connection between the late time stable attractors and the present sign shifting interaction functions because if this sign shifting nature of the present interaction functions is replaced by the unidirectional interaction functions, that means when the transfer of energy between the dark components is restricted to only in one direction\footnote{If the interaction functions $Q _{\rm I} = \Gamma (\rho_c - \rho_d)$ and $Q_{\rm III} = \Gamma (\rho_c - \rho_d - \frac{\rho_c \rho_d}{\rho_c + \rho_d})$, are
replaced by $\widetilde{Q}_{\rm I} = \Gamma (\rho_c + \rho_d)$ and $\widetilde{Q}_{\rm III} = \Gamma (\rho_c + \rho_d + \frac{\rho_c \rho_d}{\rho_c + \rho_d})$ respectively (note that $\widetilde{Q}_{\rm I}$ and $\widetilde{Q}_{\rm III}$ represent unidirectional interaction functions), then the critical points $A_0$, $C_0$ do not appear in these new unidirectional interacting scenarios.  In a similar fashion, for $Q_{\rm II}$, $Q_{\rm IV}$ and $Q_{\rm V}$, if we impose that the coupling parameters will have the opposite signs instead of the same signs which is essential for the sign shifting nature, then the late time stable attractors  $B_0$, $D_0$ and $E_0$ do not appear in the respective unidirectional interacting scenario.  } then the late time stable attractors, namely, $A_0$, $B_0$, $C_0$, $D_0$ and $E_0$ do not appear within these unidirectional interaction functions. However, in the proposed sign shifting interaction models, $A_0$, $B_0$, $C_0$, $D_0$ and $E_0$ appear at the transitional point where $Q (\rho_c, \rho_d)$ changes its sign. It is an interesting feature of the proposed sign shifting  interacting models and it is a subject for further investigations.

The case with dynamical DE equation of state presents a more general interacting scenario. The present choice of $w_d$ (eqn. (\ref{dynamical-eos-special})) covers both the phantom (for $A> 0$, equivalently, $\nu >0$) and non-phantom (for $A< 0$, equivalently, $\nu <0$) regimes. In terms of the number of  critical points, each sign shifting interacting model with dynamical phantom case (i.e. $\nu >0$) has one extra critical point compared to the corresponding sign shifting interacting model with dynamical quintessence case ($\nu <0$). Again, similar to the constant $w_d$ case, here too, we observe that if the sign shifting nature of the interaction functions is replaced by the unidirectional interaction functions, the late time stable attractors, namely, $\bar{A}_0$, $\bar{B}_0$, $\bar{C}_0$, $\bar{D}_0$ and $\bar{E}_0$ do not 
appear in this case, but in the context of the proposed sign shifting interaction models these late time stable attractors arise at the transitional point, that means where $Q (\rho_c, \rho_d)$ changes its sign. Specifically, we have the following observations:  

\begin{enumerate}
\item  Dynamical phantom:  In all the sign shifting interacting dynamical phantom scenarios, we find one matter dominated era (unstable in nature) representing a decelerating phase, late time stable attractors in which one attractor is completely DE dominated ($\Omega_d = 1$) and other attractor allows the concurrence of DE and DM. Moreover, we noticed that only Model IV in this series admits one global attractor $\bar{D}_1$ provided that the dimensionless coupling parameters satisfy $\alpha > \beta >0$. According to the existing literature on the interacting dynamical phantom scenarios \cite{Guo:2004xx,Gumjudpai:2005ry,Chen:2008ft,Shahalam:2017fqt}
concurrent existence of the matter dominated, only DE dominated ($\Omega_d =1$), and the co-existence of DE and DM ($\Omega_d \neq 0$, $\Omega_c \neq 0$), as we observed within  the context of present sign shifting interacting models, is very rare. For example, 
even though  some specific interacting models exhibit the matter dominated phase \cite{Chen:2008ft,Shahalam:2017fqt} but the simultaneous occurrence of the DE dominated stable attractor ($\Omega_d =1$) and the stable late time scaling attractor corresponding to an accelerating phase of the universe has not been found.

\item Dynamical quintessence: We find that all the scenarios in this category admit the matter dominated phase which is unstable in nature and it corresponds to a past decelerating phase. But, unlike in the phantom interacting scenario, here only one late time stable attractor is allowed which is global in nature (see Table~\ref{tab:summary}) and this critical point allows the existence of both DE and DM. The existence of the matter dominated phase within the present sign shifting models  is interesting because such phase is not so common in a variety of interaction models when $w_d$ lies in the quintessence regime, see for instance  \cite{Gumjudpai:2005ry,Boehmer:2008av,Chen:2008pz,Boehmer:2009tk,Bernardi:2016xmb}. 

\end{enumerate}

Based on the outcomes of the present article, it is evident that the sign shifting interaction models are quite appealing.  The results further emphasize that there should not have any particular reason to prefer only the unidirectional interaction functions in the context of interacting DE,  rather, the bidirectional interaction functions are quite promising and they deserve further attention.  
Specifically, the analysis with dynamical $w_d$ within these interaction models is very promising but such analysis is rare in the literature.

\section{Acknowledgments}
The authors thank the referees for their time to read our article and for giving some useful comments that helped us to improve the article.  
SH acknowledges the financial support from the University Grants Commission (UGC), Govt. of India (NTA Ref. No: 201610019097). JdH is supported by the Spanish grant 
PID2021-123903NB-I00 funded by 
MCIN/AEI/10.13039/501100011033 and by ``ERDF A way of making Europe''.  TS and SP acknowledge the financial support from the Department of Science and Technology (DST), Govt. of India under the Scheme   ``Fund for Improvement of S\&T Infrastructure (FIST)'' (File No. SR/FST/MS-I/2019/41).

\bibliography{biblio}
\end{document}